\documentclass[12pt,a4paper]{article}
\usepackage[utf8]{inputenc}
\usepackage{scalerel,stackengine,amsmath}
\usepackage{amsfonts}
\usepackage{amssymb}
\usepackage{hyphenat}
\usepackage{enumitem}
\usepackage{dsfont}
\usepackage[left=2cm,right=2cm,top=2.5cm,bottom=2.5cm]{geometry}
\usepackage{cite}
\usepackage{slashed}
\usepackage{hyperref}
\usepackage{graphicx}
\usepackage{caption}
\usepackage{float}
\usepackage{mathtools}

\usepackage{array}
\usepackage{physics}
\usepackage{float}
\usepackage{mathtools}
\usepackage{rotating}
\usepackage{amssymb}
\newcommand{\citet}{\cite} 
\usepackage{braket}
\usepackage[export]{adjustbox}
\usepackage{mathtools}
\usepackage{mathrsfs}  

\usepackage{adjustbox}
\usepackage{graphicx}
\usepackage{slashed}
\usepackage{mathrsfs}

\numberwithin{equation}{section}
\usepackage{lmodern}
\usepackage{bbold}
\usepackage{mmacells}
\usepackage[symbol]{footmisc}

\mmaDefineMathReplacement[≤]{<=}{\leq}
\mmaDefineMathReplacement[≥]{>=}{\geq}
\mmaDefineMathReplacement[≠]{!=}{\neq}
\mmaDefineMathReplacement[→]{->}{\to}[2]
\mmaDefineMathReplacement[⧴]{:>}{:\hspace{-.2em}\to}[2]
\mmaDefineMathReplacement{∉}{\notin}
\mmaDefineMathReplacement{∞}{\infty}

\mmaSet{
  morefv={gobble=2},
  linklocaluri=mma/symbol/definition:#1,
  morecellgraphics={yoffset=1.9ex}
}

\renewcommand{\b}{b}
\renewcommand{\c}{\gamma}
\renewcommand{\d}{\delta}

\newcommand{\st}{{\fontfamily{cmss}\selectfont\text{sT}}}
\newcommand{\tran}{{\fontfamily{cmss}\selectfont\text{T}}}

\begin{document}

\begin{flushright}
June 2024
\end{flushright}

\bigskip

\begin{center}
{\bf {\LARGE Supergeometric Quantum Effective Action}}
\end{center}

\vspace{5mm}

\begin{center}
{\large Viola Gattus\footnote[1]{E-mail address: {\tt
      viola.gattus@manchester.ac.uk}} and Apostolos
    Pilaftsis\footnote[2]{E-mail address: {\tt 
apostolos.pilaftsis@manchester.ac.uk}}}\\[3mm] 
{\it Department of Physics
and Astronomy, University of
Manchester,\\ Manchester M13 9PL, United Kingdom}
\end{center}
\vspace{2cm}

\centerline{\bf ABSTRACT}
\vspace{2mm}

\noindent
Supergeometric Quantum Field Theories (SG-QFTs) are theories that go beyond the standard supersymmetric framework, since they allow for general scalar-fermion field transformations on the configuration space of a supermanifold, without requiring an equality between bosonic and fermionic degrees of freedom. After  revisiting previous considerations, we extend them by calculating the one-loop effective action of minimal SG-QFTs that feature non-zero fermionic curvature in two and four spacetime dimensions. By employing an intuitive approach to the Schwinger--DeWitt heat-kernel technique and a novel field-space generalised Clifford\- algebra, we derive the ultra-violet structure of characteristic effective-field-theory (EFT) operators up to four spacetime derivatives that emerge at the one-loop order and are of physical interest${}$. Upon~minimising the impact of potential ambiguities due to the so-called multiplicative anomalies, we find that the EFT interactions resulting from the one-loop super\-geometric effective action are manifestly diffeomorphically invariant in configuration space. The extension of our approach to evaluating\- higher-loops of the super\-geometric quantum effective action is described. The emerging landscape of theoretical and phenomenological directions for further research of SG-QFTs is discussed.

\vspace{1cm}

\newpage
\tableofcontents
\newpage

\thispagestyle{empty}
\newpage

\section{Introduction}

Covariant and differential geometric methods have been playing a prominent role in driving the development of Quantum Field Theories (QFTs)~\cite{DeWitt:1967ub}. In addition to addressing issues of gauge covariance in effective\- quantum actions~\cite{Gaillard:1985uh,Pilaftsis:1996fh,Cornwall:2010upa,Binosi:2009qm}, these methods were employed to compute transition amplitudes of chiral loops in a manifestly reparametrisation invariant manner on the entire configuration space of fields and spacetime coordinates~\cite{Honerkamp:1971sh,Ecker:1972tii}. The same methods have also been utilised to formulate non-linearly realised supersymmetric theories~\cite{Alvarez-Gaume:1981exa}, as well as derive the Effective Field Theory~(EFT) limit of heterotic string theories~\cite{Dixon:1990pc}. Beyond the classical-action level, these covariant differential-geometric methods have been further developed by Vilkovisky and DeWitt~(VDW)~\cite{Vilkovisky:1984st,DeWitt:1985sg}, with the aim to address the issue of gauge-fixing parameter independence in gauge and quantum gravity theories. This issue was subsequently the subject of numerous other studies~\cite{Barvinsky:1985an,Ellicott:1987ir,Burgess:1987zi,Odintsov:1989gz}. 

More recently, related geometric effective actions were put forward as a potential solution to the so-called quantum frame problem~\cite{Kamenshchik:2014waa,Burns:2016ric,Karamitsos:2017elm} that haunts early studies of cosmic inflation. In this respect, an obstacle concerning the uniqueness of the path-integral measure of the VDW effective action was addressed beyond the tree level~\cite{Finn:2019aip,Falls:2018olk}. Finally, similar geometric techniques were employed to analyse new-physics phenomena within the framework of Effective Field Theories (EFTs) beyond the Standard Model~SM~\cite{Jenkins:2023bls, Alminawi:2023qtf, Craig:2023hhp, Assi:2023zid, Alonso:2016oah, Nagai:2019tgi, Helset:2020yio, Cohen:2021ucp, Talbert:2022unj, Helset:2022tlf, Fumagalli:2020ody, Cohen:2023ekv, Loisa:2024xuk}, also known as SMEFT~\cite{Buchmuller:1985jz, Grzadkowski:2010es, Dedes:2021abc, Alonso:2015fsp, Alonso:2023upf} (for a recent review, see~\cite{Isidori:2023pyp}).

A major theoretical difficulty that was encountered in the formulation of geometric effective actions has been the proper inclusion of fermionic fields as independent chart variables in 
the path-integral configuration space~\cite{DeWitt:1985sg}. Unlike supersymmetric theories in which fermions appear as tangent vectors in a field-space whose geometry is 
determined by their bosonic counter\-parts~\cite{Alvarez-Gaume:1981exa}, the theories that were initially presented in~\cite{Finn:2020nvn,Gattus:2023gep,Gattus:2024spj} and are of interest to us treat scalars and fermions as independent field coordinates and so allow for general scalar-field transformations on the configuration space of a supermanifold. Since fermions are Grassmannian variables, their consistent description on this more general space would necessitate extensive use of mathematical techniques from  differential {\em supergeometry} (SG) on {\em super\-manifolds}~\cite{DeWitt:2012mdz}. Hence, the theories that we will be considering here will be called Supergeometric Quantum Field Theories (SG-QFTs)~\cite{Gattus:2023gep,Gattus:2024spj}, and they should not be confused with the conventional supersymmetric theories.

In this paper we extend previous considerations and present, for the first time, analytic expressions for the one-loop effective action of minimal SG-QFT models that feature non-zero fermionic curvature. To facilitate our computations, we employ a new and more intuitive approach to the Schwinger–DeWitt heat-kernel technique 
based on the Zassenhaus formula. Furthermore, we use a novel field-space generalised Clifford algebra which enables us to derive the ultra-violet~(UV) structure of representative EFT operators up to four spacetime derivatives that arise at the one-loop level. In particular, we find EFT operators that have new characteristic\- properties, since they describe magnetic-moment-type transitions between fermions induced by a non-zero field-space curvature.
In our computations, we had to minimise the impact of potential ambiguities due to the so-called multiplicative anomalies by requiring that the well-known results be obtained in the flat field-space limit. An important finding of our study is that the EFT interactions resulting from the one-loop Super-Geometric Quantum Effective Action (SG-QEA) are manifestly diffeo\-morphically invariant in both configuration and field space. The  method that we present here is not limited to one loop, but we describe how it can be straightforwardly extended to evaluate higher-loops of the SG-QEA.

The layout of the paper is as follows. Section~\ref{sec:SGQFT} outlines basic aspects of the notion of Supergeometry in QFT. 
Specifically, Section~\ref{sec:SGQFT} provides a covariant formulation of a generic scalar-fermion SG-QFT, up to second order in spacetime derivatives of the fields. The respective Lagrangian of the SG-QFT is written down in terms of  model functions with well-defined transformation properties under the Lorentz group and field-space reparametrisations. The same section  presents our approach to deriving the {\it supermetric} that governs the SG-QFT. In Section~\ref{sec:CovInts}, we extend the old geometric  formalism of~\cite{Ecker:1972tii} which was applied to bosons only to the more general case of a SG-QFT that includes fermions.
On the basis of this formalism, we derive covariant interactions of general scalar-fermion theories, in terms of the super-Riemannian tensor in the configuration space. The covariant two-point correlation functions, along with the respective covariant propagators, are essential elements in the computation of the SG quantum effective action which will be presented in the subsequent sections. In this context, higher-order correlation functions, along with their symmetry properties in configuration space, are given in~Appendix~\ref{app:HighPointCov}.

In Section~\ref{sec:CovSFaction} we briefly review the basic features of the SG-QEA by extending the VDW formalism in order to include both bosonic and fermionic degrees of freedom. In addition, we derive a master equation~\eqref{eq:masterSGQEA} where all-orders SG-QEA can be recursively determined. Furthermore, we show how the Zassenhaus formula can be conveniently exploited,  within the context of the Schwinger–DeWitt heat-kernel technique, in order to compute the one-loop SG-QEA in  configuration space. Section~\ref{sec:CovFlatAction} gives a practical demonstration of its use by recovering known results for a minimal Yukawa theory, as well as for a more involved scalar-fermion theory, in a flat field-space at the one-loop order. Technical details pertaining to the latter case have been relegated to Appendix~\ref{app:DoubleHKM}.   We note that all our computations are performed within the Dimensional Regularisation (DR) scheme in the coordinate space, rather than in the usual momentum space. Therefore, useful formulae of the DR scheme in the coordinate space are collected in Appendix~\ref{app:DRxspace}.

In Section~\ref{sec:CovScalarCurv}, we apply our intuitive heat-kernel method to bosonic theories that exhibit non-vanishing field-space Riemannian curvature. Such a computation of the effective action gets facilitated by taking the field-space covariant heat kernel that we derive for a general frame to a local (flat) frame, and then transforming the results back to the global field-space curved frame through general covariance. This method becomes rather advantageous, since all covariant EFT operators emerge automatically in a systematic manner, unlike other more laborious approaches that utilise 't Hooft's matching procedure~\cite{tHooft:1973bhk}.  In this respect, we clarify that higher-derivative geometric EFT interactions in the configuration space can be systematically translated into EFT operators acting on the usual field space in a manifestly covariant manner.
 
In Section~\ref{sec:CovFermionAction} we calculate the one-loop effective action for a minimal SG-QFT model that future non-zero fermionic curvature. Moreover, we develop a novel Clifford algebra, where the role of the Dirac $\gamma^\mu$-matrices is assumed by field-dependent tensors, $\lambda^\mu$, and their adjugate counterparts, $\bar{\lambda}^\mu$. In fact, $\lambda^\mu$ and $\bar{\lambda}^\mu$ satisfy a field-space generalised Clifford algebra, which can be used to bosonise the fermionic effective action in a strictly covariant manner in the configuration space. In order to minimise the potential impact of the so-called multiplicative anomalies due to the bosonisation of the fermionic action, we require that the known UV divergences should be recovered in the flat field-space limit of the theory. In the same section,  we present the UV structure of typical covariant EFT operators up to four spacetime derivatives that emerge from the one-loop SG-QEA in the configuration space and could be phenomenologically interesting. Section~\ref{sec:CovFermionAction} concludes with a subsection outlining potential approaches that could enable us to go beyond the minimal supergeometric framework that we will be studying~here. Finally, Section~\ref{sec:Concl} summarises the main findings of our study and discusses new directions for future research.

\setcounter{equation}{0}
\section{Supergeometry in Quantum Field Theory}\label{sec:SGQFT}

In this section we give a short review on the theory of supermanifolds~\cite{dewitt1992supermanifolds} which is relevant to the construction of SG-QFTs that are formulated on a scalar-fermion field space~\cite{Finn:2020nvn}. 

For an SG-QFT with $N$ real scalars and $M$ Dirac fermions, one may assume that all these fields live on a supermanifold of $(N|2d M)$ dimensions in $d$ spacetime dimensions. Hence, for $d=4$ dimensions, the supermanifold chart may be represented by the $(N+8M)$-dimensional column vector
\begin{equation}
   \label{eq:PhiChart}
  \boldsymbol{\Phi}\ \equiv\ \left\{\Phi^{a}\right\}\ =\ \left(
  \begin{array}{c}\phi^{A}\\ \psi^{X}\\
  \overline{\psi}{}^{Y\:\tran}
  \end{array}
  \right)\,.
\end{equation}
Hereafter, we use lowercase Latin letters to denote configuration-space indices, whilst capitalised Latin letters are reserved for field-space indices, i.e.~$\Phi^a \equiv \Phi^A(x_A)$. For the ease of notation,  no distinction will be made between bosonic and fermionic field space indices. The nature of the index can always be inferred from the context considered.

A field reparametrisation now corresponds to a diffeomorphism on the supermanifold,
\begin{equation}
   \label{eq:PhiDiff}
    \Phi^{a}\ \rightarrow\ \widetilde{\Phi}^{a}\, =\, \widetilde{\Phi}^{a}(\boldsymbol{\Phi})\;.
\end{equation}
The above field transformation is taken to be ultralocal depending only on $\Phi^a$ and not  on any spacetime derivatives of fields, like $\partial_\mu\Phi^a$, in line with the VDW formalism~\cite{Vilkovisky:1984st,DeWitt:1985sg}. For simplicity, we will not consider possible relaxation of this restriction 
here by adopting Finslerian-type geometries in the field space \cite{Finn:2019aip,Craig:2023hhp,Alminawi:2023qtf,
Kluth:2023sey}.

Up to second order in $\partial_{\mu} \Phi^{a}$, the 
field-space diffeomorphically invariant Lagrangian $\mathcal{L}$ of a general SG-QFT may covariantly be written down as~\cite{Finn:2020nvn}
\begin{equation}
   \label{eq:LSGQFT}
    \mathcal{L}\ =\ \frac{1}{2} g^{\mu \nu} \partial_{\mu} \Phi^{A}\:_{A} k_{B}({\bf \Phi})\: \partial_{\nu} \Phi^{B}\: +\: \frac{i}{2}\, \zeta_{A}^{\mu}({\bf \Phi})\: \partial_{\mu} \Phi^{A}\: -\: U({\bf \Phi})\,.
\end{equation}
Note that $\mathcal{L}$ can be expressed in terms of the three model functions: (i)~a rank-2 field-space tensor, $_{A} k_{B}({\bf \Phi})$, (ii)~a mixed spacetime and field-space vector, $\zeta_{A}^{\mu}({\bf \Phi})$, and (iii)~$U({\bf \Phi})$,  a zero-grading scalar describing the Higgs potential and Yukawa interactions. In this minimal class of SG-QFTs under study, we assume the existence of one global (defining) frame in which $_{A} k_{B}$ vanishes when either or both indices are fermionic. The three model functions can be extracted unambiguously from the Lagrangian, according to the prescription outlined in \cite{Finn:2020nvn, Gattus:2023gep,Gattus:2024spj}.

The field-space metric of the supermanifold can be derived uniquely from the action, $S$, associated with the Lagrangian \eqref{eq:LSGQFT}. To do so, we must first construct a pure field-space vector~$\zeta_A$ to be projected out from the model function, $\zeta^{\mu}_{\:A}$. This procedure was described in great detail in~\cite{Gattus:2023gep,Gattus:2024spj}. Here, we only limit ourselves to state the relevant projection operation,
\begin{equation}
   \label{eq:projzeta}
     \zeta_{B}^{\mu} \:\:^B\left(\overleftarrow{\Sigma}_{\mu}\right)_{A}\: =\: \zeta_{A}\, ,
\end{equation}
where
\begin{equation}
   \label{eq:Sigmamu}
   \overleftarrow{\Sigma}_{\mu}  =\frac{1}{d}\left(\begin{array}{cc}
\overleftarrow{\delta}/\delta \gamma^{\mu} & 0 \\
0 & \Gamma_{\mu}
\end{array}\right)\;, 
\end{equation}
with $\Gamma^{\mu}=\text{diag}\big(\gamma^{\mu}\,, \gamma^{\mu\, \tran} \big)$. As can be seen from~\eqref{eq:projzeta}, the bosonic part of $\zeta_{A}$ can be extracted by differentiating $\zeta^{\mu}_{\:A}$ with respect to $\gamma^\mu$ (or $\sigma^\mu$). Instead, the fermionic components of $\zeta_{A}$ are obtained by appropriately contracting $\zeta^{\mu}_{\:A}$ with $\Gamma^\mu$. This projection method can be applied to a wide range of SG-QFTs, as discussed in~\cite{Gattus:2023gep,Gattus:2024spj}.

After we have determined $\zeta_A$ following the above procedure, we can then use it to construct the rank-2 field-space tensor,
\begin{equation}
  \label{eq:lambda}
   {}_{A} \lambda_{B}\ =\ \frac{1}{2}\, \Big({}_{A,}\zeta_{B}\: -\: (-1)^{A+B+AB}  \:_{B,}\zeta_{A}\Big)\; ,
\end{equation}
which is anti-supersymmetric and singular in the presence of scalar fields. Here and in the following, we employ the compact convention of index notation by DeWitt~\cite{DeWitt:2012mdz}, according to which exponents of $(-1)$ represent the grading of the respective quantities and do not participate in index summation. The grading can be~$0$ or~$1$ for commuting or anticommuting fields, respectively. Furthermore, only adjacent indices can be contracted and factors of $(-1)$ must be introduced when the position of two indices is exchanged.

The scalar contribution arising from the model function $_{A} k_{B}$ can be combined with ${}_{A} \lambda_{B}$ to form the new rank-2 field-space tensor,
\begin{equation}
   \label{eq:Lambda}
_{A} \Lambda_{B}\ =\:  _{A} k_{B} \:+\:  _{A} \lambda_{B}\; .
\end{equation}
Note that $_{A} \Lambda_{B}$ cannot play the role of the supermetric, since it is not exactly supersymmetric; it has a mixed symmetry. However, if the local form of $_{A} \Lambda_{B}$ is known, one may still be able to compute the field-space vielbeins $_A e^{\widehat{B}}$~\cite{Schwinger:1963re, Yepez:2011bw}, where a hat ($\:\widehat{}\:$) on a field-space index signifies {\em local-frame} coordinate. These can in turn be used to compute the supermanifold metric $_A G_B$ from the local field-space metric ${}_{\widehat{A}} H_{\widehat{B}}$ as follows~\cite{Finn:2020nvn}:
\begin{equation}
  \label{eq:aGb}
    _A G _B \:=\: _A e^{\widehat{M}}\:\:_{\widehat{M}} H_{\widehat{N}}\:\:^{\widehat{N}}e^{\st}_B\;,
\end{equation}
where the local field-space metric is given by
\begin{equation}
    \label{eq:aHb}
 { }_{\widehat{A}} H_{\widehat{B}} \equiv\left(\begin{array}{ccc}
{\bf 1}_{N} & 0 & 0 \\
0 & 0 & {\bf 1}_{dM}  \\
0 & -{\bf 1}_{dM} & 0 
\end{array}\right)   
\end{equation}
and the superscript $\st$ denotes the operation of supersymmetrisation. Observe that the rank-2 field-space tensor ${}_A G_B$ is qualified to be the {\it supermetric} of the field-space supermanifold, since it is supersymmetric by construction, i.e. 
\begin{equation}
 {}_A G_B\: =\: ({}_A G_B)^{\sf sT}\: =\: (-1)^{A+B+AB} {}_B G_A\;.   
\end{equation}
In configuration space, the supermetric and the field-space vielbeins are generalised as
\begin{equation}
   \label{eq:CSGmetric}
    { }_{a} G_{b}\ =\ { }_{A} G_{B}(x_A)\; \delta(x_A - x_{B})\;,\qquad
    _a e^{\hat{b}}\ =\ _A e^{\widehat{B}}(x_A) \;
    \delta(x_A - x_{\widehat{B}})\;.
\end{equation}
where $\delta(x_{A} - x_{B})$ is the $d$-dimensional $\delta$-function. The configuration-space Riemann tensor is computed from the Christoffel symbols
$\Gamma^{a}_{\:\:b c}$~\cite{DeWitt:2012mdz}, {\it viz.}
\begin{equation}
\begin{aligned}
       \label{eq:Riemann}
R^a_{\:\:b c d}\: &=\: -\,\Gamma^{a}_{\:\:b c,d}\: +\: (-1)^{c d}\: \Gamma^{a}_{\:\:b d,c}\: +\:(-1)^{c(m+b)}\: \Gamma_{\:\:m c}^{a} \Gamma_{\:\:b d}^{m}\: -\: (-1)^{d(m+b+c)}\: \Gamma_{\:\:m d}^{a} \Gamma_{\:\:b c}^{m}\; \\
&=\: R^A_{\:\:BCD}\:\delta(x_A - x_{B})\:\delta(x_A - x_{C})\:\delta(x_A - x_{D})\:.
\end{aligned}
\end{equation}

Finally, we introduce the useful operations of superdeterminant and supertrace. The entries of any $(N+M)\times (N+M)$-dimensional rank-2 field-space tensor (also known as supermatrix) may be arranged in the following block structure:
\begin{equation}
  \label{eq:Mblock}
    {\cal M}\ =\ \left(\begin{array}{cc}
        A & C \\
        D & B
    \end{array}\right),
\end{equation}
where $A$ and $B$ are $N\times N$ and $M\times M$ matrices with commuting entries, while $C$ and $D$ are $N\times M$ and $M \times N$ dimensional matrices of anticommuting numbers, respectively. For such a matrix, the superdeterminant may be defined in two equivalent ways:
\begin{equation}
   \label{eq:sdetM}
\operatorname{sdet}{\cal M}\ =\ \frac{\operatorname{det}A}{\operatorname{det}(B-DA^{-1}C)}\ =\ \frac{\operatorname{det}(A-CB^{-1}D)}{\operatorname{det}B}\; ,
\end{equation}
where $\operatorname{det}$ is the ordinary matrix determinant.

The supertrace is defined only for mixed rank-2 tensors, which have a left index as a subscript and right index as a superscript, or vice-versa. If such a mixed rank-2 tensor is represented by a matrix ${\cal M}$ with the block structure as given in~\eqref{eq:Mblock}, its supertrace is evaluated as
\begin{equation}
    \operatorname{str}{\cal M}\ =\  \operatorname{tr}A\: -\:\operatorname{tr}B\,,
\end{equation}
where $\operatorname{tr}A$ and $\operatorname{tr}B$ indicate ordinary traces of the matrices $A$ and $B$. 
The supertrace is invariant under supertransposition and cyclic permutations, in close analogy to the usual trace properties for commuting numbers. All the aforementioned definitions and identities will prove valuable for the evaluation of SG-QEAs (cf.~Section~\ref{sec:CovSFaction}).

\section{Covariant Interactions for Scalar-Fermion Theories}\label{sec:CovInts}

In this section we will extend the previous considerations in~\cite{Gattus:2023gep,Gattus:2024spj} and present the methodology for deriving two- and higher-point covariant interactions for a general SG-QFT that includes both scalars and fermions. We will also give explicit analytic results for the two lowest-point correlation functions, such as the covariant scalar-fermion propagator and the three-field interactions. Further analytic expressions for covariant higher-point vertices may be found in Appendix~\ref{app:HighPointCov}.

To derive the covariant inverse propagator, three- and higher-point vertices in a relatively compact way, we adopt methodology and notation first introduced in \cite{Ecker:1972tii}. To this end, we note that the covariant functional differentiation of $\partial_\mu \Phi^a$ with respect to $\Phi^b$ in the configuration space is given by
\begin{equation}
  \label{eq:Dmuab}
    (\partial_\mu \Phi^a)_{;b}\ =\ \big(\delta^A_{\:\:B}\:\partial_{\mu}^{(A)} +\Gamma^{A}_{\:\:BM}\:\partial_\mu\Phi^M\big)\,\delta(x_A-x_B)\ \equiv\ (D_\mu)^a_{\:\:b}\; .
\end{equation}
In the above, the semicolon (;) stands for  covariant functional differentiation in configuration space, 
$\partial_{\mu}^{(A)} \equiv \partial/\partial x^\mu_A$,
and $\Gamma^{A}_{\:\:BC}$ are the field-space Christoffel symbols.
We should observe that in this covariant formulation $(D_\mu)^a_{\:\:b}$ as defined in~\eqref{eq:Dmuab} is strictly not a differential operator, but a mixed spacetime and field-space tensor in configuration space. Nevertheless, for later convenience, we term it abusively as such in this work (see also our discussion in Section~\ref{sec:CovScalarCurv}).

In the covariant formalism of~\cite{Ecker:1972tii}, an important role plays the identity, 
\begin{equation}
   \label{eq:DmuRabc}
     (D_\mu)_{ab;c} \:=\:R_{abcm}\:\partial_\mu\Phi^m\,,
\end{equation}
where $(D_\mu)_{ab} \:=\: G_{am}\:(D_\mu)^m_{\:\:b}$. Most remarkably, we were able to prove that the Ecker--Honerkamp~(EH) identity~\eqref{eq:DmuRabc} will still hold true within a SG-QFTs in which fermionic fields are included. A detail proof of the EH identity~\eqref{eq:DmuRabc} within the context of supermanifolds is given in Appendix~\ref{app:HighPointCov}.
Furthermore, the differential operator, $(D_\mu)_{ab}$, satisfies the following (anti-)symmetric properties:
\begin{equation}
   (D_\mu)_{ab}=-(-1)^{ab}(D_\mu)_{ba}\:,\qquad\qquad
   ^a(D_\mu)_{b}=-(-1)^{b(a+1)}\:_b(D_\mu)^{a}\:.
\end{equation}
Finally, in terms of the field-space vielbeins given in~\eqref{eq:CSGmetric}, another equivalent  representation 
of~${}^a(D_\mu)_{b}$ in field-space can be obtained,
\begin{equation}
   \label{eq:DmuabVB} 
(D_\mu)^A_{\ B} = \:^A e_{\widehat{A}}\:\:(D_\mu)^{\widehat{A}}_{\ \widehat{B}}\:\:^{\widehat{B}}e^{\st}_{\:\:B}\:  
\, =\, ^A e_{\widehat{A}}\left[{}^{\widehat{A}}\delta_{\widehat{B}}\:\partial_\mu^{(\widehat{A})} + (-1)^C\:\partial_\mu \Phi^{C}\:\:\omega_{C\:\:\:\:\widehat{B}}^{\:\:\widehat{A}}\right]\:^{\widehat{B}}e^{\st}_{\:\:B}\; ,
\end{equation}
where 
$\partial^{(\widehat{A})}_\mu \equiv \partial/\partial x^\mu_{\widehat{A}}\,$, and 
$\omega_{C\:\:\:\:\widehat{B}}^{\:\:\widehat{A}}$ is the {\em spin connection} in the field space. More details on the role of the field-space spin connection are given in Appendix~\ref{app:spin_connection}. Here we only 
quote its definition in terms of field-space vielbeins and metric connection coefficients, 
\begin{equation}
     \label{eq:spinomega}
\omega_{C\:\:\:\:\widehat{B}}^{\:\:\widehat{A}}\: =\: (-1)^{C(A+\widehat{A})}\left(-\:^{\widehat{A}} e^{\st}_{\:\:A\:,C}\:+\;\:^{\widehat{A}} e^{\st}_{M}\;\Gamma^{M}_{\:\:AC}\right){}^{A}e_{\widehat{B}}\;.
\end{equation}
As we will see in Section~\ref{sec:CovScalarCurv}, the representation of~${}^A(D_\mu)_{B}$ in~\eqref{eq:DmuabVB} will prove useful to find an expression for the configuration-space heat kernel in the {\em global curved frame}.

Given the action $S$ of the Lagrangian~\eqref{eq:LSGQFT}, we can now calculate the equation of motion of the fields $\Phi^a$ (with $S_a\equiv S_{;a}$ and $U_a \equiv U_{;a}$),
\begin{equation}
   \label{eq:Sa}
    S_{a}\ =\ (\partial_\mu\Phi^m)\:_m k_n\: (D^\mu)^n_{\:\:a}\: +\: \frac{1}{2}(-1)^{an}(\partial_\mu\Phi^m)\:_m k_{n;a}\: (\partial^\mu\Phi^n)\: +\: i(-1)^a\:_a\lambda^{\mu}_{m}\:(\partial_\mu\Phi^m)\: -\: U_a\:.
\end{equation}
In~\eqref{eq:Sa}, we have followed~\cite{Gattus:2023gep,Gattus:2024spj} and defined the mixed tensor,
\begin{equation}
  \label{eq:lambdamu}
   {}_{a} \lambda^\mu_{b}\ \equiv\ \frac{1}{2}\, \Big({}_{a,}\zeta^\mu_{b}\: -\: (-1)^{a+b+a b}  \:_{\b,}\zeta^\mu_{a}\Big)\; .
\end{equation} 
This quantity is a proper spacetime vector and rank-2 field-space tensor, which turns out to be anti-supersymmetric.  
It can be deduced from $\zeta^\mu_{a}$ via ordinary functional differentiation, because the Christoffel symbols cancel.

To significantly reduce the length of our analytic expressions, we introduce the following conventions for the operations of symmetrisation and cyclic permutation of configuration-space indices: 
\begin{equation}
   \label{eq:operations}
\begin{aligned}
       [abc]&=abc+(-1)^{ab}bac+(-1)^{c(a+b)}cab \qquad\hspace{6mm}\text{(symmetrisation)}\\
    (abc)&=abc+(-1)^{a(b+c)}bca+(-1)^{c(a+b)}cab\qquad\text{(cyclic permutation)}
\end{aligned}
\end{equation}
We note that repeated indices are excluded from the operations mentioned above.
With the above conventions in mind, the covariant inverse propagator is found to be
\begin{equation}
   \label{eq:Sab}
\begin{aligned}
S_{ab}\:=\:& (-1)^{am}(D_\mu)^m_{\:\:a}\:_m k_n\:(D^\mu)^n_{\:\:b}+i(-1)^{a}\left(\:_a\lambda^{\mu}_{m}(D_\mu)^m_{\:\:b}+(-1)^{bm}\:_a\lambda^{\mu}_{m;b}\:\partial_\mu\Phi^m\right)-U_{ab}\\
& +\,\partial_\mu\Phi^m\:\left(\:_m k_n\:R^{n}_{\:\:abp}\:\partial^\mu\Phi^p+(-1)^{an}\:_m k_{n;[a}\:(D^\mu)^n_{\:\:b]}+\frac{1}{2}(-1)^{n(a+b)}\:_m k_{n;ab}\:\partial^\mu\Phi^n\right)\,.
\end{aligned}
\end{equation}
Here and in the following, we drop the semicolon (;) and adopt the abbreviation: $S_{abc\dots}\equiv S_{;abc\dots}$, for covariant functional derivatives that act on the classical action~$S$, as well as for the scalar potential, i.e.~$U_{abc\dots}\equiv U_{;abc\dots}$.
Evidently, because the action $S$ is a scalar, the covariant inverse propagator $S_{ab}$ turns out to be super\-symmetric. 
In the homogeneous and flat field-space limit, $\partial_\mu \Phi^a \!\to\! 0$ and $R_{abcd}\!\to\!0$, only the first two terms on the RHS of~\eqref{eq:Sab} will survive, besides $U_{ab}$. 

We may proceed in a similar fashion and calculate the three-point interaction vertex by taking a third field-space covariant derivative of the action $S$, i.e.~$S_{abc}$. By making use of the supersymmetric property of the rank-2 tensor $_a k_b$ and the conventions of~\eqref{eq:operations}, the three-vertex can be written as
\begin{equation}
   \label{eq:Sabc}
\begin{aligned}
  S_{abc}\:=\:&\  \partial_\mu\Phi^m\left((-1)^{cp}\:_m k_n\:R^{n}_{\:\:abp;c}+(-1)^{an}\:_m k_{n;[a}\:R^{n}_{\:\:bc]p}+\frac{1}{2}(-1)^{p(a+b+c)}\:_m k_{p;abc}\:\right)\partial^\mu\Phi^p\\
   &\ +(-1)^{am}(D_\mu)^m_{\:\:[a}\left( \:_mk_n\:R^{n}_{\:\:bc]p}+(-1)^{p(b+c)}\:_m k_{p;bc]}\right)\partial^\mu\Phi^p+\partial_\mu\Phi^m\:_m k_n\:R^{n}_{\:\:abp}(D^\mu)^p_{\:\:c}\\
   &\ +(-1)^{am+bn}(D_\mu)^m_{\:\:(a}\:_m k_{n;b}\:(D^\mu)^n_{\:\:c)}+i(-1)^{a}\left(\:_a\lambda^{\mu}_{m}\:R^{m}_{\:\:bcp}\partial_\mu\Phi^p+(-1)^{bm}\:_a\lambda^{\mu}_{m;[b}(D_\mu)^m_{\:\:c]}\right.\\
   &\ \left.+(-1)^{m(b+c)}\:_a\lambda^{\mu}_{m;bc}\partial_\mu\Phi^m\right)\: -\: U_{abc}\:.
\end{aligned}
\end{equation}
It should be noted that in the homogeneous limit $\partial_{\mu}{\bf \Phi}\to 0$, we recover the expressions for the vertices of the fermionic sector presented in~\cite{Gattus:2023gep,Gattus:2024spj}. 

The methodology that we established in this section can be used to compute higher-point correlation functions and express them in a strictly manifest covariant manner. However, the above exercise has also shown that such a task becomes increasingly more and more complex, yielding lengthier expressions. In  Appendix~\ref{app:HighPointCov} we display the four-point vertex 
which can be of direct relevance to $2\to 2$ scattering processes. 

\section{Covariant Scalar-Fermion Effective Action}\label{sec:CovSFaction}

As mentioned in the Introduction, a rigorous differential-geometric approach to computing field-space covariant quantum effective actions, well beyond the one-loop level, is provided by the so-called VDW formalism~\cite{Vilkovisky:1984st,DeWitt:1985sg,REBHAN1987832, Ellicott:1987ir}. For our purposes, it would be useful to first briefly review the basic aspects of this formalism by considering both bosonic and fermionic degrees of freedom, along the lines~\cite{Finn:2020nvn, Pilaftsis:2022las}. Such a covariant scalar-fermion effective action will also equivalently be called the Supergeometric Quantum Effective Action.

The SG-QEA, denoted as $\Gamma[{\bf \Phi}]$, may be evaluated iteratively in a loopwise $\hbar$-expansion by the following functional integral equation:
\begin{equation}\label{eq:SG-QEA}
  \exp \left(\frac{i}{\hbar} \Gamma[{\bf \Phi}]\right)\ =\: \int \sqrt{|\operatorname{sdet} G|}\left[\mathcal{D}{\bf \Phi}_q\right] \exp \left(\frac{i}{\hbar} S\left[{\bf \Phi}_q\right]+\frac{i}{\hbar} \int \d^{4} x \sqrt{-g}\: \Gamma[{\bf \Phi}]_{,a}\: \Sigma^{a}\left[{\bf \Phi},\! {\bf \Phi}_q\right]\right). 
\end{equation}
Here ${\bf \Phi}$ (${\bf \Phi}_q$)  collectively denotes the mean (quantum) fields, $G = \{ {}_a G_b\}$ is the supermetric,  and the comma (,) denotes functional differentiation with respect to a mean field. Furthermore, $\Sigma^{a}\left[{\bf \Phi},\!{\bf \Phi}_q\right]$ is defined to be a superposition of geodesic tangent vectors  $\sigma^{a}\left[{\bf \Phi},\!{\bf \Phi}_q\right]$ on the supermanifold, 
\begin{equation}
    \Sigma^{a}\left[{\bf \Phi},\!{\bf \Phi}_q\right]\: =\: \left(C^{-1}[{\bf \Phi}]\right)^{a}{ }_{b}\:\: \sigma^{b}\left[{\bf \Phi},\! {\bf \Phi}_q\right].
\end{equation}
The rank-2 tensor $C_{\:\:b}^{a}[{\bf \Phi}]$ may be determined by extending Vilkovisky's condition of vanishing tadpoles \cite{Vilkovisky:1984st,DeWitt:1985sg} to curved field-spaces, i.e. 
\begin{equation}
   \label{eq:Sigma}
    \big<\, \Sigma^{a}\left[{\bf \Phi},\!{\bf \Phi}_q\right]\big>_\Sigma\, =\ 0\,.
\end{equation}
The latter restriction ensures that the effective action can be calculated perturbatively as a sum of 1PI graphs \cite{Burgess:1987zi}. By imposing~\eqref{eq:Sigma} on~\eqref{eq:SG-QEA}, one is able to find the functional form of $C_{\:\:b}^{a}[{\bf \Phi}]$ perturbatively in powers of the tangent vector~$\sigma^{a}\left[{\bf \Phi},\! {\bf \Phi}_q\right]$\cite{DeWitt:1985sg}:
\begin{equation}
   \label{eq:Cab}
    C_{\:\:b}^{a}[{\bf \Phi}]\: =\: \left\langle\sigma^{a}_{\:\:;b}\left[{\bf \Phi},\! {\bf \Phi}_q\right]\right\rangle_\Sigma\: =\:
    \left\langle\delta^a_{\:\:b}\,-\,\frac{1}{3}(-1)^{m(b+n)}R^{a}_{\:\:mbn}[{\bf \Phi}]\,\sigma^m\left[{\bf \Phi},\!{\bf \Phi}_q\right]\,\sigma^n\left[{\bf \Phi},\! {\bf \Phi}_q\right]+\ldots\right\rangle_\Sigma .
\end{equation}

After performing an $\hbar$ expansion of the SG-QEA given in~\eqref{eq:SG-QEA}, the following one- and two-loop expressions are obtained~\cite{Finn:2020nvn}:
\begin{align}
   \label{eq:Gamma1loop}
   \Gamma^{(1)}[{\bf \Phi}]\ &=\ \frac{i}{2}\ln \operatorname{sdet}\big({}^{a} S_{b}\big)\ =\ \frac{i}{2} \operatorname{str}\big(\!\ln {}^{a} S_{b}\big) \:,\\
    \label{eq:Gamma2loop}
    \Gamma^{(2)}[{\bf \Phi}]\ &=\ -\frac{1}{8}S_{\{a b c d\}}\,\Delta^{dc}\Delta^{ba}\: +\: \frac{1}{12}(-1)^{bc + m(b + d)}\:S_{\{m c a\}}\,\Delta^{a b}\Delta^{cd}\Delta^{m n}\,{}_{\{ n d b\}}S\:, 
\end{align}
where  $\:_{a} S_{b} \equiv\!\:_{a} \overrightarrow{\nabla}S\overleftarrow{\nabla}_{b} \equiv {}_{a}\Delta^{-1}_{b}$ and $\Delta^{ab}$ is its inverse, which is determined by the equation: ${\Delta^{ac}\:_{c}S_{b}=\:^{a}\delta _{b}}$. Note that the frame-covariant  propagator, $\Delta^{ab}$, and its inverse,~${}_{a}\Delta^{-1}_{b}$, are both supersymmetric. In \eqref{eq:Gamma2loop}, the notation $\{\ldots\}$ denotes the operation of super-symmetrisation of all indices enclosed \cite{Finn:2020nvn},
\begin{equation}
    \{a_1 \ldots a_2\}\: =\: \frac{1}{n!}\sum_{P}(-1)^P P[a_1 \ldots a_2]\,,
\end{equation}
where $P$ runs over all possible permutations of the $n$ indices and the factor $(-1)^P$ yields $-1$ for odd number of swaps between fermionic indices and $+1$ otherwise. 
Note that both \eqref{eq:Gamma1loop} and \eqref{eq:Gamma2loop} are superscalars, as one would expect. The presence of the pre-factor  $(-1)$ in the second term of \eqref{eq:Gamma2loop} does not spoil covariance, but it is consistent with the convention that only adjacent pairs of indices can be contracted in the standard fashion. 

\subsection{Master Equation for All Loop Orders Effective Action}\label{subsec:Master}

While the one- and two-loop order quantum actions, \eqref{eq:Gamma1loop} and \eqref{eq:Gamma2loop}, can readily be  obtained from a perturbative expansion of the implicit equation~\eqref{eq:SG-QEA}, it is still useful to derive a recursive master equation for determining the SG-QEA to all loops. To this end, we follow the One-Particle-Irreducible (1PI) effective action method outlined in \cite{Kim:2006th} by extending it appropriately to incorporate fermionic degrees of freedom while maintaining configuration-space covariance. As we will see below, this method descends from the so-called Two-Particle-Irreducible (2PI) effective action as introduced in~\cite{Cornwall:1974vz}.

We start by covariantly generalising the instrumental identity that involves two functional derivatives of the effective action:
\begin{equation}
   \label{eq:aGamb}
    _{a}\Gamma_{m}\:\Gamma^{mb}\ =\ -\:_{a}\delta^{b}\,,
\end{equation}
where $\:_{a} \Gamma_{b} \equiv\!\:_{a} \overrightarrow{\nabla}\Gamma\overleftarrow{\nabla}_{b}$ and its inverse $\Gamma^{ab}$ which is the full frame-covariant Feynman propagator. 
On the other hand, by making use of the 1PI--2PI equivalence of the respective effective actions at their extrema at the one-loop level and beyond~\cite{Cornwall:1974vz}, $\Gamma^{ab}$ that enters~\eqref{eq:aGamb} can be equivalently calculated through the $(n-1)$-loop order as~\cite{Kim:2006th}: 
\begin{equation}
   \label{eq:Gaminvab}
    \Gamma^{ab}\: \equiv\:  \sum_{k=0}^{n-1}\Gamma^{(k)\, ab}\ =\ 2i\,\sum^n_{l=1}\,
    \Gamma^{(l)}\,\frac{\overleftarrow{\delta}}{\delta\,{}_a \Delta^{-1}_b}\; ,
\end{equation}
where the covariant inverse propagator $\:_{a} \Delta^{-1}_{b}$ is defined after~\eqref{eq:Gamma2loop}. Substituting \eqref{eq:Gaminvab} in \eqref{eq:aGamb} and using the identity
\begin{equation}
    \Delta^{n m}\frac{\overleftarrow{\delta}}{\delta\, {}_a \Delta^{-1}_{b}}\, =\, -(-1)^{m(a+b)}\,\Delta^{n a}\,\Delta^{b m}\;,
\end{equation}
in order to change the variable of functional differentiation from $_a \Delta^{-1}_{b}$ to its inverse $\Delta^{ab}$ by virtue of the chain rule, we eventually arrive at the important relation, 
\begin{equation}
   \label{eq:DdD}
    2 i \sum^n_{l=1}  \Gamma^{(l)}\frac{\overleftarrow{\delta}}{\delta \Delta^{ab}}=(-1)^{ab +b}\:_b\Delta^{-1}_{m}\:\Gamma^{mn}\:_n\Delta^{-1}_{b}\,.
\end{equation}
Yet, the $(n-1)$th loop-order term, $\Gamma^{(n-1)\,ab}$ (with $n>1$), may also be written down as a power series expansion of propagators~$\Delta^{ab}$ and 1PI self-energies as follows (appropriate covariant index contractions implied):
\begin{equation}
  \label{eq:GammaDelta}
\Gamma^{(n-1)\,ab}\ =\ {}^a\big[\,\Delta\sum^{n-1}_{k=1}(-1)^k\, \Pi^{(l_1)}\Delta\,\Pi^{(l_2)}\dots \Delta \,\Pi^{(l_k)}\Delta\,\big]{}^{b}\, ,
\end{equation}
with the constraint $l_1+l_2+\dots +l_k = n-1$, for
all possible combinations of $l_i$-loop 1PI self-energies $\Pi^{(l_i)}_{a b}\equiv \Gamma^{(l_i)}_{ab}$ (with $i=1,2,\dots,k$) that appear on the RHS of~\eqref{eq:GammaDelta}.
Notice that we have: $\Gamma^{(0)\,ab} = \Delta^{ab}$, at zeroth 
loop order, or equivalently at tree level.

We may now insert \eqref{eq:GammaDelta} into \eqref{eq:DdD} and 
utilise the well-known Jacobi's formula which is also applicable for superdeterminants,
\begin{equation}
  \label{eq:Jacobi}
    \delta \ln \operatorname{sdet}{\cal M}\ =\ \operatorname{str}({\cal M}^{-1}\,\delta {\cal M})\,.
\end{equation}
This in turn gives rise to 
\begin{equation}
\operatorname{sdet}\Delta^{-1}\frac{\overleftarrow{\delta}}{\delta \Delta^{ab}}\ =\: -\,\big(\operatorname{sdet}\Delta^{-1}\big)\:\Delta^{-1}_{ab}\, .
\end{equation}
Then, the following {\em master functional differential equation} for evaluating the 1PI SG-QEA to all loop orders can be derived:
\begin{equation}
   \label{eq:masterSGQEA}
     2 i\, \bigg(\Gamma^{(n)}\frac{\overleftarrow{\delta}}{\delta \Delta^{a b}}\bigg)\, \delta\Delta^{ b a}\: =\: 
     2 i\, \big(\Gamma^{(n)}\overleftarrow{\delta}\big)\: =\: \operatorname{str}\bigg[\,\Delta^{-1}\sum^{n-1}_{k=1}(-1)^k\, \Delta\,\Pi^{(l_1)}\Delta\, \Pi^{(l_2)}\Delta\, \dots \Pi^{(l_k)}\,\big(\delta\Delta\big)\,\bigg]\,,
\end{equation}
which holds true for all $n \geq 2$, provided $l_1+l_2+\dots +l_k = n-1$ for all terms of summation in~\eqref{eq:masterSGQEA}. For $n=1$, the RHS of~\eqref{eq:masterSGQEA} equates to: $\operatorname{str}[\Delta^{-1}\,\delta\Delta ]$. Thus,
we get the one-loop SG-QEA given in~\eqref{eq:Gamma1loop} upon functional integration of~\eqref{eq:masterSGQEA} by means of Jacobi's formula~\eqref{eq:Jacobi}.   In~this way, the master equation~\eqref{eq:masterSGQEA} allows for the $n$th loop-order term of the  effective action to be recursively obtained from covariant 1PI self-energies of loop order $(n-1)$ and lower.

As an illustrative example, we will now demonstrate how the two-loop SG-QEA in~\eqref{eq:Gamma2loop} can be evaluated from the master equation~\eqref{eq:masterSGQEA}. Thus, we consider the case $n=2$ in~\eqref{eq:masterSGQEA}. This yields
\begin{equation}\
   \label{eq:dGamma2dD}
     2i \,\bigg(\Gamma^{(2)}\frac{\overleftarrow{\delta}}{\delta \Delta^{a b}}\bigg)\, \delta\Delta^{ba} = -(-1)^c \:_c\left(\Pi^{(1)}\,\delta\Delta\right)^c = -(-1)^c \:_c\Pi^{(1)}_d\,\delta\Delta^{dc}.
\end{equation}
where the one-loop proper self-energy can be computed to be
\begin{equation}
  \label{eq:Pi1cd}
    _c\Pi_{d}^{(1)}\ =\ \frac{i}{2}(-1)^m
    \Big[\:{}_{c;}\Delta^{mn}\:_n\Delta^{-1}_{m;d}\: +\: (-1)^{c(m+n)}\Delta^{mn}\:_{c;n}\Delta^{-1}_{m;d}\,\Big].
\end{equation}
It is then easy to verify that the following identity holds true
\begin{equation}
  \label{eq:cDab}
    _{c;}\Delta^{a b}\ =\ -(-1)^{c(a+m)}\Delta^{am}\:_{c;m}\Delta^{-1}_{n}\,\Delta^{nb}.
\end{equation}
Substituting \eqref{eq:cDab} in \eqref{eq:Pi1cd} and the resulting expression of $_c\Pi_{d}^{(1)}$ back in~\eqref{eq:dGamma2dD}, we find
\begin{equation}
   \label{eq:deltaGamma2}
\begin{aligned}
2 i\, \big(\Gamma^{(2)}\overleftarrow{\delta}\big)\ =&\
-(-1)^{m + c}\:\frac{i}{2}\,\Big[-(-1)^{c(m+p)}\Delta^{mp}\:_{c;p}\Delta^{-1}_q\:\Delta^{qn}\:_n\Delta^{-1}_{m;d}\\
&\ +(-1)^{c(m+n)}\Delta^{mn}\:{}_{c;n}\Delta^{-1}_{m;d}\,\Big]\,\delta\Delta^{dc}\;.
\end{aligned}
\end{equation}
Proceeding as in~\cite{Kim:2006th}, we may identify: ${}_a\Delta^{-1}_{b;c} = {}_aS_{bc}$ and ${}_{a;b}\Delta^{-1}_{c;d} = {}_{ab}S_{cd}$, and then replace these expressions back in~\eqref{eq:deltaGamma2}.
Finally, after integrating the RHS of~\eqref{eq:deltaGamma2} functionally over~$\Delta^{dc}$ and some rearranging of superindices~\cite{DeWitt:2012mdz}, we recover~\eqref{eq:Gamma2loop} upon supersymmetrisation of  indices.

\subsection{The Heat Kernel Method and Schwinger's Proper Time}

The one-loop effective action~\eqref{eq:Gamma1loop} may be computed explicitly in the coordinate or $x$-space, with the help of the well-known heat-kernel method. This method employs the Schwinger’s proper time parameter $t$, which enables one to represent the propagator $\Delta$ of a field in the $x$-space, or the logarithm of its propagator, i.e.~$\ln \Delta$,  as an integral over $t$. The advantage of this method is that the UV divergences in an operator expansion of the effective action show up explicitly as $1/\varepsilon$-poles in the DR scheme, after integrating over $t$. These UV poles arise from the lower limit of the integration when $t\to 0^+$. The standard heat kernel method relies on an iterative solution of a Schr\"odinger-type equation as powers of~$t$ which give all higher-order operators of the one-loop effective action up to a given energy dimension. Here we will only briefly review the standard heat-kernel approach, but the interested reader is encouraged to consult the textbook~\cite{Zinn-Justin:2002ecy} for more details. 

Let us give some basic results from a simple QFT model with a real scalar field~$\phi$, using the heat-kernel method. The action of the model in $d$ spacetime dimensions is     
\begin{equation}
  \label{eq:Sscalar}
    S\ =\ \int d^dx\, \Big[\,\frac{1}{2}(\partial_{\mu}\phi)^2\: -\:\frac{1}{2}m^2 \phi^2\: -\: U(\phi(x))\,\Big]\,,
\end{equation}
where $U$ is a generic scalar potential which is a polynomial of $\phi$ that contains a  mass term, $m^2\phi^2$,  and possibly higher-order interactions, $\lambda_n\phi^n$, when $n>2$. We may now require that the lowest order heat kernel for this theory be the tree-level inverse propagator, 
\begin{equation}
   \label{eq:Kxy}
    K(x,y)\ =\ \frac{\delta^2 S}{\delta\phi(x)\,\delta \phi (y)}\ =\ \left(-\partial^2_x\,-\, m^2\right)\;\delta (x-y)\, ,
\end{equation}
where $\delta (x- y)$ is the usual Dirac-delta function in $d$ dimensions. For such a theory, the one-loop effective action up to order $t^3$ in the heat-kernel expansion is found to be~\cite{Zinn-Justin:2002ecy}
\begin{equation}
   \label{eq:QEAscalar}
\begin{aligned}    
     \Gamma^{(1)}\ =&\ \frac{i}{2}\operatorname{tr}\ln \big(\!-\partial^2\,-\, m^2\, -\, V\big)\\
     =&\ -\frac{i}{2}\int d^dx \int_{0}^{\infty}\frac{dt}{t}\frac{e^{-m^2t}}{(4\pi t)^{d/2}}\left[Vt+\frac{1}{2}V^2 t^2+\frac{1}{6}\left(V^3 -(\partial_\mu V)^2\right)t^3+\mathcal{O}(t^4)\right]\,,
\end{aligned}
\end{equation}
where the short-hand, $V\equiv U^{\prime\prime} = \partial^2U/\partial \phi^2$, was used and total derivatives were ignored in~\eqref{eq:QEAscalar}. Observe that the mass parameter $m^2$ acts as an IR regulator for the upper limit of the $t$-integral, $t\to \infty$. Upon integration over $t$, the effective action in DR is given by
\begin{equation}
   \label{eq:QEAscalarDR}
\begin{aligned}
     \Gamma^{(1)}\ =\ -\frac{i}{2}\int \frac{d^dx}{(4\pi)^{d/2}}\;\Big[&(-m^{2})^{(-1+d/2)}V\,\Gamma\Big(1-\frac{d}{2}\Big)\: +\: \frac{1}{2}(-m^{2})^{(-4+d)/2}V^2\,\Gamma\Big(2-\frac{d}{2}\Big)\\
     &+\: \frac{1}{6}(-m^{2})^{(-6+d)/2}\Big(V^3 -(\partial_\mu V)^2\Big)\,\Gamma\Big(3-\frac{d}{2}\Big)\Big]\;.
\end{aligned}
\end{equation}
In $d=4-2\epsilon$ dimensions, the first two terms up to ${\cal O}(t^2)$ on the RHS of~\eqref{eq:QEAscalar} exhibit UV divergences, while the last one is UV finite. Instead, when~$d=2-2\epsilon$, only the first term corresponding to $\mathcal{O}(t)$ in~\eqref{eq:QEAscalar} is UV divergent. We~note that the same results for the UV poles are obtained by standard diagrammatic methods, for a scalar QFT defined by the action~$S$ in~\eqref{eq:Sscalar}. The explicit analytic form of such UV poles will be given in Section~\ref{sec:CovFlatAction} for minimal scalar-fermion QFT models, by employing a more intuitive method based on the Zassenhaus formula which we discuss in the next subsection.

\subsection{Zassenhaus Formula in Effective Action}\label{subsec:ZFaction}

In order to determine the heat kernel for a scalar-fermion QFT in a curved field-space, one often has to deal with so-called \textit{non-minimal} operators which have non-Laplacian form~\cite{Barvinsky:1985an,Barvinsky:2021ijq}. In this case, it becomes less obvious how to calculate the polynomial coefficients of the usual heat kernel expansion in a systematic way. In fact, these heat-kernel coefficients, known as the Hadamard–Minakshisundaram–DeWitt or Seeley–Gilkey coefficients, contain the information about the UV structure of the theory and can be evaluated  straightforwardly for theories for which the inverse propagator is \textit{minimal}. Minimal operators consist of a leading quadratic differential operator, the Laplacian $\Box$, plus a lower-order differential operator,
\begin{equation}
   \label{eq:Fnabla}
    F(\partial)\ =\ -\Box \:\delta_A^B\: +\: P(\partial)\,.
\end{equation}
Here,  $\Box \equiv \partial_\mu\partial^\mu \equiv \partial^2$ is the Laplacian in $d$ spacetime dimensions and $P(\partial)$ is a polynomial that depends linearly on~$\partial_\mu$ and possibly on other field-dependent terms. 

There have been numerous studies~\cite{Alexandrov:1996gu, Avramidi:2000isc, Gusynin, Moss:2013cba, Barvinsky:2021ijq, Gusynin:1990ek, Barvinsky:1985an}, where the non-minimal or the higher-order minimal operators are expanded to have a second-order differential operator as a leading term, which allows one to compute the heat-kernel coefficients even for a curved space-time. Alternative approaches to the heat kernel expansion include the worldline formalism \cite{Ahmadiniaz:2022yam, Edwards:2022dbd, Manzo:2024gto}, or employing the Schwinger proper time formalism without recourse to perturbative expansion${}$~\cite{Abel:2023ieo}.
These other approaches do not address the main objective of our present work. For definiteness, in addition to recent studies~\cite{Alonso:2022ffe,Jenkins:2023bls}, our goal is to  evaluate the UV structure of the field-space covariant effective action directly in the coordinate space, rather in the momentum space. Furthermore, we have to deal with the issue of {\em non-minimal} heat kernels for covariant scalar-fermion theories which will be discussed in more detail in Sections~\ref{sec:CovScalarCurv} and~\ref{sec:CovFermionAction}. 

We will now demonstrate how the Zassenhaus formula can be used to derive the heat-kernel coefficients in a more intuitive manner. 
Let us therefore recall the Zassenhaus formula for two  non-commutative operators or matrices, $X$ and $Y$~\cite{Dupays:2021osq, Kimura:2017xxz}:
\begin{equation}
  \label{eq:Zass}
    \exp\big[t(X+Y)\big]\: =\: \exp(t X)\, \exp(t Y)\, \exp\bigg(\!\!-\frac{t^2}{2}[X, Y]\bigg)\, \exp\bigg[\frac{t^3}{6}\Big(2[Y,[X, Y]]+[X,[X, Y]]\Big)\bigg]\dots
\end{equation}
Note that~\eqref{eq:Zass} is akin to the more known Baker–-Campbell–-Hausdorff formula, but the latter proves to be less useful for a heat-kernel expansion. 

For illustration, let us consider the scalar QFT defined in~\eqref{eq:Sscalar}. In this case, we may consider the operators: $\widehat{X}=-\partial^2-m^2$ and $\widehat{Y} = -V$. In the heat-kernel method, the following operator appears in the one-loop effective action in the $x$-space representation:
\begin{equation}
   \label{eq:HKtxy}
    \langle x| e^{-t(-\partial^2-m^2-V)}|x'\rangle\, =\, \int d^dy\, \langle x| e^{t(\partial^2+m^2)}|y\rangle\,\langle y|e^{tV}e^{-\frac{t^2}{2}[\partial^2,V]}e^{\frac{t^3}{6}\left(2[V,[\partial^2,V]]+[\partial^2,[\partial^2,V]]\right)}e^{\mathcal{O}(t^4)}|x'\rangle\,,
\end{equation}
where we have used the Zassenhaus formula~\eqref{eq:Zass} up to order~$t^3$, and inserted conveniently the resolution of the unit operator, ${\bf \hat{1}} \equiv \int d^dy\, |y\rangle\,\langle y|$, in the same $x$-space representation. By~doing${}$~so, we have deviated from the standard approach by separating the leading heat kernel factor in~\eqref{eq:HKtxy},
given by~\cite{Zinn-Justin:2002ecy}
\begin{equation}
   \label{eq:Gin}
    \langle x| e^{-t(-\partial^2-m^2)}|y\rangle\ =\   \frac{e^{-\frac{1}{4t}(x-y)^2+m^2t}}{(4\pi t)^{d/2}}\;, 
\end{equation}
from the remaining operators that are sandwiched within the expectation value: $\langle y| \dots |x\rangle$. We~should clarify here that unlike in  its $x$-space representation, an operator, like $\widehat{X}$ or $\widehat{Y}$, does not include a $d$-dimensional delta-function, e.g. 
\begin{equation}
  \label{eq:Xrep}
    \langle x|\widehat{X}|y\rangle\ =\ X(x\,,y)\ =\ \big(\!-\partial_x^2-m^2\big)\, \delta (x-y)\, .
\end{equation}
We should therefore remark that the $x$-space representation of $\widehat{X}$ in~\eqref{eq:Xrep} is fully equivalent to the tree-level inverse propagator $K(x,y)$ in~\eqref{eq:Kxy} which is obtained by taking two functional derivatives of the effective action $S$ with respect to the scalar field $\phi (x)$.

It is now not difficult to carry out the integral over the $y$-coordinate in~\eqref{eq:HKtxy}. Up to total spacetime derivatives, we find to order~$t^3$, 
\begin{eqnarray}
  \label{eq:HKt3}
\langle x| e^{-t(-\partial^2-m^2-V)}|x'\rangle \!\!&=&\!\! 
    \frac{e^{-\frac{1}{4t}(x-x')^2+m^2t}}{(4\pi t)^{d/2}}\bigg[1+Vt+\frac{1}{2}V^2 t^2+\frac{1}{6}V^3t^3\bigg] \bigg[1-\frac{t^2}{2}\Big(\partial^2 V + 2 (\partial^\mu V ) \partial_\mu\Big)\bigg]\nonumber\\
    &&\times\, \bigg[1-\frac{2}{3}(\partial_\mu V )^2 t^3\bigg]\;,
\end{eqnarray}
where all spacetime derivatives are with respect to the position~$x$ and act as usual from left to right. After subtracting its non-interacting part ($V=0$), 
the one-loop effective action $\Gamma^{(1)}$ may be computed by taking the functional trace of~\eqref{eq:HKt3} and including a pre-factor $(i/2)$ according to \eqref{eq:Gamma1loop}, i.e.
\begin{equation}
    \label{eq:QEA-HKt}
\Gamma^{(1)}\ =\ -\frac{i}{2}\, \int d^dx\, \int_0^\infty\frac{dt}{t}\,\int d^dx'\, \delta (x'-x)\, \langle x| \big( e^{-t(-\partial^2-m^2-V)} - e^{-t(-\partial^2-m^2)}\big)|x'\rangle\; .
\end{equation}
Plugging~\eqref{eq:HKt3} in~\eqref{eq:QEA-HKt} and integrating first over $x'$ and then over $t$ yields~\eqref{eq:QEAscalarDR} to order~$t^3$, as it should be expected. It is important to notice that the origin of all heat kernel coefficients becomes evident when the Zassenhaus formula is at work, namely in effective action. In the next sections, the Zassenhaus method will be employed to compute covariant one-loop effective actions that include both scalars and fermions. 

\section{Effective Actions in Flat Field Space}\label{sec:CovFlatAction}

In the previous section we discussed an intuitive heat-kernel method based on the Zassenhaus formula which has significantly aided our computation of the one-loop effective action for scalar QFTs. In this section we will present two useful applications of this Zassenhaus method to two archetypal fermionic QFT models. The first model consists of a Dirac fermion $\psi$ that couples to a background real scalar field $\phi$ through a Yukawa interaction. The second model extends the previous one by including further interactions of a dynamical scalar field $\phi$ to the Dirac fermion~$\psi$, beyond those contained in the Yukawa sector.

\subsection{Minimal Yukawa Theory}\label{subsec:Yukawa}

As a warm-up exercise, let us consider a simple Yukawa theory
described by the action,
\begin{equation}
  \label{eq:SYukawa}
    S\ =\ \int d^d x \left[\,\frac{i}{2}\left(\bar{\psi} (\slashed{\partial}\psi)\: -\: (\slashed{\partial}\bar{\psi})\,\psi\right)-
    \big(m_f+Y(\phi)\big)\,\bar{\psi}\psi\,\right]\,.
\end{equation}
To recast the fermionic kernel into the minimal operator form of \eqref{eq:Fnabla}, we utilise the well-known procedure of \textit{bosonisation}. This procedure consists of taking advantage of the symmetric property of the Dirac operator spectrum, so as to effectively `square' the inverse of the fermion propagator,
\begin{equation}
   \label{eq:DiracSq}
    \big(i\slashed{\partial}-m_f\big)\,
    \big(i\slashed{\partial}+m_f\big)
    \: =\: -\partial_\mu\partial^\mu-m_f^2\;.
\end{equation}
The above bosonisation property is useful, since it can be used to obtain an explicit $x$-space representation for the fermionic propagator, 
\begin{equation}
   \label{eq:Fpropx}
\begin{aligned}
   \langle x|\,\big(i\slashed{\partial}-m_f\big)^{-1}\, |x'\rangle\  &=\ \int d^dy\: \langle x| \big(i\slashed{\partial} + m_f\big) |y\rangle\, 
 \langle y|\big(-\partial_\mu\partial^\mu-m_f^2\big)^{-1}\,|x'\rangle \\
&=\ \int_{0}^{\infty} \frac{dt}{(4\pi t)^{d/2}}\, e^{-\frac{1}{4t}(x-x^\prime)^2}\, \left[\,m_f \:{\bf 1}_d\: -\: \frac{1}{2t}\,(x-x^\prime)_{\mu}\gamma^\mu\,\right]\,.
\end{aligned} 
\end{equation}
The linear representation~\eqref{eq:Fpropx} of the fermionic propagator turns out to be not so convenient when computing the one-loop effective action through the heat-kernel method. 
Nonetheless, \eqref{eq:Fpropx} will still be useful to evaluate
the contributions to the effective action from mixed scalar-fermion propagators, as we will see in the next subsection.

By virtue of~\eqref{eq:DiracSq}, the one-loop effective action~$\Gamma^{(1)}$ for this theory may be evaluated as follows:
\begin{equation}
  \label{eq:QEAYukawa}
 \begin{aligned} 
   \Gamma^{(1)}\ =&\ -i\,\ln \operatorname{det}\Big[\left(i\slashed{\partial}-m_f-Y(\phi)\right)\left(i\slashed{\partial}-m_f\right)^{-1}\Big]\\
   =&\ -\frac{i}{2}\,\operatorname{tr}\ln\Big\{\big[-\partial^2 -\big(m_f+Y(\phi)\big)^2+i\slashed{\partial}Y\big]\,\big(\!-\partial^2-m_f^2\big)^{-1}\Big\}\;.
 \end{aligned}   
\end{equation}
Here, the extra minus sign arises due to the closed fermion loop, and the trace is understood to act over both the spinor and coordinate spaces. In~\eqref{eq:QEAYukawa}, the effective action $\Gamma^{(1)}$ is renormalised, such that $\Gamma^{(1)} =0$ for vanishing Yukawa interactions, i.e.~for $Y(\phi ) = 0$. Moreover, we observe that only even number of $\gamma^\mu$ matrices   contribute after taking the spinorial trace.

With the aid of Schwinger's proper-time integral representation, we may write the one-loop effective action $\Gamma^{(1)}$ in~\eqref{eq:QEAYukawa} as
\begin{equation}
   \label{eq:HKYuk}
   \Gamma^{(1)}\ =\ \frac{i}{2}\,\int_{0}^{\infty}\frac{dt}{t}\operatorname{tr}\left(e^{-t\left[-\partial^2 -(m_f+Y(\phi))^2+i\slashed{\partial}Y\right]}-e^{-t\left[-\partial^2 -m_f^2\right]}\right)\, .   
\end{equation}
Following the Zassenhaus method outlined in Subsection~\ref{subsec:ZFaction}, the one-loop effective 
action of~\eqref{eq:HKYuk} that contains the UV part of the theory is found to be
\begin{equation}
   \label{eq:G1fermion}
\begin{aligned}
     \Gamma^{(1)}\ &=\ \frac{i}{2}\int d^dx\, \frac{d}{(4\pi)^{d/2}}\,\bigg\{U (-m_f^2)^{\frac{1}{2}(d-2)}\,\Gamma\Big(1-\frac{d}{2}\Big)+\frac{1}{2}\Big(U^2-(\partial_\mu Y)^2\Big)\,(-m^2_f)^{\frac{1}{2}(d-4)}\,\Gamma\Big(2-\frac{d}{2}\Big)\\
     &\quad+\frac{1}{6}\,\Big[U^3-3U(\partial_\mu Y)^2-4(U+m_f^2)(\partial_\mu Y)^2+(\partial_\mu\partial_\nu Y)^2\Big](-m^2_f)^{\frac{1}{2}(d-6)}\,\Gamma\Big(3-\frac{d}{2}\Big)\bigg\}\,,
\end{aligned}
\end{equation}
where $U \equiv 2m_fY + Y^2$. In $d=4-2\epsilon$, the UV-divergent part of the one-loop effective action is given by
\begin{equation}
  \label{eq:G1UVYukawa}
    \Gamma^{(1)}_{\rm UV}\ =\ i\int d^dx\, \frac{1}{8 \pi^2}\,
    \bigg[\, \frac{1}{\epsilon}\,m_f^2\,U\: +\: \frac{1}{2\epsilon}\Big(U^2-(\partial_\mu Y)^2\Big)\: +\: \text{finite}\, \bigg]\,.
\end{equation}
Setting now $Y=h\phi$ in \eqref{eq:G1UVYukawa} yields
\begin{equation}
  \label{eq:G1UVYukminimal}
    \Gamma^{(1)}_{\text{UV}}\ =\ i\int d^dx\, \frac{1}{8\pi^2}
    \bigg( \frac{3}{\epsilon}h^2\phi^2 m_f^{2}+\frac{2}{\epsilon}h\phi m_f^{3}+\frac{1}{2\epsilon}h^4\phi^4+\frac{2}{\epsilon}h^3\phi^3 m_f -\frac{1}{2\epsilon}h^2(\partial_\mu \phi)^2\,\bigg)\,.
\end{equation}
Including the kinetic term for the background scalar field~$\phi$,
the Lagrangian containing the UV-divergent Counter-Terms (CTs) 
associated with this minimal Yukawa model reads:
\begin{eqnarray}
\label{eq:LctY}
     \mathcal{L}_{\rm CT} \!&=&\! \frac{1}{2}\delta Z_\phi(\partial_\mu \phi)^2\, -\, \frac{1}{2}\delta m_b^2\phi^2\, +\, i\left(\delta Z_\psi \bar{\psi}\slashed{\partial}\psi\, -\, \delta Z_{\bar{\psi}}\slashed{\partial}\bar{\psi}\psi\right)\, -\, \delta m_f \bar{\psi}\psi\, -\, \delta h \bar{\psi}\psi\phi \nonumber\\
     &&-\, \frac{1}{4!}\delta\lambda_4\phi^4\, -\, \frac{1}{3!}\delta\lambda_3\phi^3\, -\, \delta_c\phi\; ,
\end{eqnarray}
with 
\begin{equation}
   \label{eq:CTYuk}
    \delta m_b^2=\frac{3h^2m_f^2 }{4 \pi^2\epsilon},\qquad 
    \delta Z_\phi=\frac{h^2}{8 \pi^2\epsilon},\qquad 
    \delta_c =\frac{hm_f^3}{4 \pi^2\epsilon},\qquad 
    \delta\lambda_3=\frac{3h^3m_f}{2 \pi^2\epsilon},\qquad
    \delta\lambda_4= \frac{3h^4}{2 \pi^2\epsilon}\; .
\end{equation}
Moreover, we have $\delta Z_\psi = \delta Z_{\bar{\psi}} = 0$.
This is a consequence of the fact that the real field $\phi$ is not dynamical in this simple Yukawa setting.

By comparing \eqref{eq:SYukawa} to \eqref{eq:LctY}, one notices that several new operators have been generated, proportional to $\delta m_b^2$, $\delta Z_\phi$, $\delta_c$, $\delta\lambda_3$ 
and $\delta\lambda_4$. This is because the minimal Yukawa model of \eqref{eq:SYukawa} misses these renormalisable operators at tree level which are not forbidden by some symmetry imposed on the theory. In the next subsection, we will therefore consider a minimal UV-complete QFT model with dynamical scalar and fermion fields.

\subsection{Minimal Scalar-Fermion Theory}\label{subsec:SFtheory}

We will now  apply the Zassenhaus formula to the heat kernel method by computing the covariant one-loop effective action of a minimal QFT model with one scalar and one Dirac fermion. Specifically, the original theory of interest to us is described by the Lagrangian~\cite{Finn:2020nvn}
\begin{equation}
   \label{eq:LagSF}
    \begin{aligned}
\mathcal{L}\ =&\ \frac{1}{2} k(\phi) \partial_\mu \phi \partial^\mu \phi\: -\: \frac{1}{2} h(\phi) \bar{\psi} \gamma^\mu \psi \partial_\mu \phi\: +\: \frac{i}{2} g(\phi) \bar{\psi} \gamma^\mu \partial_\mu \psi \\
&\ -\frac{i}{2} g(\phi) \partial_\mu \bar{\psi} \gamma^\mu \psi\: -\: Y(\phi) \bar{\psi} \psi\: -\: V(\phi)\,.
\end{aligned}
\end{equation}
Even though this model has a non-trivial supermetric ${}_aG_b$, it turns out that the resulting super-Riemannian tensor vanishes~\cite{Finn:2020nvn,Gattus:2023gep,Gattus:2024spj}, implying 
a flat field space for the theory. As a consequence, the Lagrangian~\eqref{eq:LagSF} can be brought into a canonical Cartesian form after a suitable reparametrisation of the fields,
\begin{equation}
  \label{eq:LSFminimal}
    \mathcal{L}\ =\ \frac{1}{2}(\partial_\mu \phi)^2 +\frac{i}{2}\left(\bar{\psi}\slashed{\partial}\psi-\slashed{\partial}\bar{\psi}\:\psi\right)-Y(\phi)\bar{\psi}\psi\: -\: V(\phi)\; .
\end{equation}
More details concerning the analytic form of the required field transformations for this scenario, as well as for a similar class of multi-fermion models may be found in~\cite{Gattus:2023gep,Gattus:2024spj}. 

For our simple scalar-fermion theory defined by the Lagrangian~\eqref{eq:LSFminimal}, the field-space metric is the local 4D metric~\eqref{eq:aHb} and the covariant inverse propagator $_aS_b$ has been calculated to be
\begin{equation}
  \label{eq:aSb}
    _aS_b\ \equiv\ _A S_B\, \delta(x_A - x_B)\ =\ \left(\begin{array}{ccc}
       -\partial_\mu\partial^\mu -Y^{\prime\prime}\bar{\psi}\psi-V^{\prime\prime} &  -Y^\prime \bar{\psi} & Y^\prime \psi^\tran \\
        Y^\prime \bar{\psi}\:^\tran  & 0  &(-i\slashed{\partial}+Y)^\tran\\
        -Y^\prime \psi & i\slashed{\partial}-Y & 0
    \end{array}\right)\, \delta (x_A -x_B)\; ,
\end{equation}
where all spacetime derivatives are with respect to the position~$x_A$ and the prime (${}^\prime$) indicates a $\phi$ derivative, e.g.~$Y' = \partial Y(\phi)/\partial \phi$. Notice that for a single scalar field $Y=Y^\tran$.

It proves now more convenient to use the second expression in \eqref{eq:sdetM} for the computation of the superdeterminant that appears in the one-loop SG-QEA given in~\eqref{eq:Gamma1loop}.
Hence, partitioning the matrix~\eqref{eq:aSb} according to the block structure \eqref{eq:Mblock}, the covariant one-loop effective action under evaluation becomes
\begin{eqnarray}
  \label{eq:G1SFmodel}
     \Gamma^{(1)}\!&=&\! \frac{i}{2}\operatorname{tr}\ln\Big\{\!-\partial_\mu\partial^\mu -V^{\prime\prime}-Y^{\prime\prime}\bar{\psi}\psi-Y^\prime\bar{\psi}\:\big(i\slashed{\partial}-Y\big)^{-1}\:Y^\prime\psi-Y^\prime\psi^\tran\big[\big(\!-\!i\slashed{\partial}+Y\big)^{\tran}\big]^{-1}Y^\prime\bar{\psi}\:^\tran\Big\}\nonumber\\
      && -\,\frac{i}{2}\operatorname{tr}\ln\Big(\!-\partial_\mu\partial^\mu - i\slashed{\partial}Y - Y^2\Big)
      \; .
\end{eqnarray}
When computing the above expression explicitly, we must pay attention to the fact that the first trace in~\eqref{eq:G1SFmodel} is taken over the position-space only, while the second one over both position and spinorial spaces. There is new degree of difficulty in evaluating the UV structure from~\eqref{eq:G1SFmodel}. This arises from the propagator-like operator $\left(i\slashed{\partial}-Y\right)^{-1}$ sandwiched between fermionic fields. To deal with this extra complexity, we have developed a double heat kernel technique that makes use of two proper times and a double Schwinger integral. Technical aspects of this method are given in Appendix~\ref{app:DoubleHKM}. 

For the sake of illustration, we consider the renormalisable potential, $V=\frac{1}{2}m^2_b\phi^2+\frac{\lambda}{4}\phi^4$, which corresponds to a massive scalar field. Similarly, we choose a renormalisable Yukawa model function $Y(\phi)$, i.e.~$Y =h\phi+m_f$. As before, the mass terms, $m^2_b$ and $m_f$, could be used to act as IR regulators for the integrals over Schwinger's proper time(s). Having thus specified all interactions of our theory, the one-loop effective action given in~\eqref{eq:G1SFmodel} takes on the form,
\begin{equation}
  \label{eq:G1SFren}
     \Gamma^{(1)}\ =\ \frac{i}{2}\operatorname{tr}\ln\Big(\!\!-\partial_\mu\partial^\mu -m_b^2-M\Big)\: -\: \frac{i}{2}\operatorname{tr}\ln\Big(\!-\partial_\mu\partial^\mu -m_f^2-N \Big)\,,
\end{equation}
where 
\begin{equation}
  \label{eq:Defs_M_N}
\begin{aligned}
    M\ &=\ \frac{\lambda}{2}\phi^2+h^2\bar{\psi}\left(i\slashed{\partial}-h\phi-m_f\right)^{-1}\psi+h^2\psi^\tran\left[(i\slashed{\partial}+h\phi+m_f)^\tran\right]^{-1}\bar{\psi}^\tran\,,\\
N\ &=\ ih\slashed{\partial}\phi+h^2\phi^2+2h\phi m_f\;.
\end{aligned}
\end{equation}

We should now observe that the second trace term in~\eqref{eq:G1SFren} has already been computed in the previous subsection. As for evaluating the first trace, we use the double heat kernel method outlined in Appendix~\ref{app:DoubleHKM}. By employing this method along with the Zassenhaus formula, we can determine the complete UV structure of the one-loop effective action for this scalar-fermion theory. The UV-divergent part of the effective action is given by
\begin{equation}
\begin{aligned}
    \Gamma^{(1)}_{\text{UV}}\ =&\ -\frac{i}{2}\int d^d x\frac{1}{(4\pi)^2}\frac{1}{\epsilon}\,\Big[ih^2\left(\bar{\psi}\slashed{\partial}\psi-\slashed{\partial}\bar{\psi}\psi\right)+2h^2(h\phi+m_f)\bar{\psi}\psi+\frac{\lambda}{2}m_b^2\phi^2+\frac{\lambda^2}{8}\phi^4\Big]\\
    &\ +\frac{i}{2}\int d^d x\frac{4}{(4\pi)^2}\frac{1}{\epsilon}\,\Big[3h^2\phi^2m_f^2+2h\phi m_f^3-\frac{1}{2}h^2(\partial \phi)^2+\frac{1}{2}h^4 \phi^4+2h^3 m_f \phi^3\Big]\,. 
\end{aligned}
\end{equation}
Matching this expression to the CT Lagrangian~\eqref{eq:LctY}, the following CTs in the DR scheme may be deduced: 
\begin{equation}
   \label{eq:CTSFmodel}
\begin{aligned}
    \delta Z_\phi\:&=\: \frac{h^2}{8\pi^2\epsilon}\;,\qquad\hspace{2mm} 
    \delta m_b^2\: =\: -\frac{\lambda m_b^2}{32\pi^2\epsilon}+\frac{3h^2m_f^2}{4\pi^2\epsilon}\;,\qquad \delta Z_{\psi}\: =\: -\frac{h^2}{32\pi^2\epsilon}\: =\: \delta Z_{\bar{\psi}}\;,\\
    \delta m_f\: &=\: -\frac{h^2 m_f}{16\pi^2\epsilon}\;,\qquad 
    \delta h\: =\: -\frac{h^3}{16\pi^2\epsilon}\;,\qquad\qquad\qquad\hspace{2mm} 
    \delta \lambda_4\: =\: -\frac{3\lambda^2}{32\pi^2\epsilon}+\frac{3h^4}{2\pi^2\epsilon}\;,\\
    \delta \lambda_3\: &=\: \frac{3h^3m_f}{2\pi^2\epsilon}\;,\qquad \hspace{2mm} \delta c\: =\: \frac{h^2m_f^3}{4\pi^2\epsilon}\;.
\end{aligned}
\end{equation}

Before concluding this section, two remarks are in order. First, the CTs computed from the covariant one-loop effective action for this scalar-fermion theory agree perfectly well with those stated in~\eqref{eq:CTYuk} for the minimal Yukawa model, in the two applicable limits: $\lambda \to 0$ and ${\delta Z_\psi = \delta Z_{\bar{\psi}} \to 0}$. Second, there is no need to enforce a bosonisation procedure when the one-loop SG-QEA~\eqref{eq:Gamma1loop} is used. Instead, the required bosonisation is rather an outcome of the symmetric construction of the supermanifold chart given in~\eqref{eq:PhiChart}. In the next sections, we will apply the heat kernel techniques developed here to calculate SG-QEAs for theories with non-zero field-space curvature.

\section{Covariant Scalar Effective Action with Curvature}\label{sec:CovScalarCurv}

So far, we have analysed scenarios with vanishing field-space curvature with the aim to showcase our heat kernel approach using the Zassenhaus formula. Here, we will apply this approach to covariant scalar QFTs, or scalar SG-QFTs, which have no fermions but a non-zero Riemannian curvature in the configuration space.  

In the absence of fermions, the field-space tensor $_Ak_B$ that occurs in the Lagrangian~\eqref{eq:LSGQFT} for this scalar SG-QFT  plays the role of the supermetric~\cite{Ecker:1972tii}, i.e.~$_aG_b\equiv {}_Ak_B\, \delta(x_A - x_B)$. Since covariant field derivatives acting on the supermetric $_aG_b$ and its inverse $G^{ab}$ vanish, the mixed-ranked tensor $^aS_b$ is readily determined from the inverse propagator $_aS_b$ as
\begin{equation}
   \label{eq:aupSbdown}
     ^aS_b\ =\ G^{am}\, _mS_b\ =\ -(D_{\mu})^{a}_{\ m}\,(D^{\mu})^{m}_{\ b}-R^{a}_{\:\:mbp}(\partial_{\mu}\phi^m)(\partial^{\mu}\phi^p)-U^a_{\ b}\; ,
\end{equation}
where $(D_{\mu})^{a}_{\ b}$ was defined in \eqref{eq:Dmuab} and $U^a_{\ b} \equiv U^{;a}_{\ ;b}$. For such a theory with bosons only, the leading differential operator within $^aS_b$ contains a Laplacian, and so it has the minimal form of~\eqref{eq:Fnabla}. We may therefore represent the configuration-space rank-2 tensor $(D_{\mu})^a_{\ b}$ in a Hilbert space  of single off-shell states as follows: 
    \begin{equation}
   \label{eq:aDmub}
(D_\mu)^a_{\ b}\ =\ 
\Big(\delta^A_{\ B}\:\frac{\overrightarrow{\partial}}{\partial x^\mu_A}\: +\: \Gamma^{A}_{\:\:BM}[\boldsymbol{\phi} (x_A)]\:\partial_\mu\phi^M(x_A)\Big)\,\delta(x_A-x_B)\ =\
\langle x_A| (D_\mu)^A_{\ B} |x_B\rangle\; ,
\end{equation}
with 
\begin{equation}
   \label{eq:ADmuB}
(D_\mu)^A_{\ B}\ =\ \delta^A_{\ B}\: \frac{\overrightarrow{\partial}}{\partial x^\mu}\; +\; \Gamma^{A}_{\ BM}[\mbox{\boldmath $\phi$} (\hat{x})]\; \partial_\mu\phi^M(\hat{x})\; .
\end{equation}
Evidently, $(D_\mu)^A_{\ B}$ is a field-dependent differential operator and acts as usual from left to right on a single-state Hilbert space, which obeys the following properties:
\begin{equation}
   \label{eq:1Hilbert}
\langle x_A |x_B \rangle\ =\ \delta (x_A - x_B)\,,\quad
\langle p |x \rangle\ =\  e^{ip_\mu x^\mu}\,,\quad
\hat{x}^\mu|x_A\rangle = x^\mu_A |x_A\rangle\,,\quad
\hat{p}_\mu\ =\ i \frac{\overrightarrow{\partial}}{\partial x^\mu}\;.
\end{equation}
On the basis of these properties, it is easy to verify the validity of~\eqref{eq:aDmub}. It is important to stress here that the representation of the differential operator in~\eqref{eq:ADmuB} is fully consistent with our earlier description of such operators given in~\eqref{eq:Xrep}. Therefore, as a consequence of this representation, the product of two such operators can be written equivalently in several different ways as
\begin{equation}
  \label{eq:DDab}
    (D_{\mu})^{a}_{\:\:m}(D^{\mu})^{m}_{\:\:b}\ =\  \int_{x_M}\langle x_A|  (D_{\mu})^{A}_{\:\:M}|x_M\rangle\, \langle x_M| (D^{\mu})^{M}_{\:\:B}|x_B\rangle\ =\
    \langle x_A|  (D_{\mu})^{A}_{\:\:M}(D^{\mu})^{M}_{\:\:B}|x_B\rangle\; .
\end{equation}
In the above, we abbreviated a spacetime integral as, $\int_{x_M}\equiv \int d^dx_M$, which will be adopted from now on.

In a curved field space, the operator in~\eqref{eq:DDab} represents the covariant field-space Laplacian. By making use of the representation~\eqref{eq:DmuabVB}, it can be expressed through field-space vielbeins in terms of the local-frame Laplacian $(\widehat{D}^2)^{\widehat{A}}_{\ \widehat{B}} \equiv (D_\mu)^{\widehat{A}}_{\ \widehat{M}}\, (D^\mu)^{\widehat{M}}_{\ \,\widehat{B}}$ (which contains the 
ordinary Laplacian $(\partial^2)^{\widehat{A}}_{\ \widehat{B}} \equiv \delta^{\widehat{A}}_{\ \widehat{B}}\, \partial^2$) as 
\begin{equation}
  \label{eq:HKlead}
(D^2)^A_{\ B}\,\equiv\,(D_{\mu})^{A}_{\ M}\, (D^{\mu})^{M}_{\ B}\ =\ 
{}^A e_{\widehat{A}}\: (\widehat{D}^2)^{\widehat{A}}_{\ \widehat{B}}\; 
{}^{\widehat{B}}e^{\st}_B\; .
\end{equation}

Similarly, with the help of field-space vielbeins, one can write higher powers of $D^2$, $(D^2)^n$ (with $n\geq 1$), in terms of $(\widehat{D}^2)^n$. Taking all these properties into account, 
the covariant heat kernel of~$D^2$ is found to be
\begin{equation}
  \label{eq:HKD2}
    \langle x_A|  e^{tD^2}|x_B\rangle\ =\ 
    \langle x_A| {}^A e_{\widehat{A}}(\hat{x})\;
{}^{\widehat{A}}\big( e^{t\widehat{D}^2}\big)_{\widehat{B}}\:
{}^{\widehat{B}}e^{\st}_B (\hat{x})
|x_B\rangle\: =\: K^A_{\ B}\: \frac{e^{-(x_A-x_B)^2/{4t}}}{(4\pi t)^{d/2}}\ +\ W^A_{\ B} \,,
\end{equation}
where 
\begin{equation}
  \label{eq:KAB}
K^A_{\ B}\ \equiv\ {}^A e_{\widehat{A}} (x_A)\: {}^{\widehat{A}}e^{\st}_B (x_B)\; 
\end{equation}
and $W^A_{\ B}$ is a remainder that depends on the field-space spin connection given in~\eqref{eq:spinomega}. Further technical details are given in Appendix~\ref{app:spin_connection}. For simplicity, we do not consider contributions from the spin-connection in this work, but we set $W^A_{\ B} = 0$. This could be partially justified by the fact that thanks to the first equality in~\eqref{eq:1Hilbert}, $W^A_{\ B} \to 0$ in the UV limit for the proper time~$t$, corresponding to $t\to 0$ in~\eqref{eq:HKD2}. 
In the coincident limit $x_A \to x_B$, we have $K^A_{\ B}\to \delta^A_{\ B}$, as a consequence of~\eqref{eq:aGb}. Thus, \eqref{eq:HKD2} reduces to the well known result of the heat kernel in the {\em local frame}. Also, it is not difficult to show that the covariant heat kernel in~\eqref{eq:HKD2} is a solution to the covariant heat diffusion equation,
\begin{equation}
   \label{eq:heat}
  \frac{\partial}{\partial t}\,  \langle x_A|  e^{tD^2}|x_B\rangle\  =\   \langle x_A|\,  D^2\, e^{tD^2}|x_B\rangle\ =\   \langle x_A|\, e^{tD^2} D^2|x_B\rangle\,.
\end{equation}
Since we are interested in the UV structure of the covariant effective action in this work, we will only write down the leading field-independent term in~\eqref{eq:HKD2} proportional to $\delta^A_{\ B}$, which dominates in the coincident limit: $x_A \to x_B$. In other words, we omit the covariant field-space tensor $K^A_{\ B}$ in~\eqref{eq:KAB} in all our covariant computations from now on, unless its explicit use is required. We only note that the presence of $K^A_{\ B}$ allows us to easily switch frames, i.e.~from global to local frame and vice versa, in~order to maintain global covariance in all analytic results that we will present in this and the next sections.  

To facilitate our computation that follows below, we introduce the field-space matrix shorthands, 
\begin{equation}\label{eq:R_and_U}
    R\,(\partial\phi)^2\: \equiv\: R^{A}_{\:\:MBP}\,(\partial_{\mu}\phi^M)(\partial^{\mu}\phi^P)\,,\qquad 
    U\: \equiv\: U^A_{\ B}\;  .
\end{equation}
We may now employ the Zassenhaus formula in the heat kernel to order $t^2$ in the exponent and express the one-loop SG-QEA stated in~\eqref{eq:Gamma1loop} in the following way:
\begin{equation}
   \label{eq:G1ScalarCurv}
\begin{aligned}
    \Gamma^{(1)}[\phi]\ &=\ \text{tr}\, \Big(\ln {}^aS_b[\phi] - \ln {}^aS_b[0]\Big)\\
    &\hspace{-1cm}=\ -\frac{i}{2}\int_{x_A,x_B}\int^{\infty}_{0}\frac{dt}{t} \frac{e^{-(x_A-x_B)^2/{4t}}}{(4\pi t)^{d/2}}\langle x_B |\big( e^{t (R (\partial \phi)^2 + U )}\:e^{-\frac{t^2}{2}[D^2\:,\:R (\partial\phi)^2+U]} - 1\big)|x_A \rangle\,,
\end{aligned}
\end{equation}
where we now trace not only over position but also over field-space indices. To evaluate explicitly the covariant effective action $\Gamma^{(1)}$ in~\eqref{eq:G1ScalarCurv}, only the leading term of the kernel~\eqref{eq:HKD2} was taken into account, while all other terms are left generally covariant. This ensures that field-space covariance is maintained in all our subsequent operations. 

We may now compute the commutators that occur in \eqref{eq:G1ScalarCurv} in the standard fashion, e.g. 
\begin{equation}
  \label{eq:D2Rcom}
    [D^2\:,\:R (\partial\phi)^2+U]\ =\ D^2\left( R\, (\partial\phi)^2+U\right)\: -\: \left( R\,(\partial\phi)^2+U\right)D^2\,.
\end{equation}
Given that in the DR scheme the UV divergences emanate from $\mathcal{O}(t)$ and $\mathcal{O}(t^2)$ terms in~${d=4-2\epsilon}$ dimensions, we expand the exponentials on the RHS of~\eqref{eq:G1ScalarCurv} through this order to obtain 
\begin{eqnarray}
  \label{eq:G1CovScalt2}
   \Gamma^{(1)} &=& -\frac{i}{2}\int_{x_A,x_B}\int^{\infty}_{0}\frac{dt}{t} \frac{e^{-(x_A-x_B)^2/{4t}}}{(4\pi t)^{d/2}}\:\Big[
t\left(R_{MN}\:\partial_{\nu}\phi^M\:\partial^{\nu}\phi^N+U^{A}_{\:\:A}\right)
   +\frac{t^2}{2}U^{A}_{\:\:M}U^{M}_{\:\ A}\nonumber\\[1mm]
    &&+\frac{t^2}{2}
\left(R^{A}_{\:\:MPN}\:\partial_{\nu}\phi^M\:\partial^{\nu}\phi^N\:U^{P}_{\:\:A} + U^{A}_{\ P}R^{P}_{\ MAN}\:
\partial_{\nu}\phi^M\:\partial^{\nu}\phi^N\right)\\[1mm]
    &&+\frac{t^2}{2}
R^{A}_{\:\:MPN}\:\partial_{\mu}\phi^M\:\partial^{\mu}\phi^N\:
R^{P}_{\:\:SAT}\:\partial_{\nu}\phi^S\:
\partial^{\nu}\phi^T\:\Big]\,\delta(x_B-x_A)\ +\ \Delta\Gamma^{(1)}\,,\nonumber
\end{eqnarray}
with
\begin{equation}
  \label{eq:DeltaG1Scalar}
\begin{aligned}
   \Delta\Gamma^{(1)}&=-\frac{i}{2}\int_{x_A,x_B}\int^{\infty}_{0}\frac{dt}{t} \langle x_A|  e^{tD^2}|x_B\rangle\:\frac{t^2}{2}\,  \langle x_B |\big(R^{A}_{\:\:MPN}\:\partial_{\nu}\phi^M\:\partial^{\nu}\phi^N+U^{A}_{\:\:P}\big)(D_{\mu})^{P}_{\:\:C}(D^{\mu})^{C}_{\:\:A}\\
   &\hspace{6cm}-(D_{\mu})^{A}_{\:\:C}(D^{\mu})^{C}_{\:\:P} \left(R^{P}_{\:\:MAN}\:\partial_{\nu}\phi^M\:\partial^{\nu}\phi^N\:+U^{P}_{\:\:A}\right)|x_A \rangle\;.
\end{aligned}
\end{equation}
In the covariant one-loop effective action $\Gamma^{(1)}$ in~\eqref{eq:G1CovScalt2}, we have isolated an extra contribution, $\Delta\Gamma^{(1)}$ given in~\eqref{eq:DeltaG1Scalar}, which contains derivatives of $\delta(x_B-x_A)$. 
However, these extra terms cancel against each other, after exploiting the property of $D^2$ in the diffusion equation~\eqref{eq:heat} under the configuration-space trace. 
In fact, it is easy to see that $D^2$ is acting on the heat kernel from both its right and left side in~\eqref{eq:DeltaG1Scalar}, but with an extra negative sign arising from the commutator in~\eqref{eq:D2Rcom}. As a consequence, $\Delta \Gamma^{(1)}$ vanishes identically, i.e.~$\Delta \Gamma^{(1)} = 0$.

We now integrate over~$t$ using the integral identity
\begin{equation} \label{eq:Int_tn}
    \int_{0}^\infty \frac{dt}{t}\: \frac{e^{-(x_A-x_B)^2/{4t}}}{(4 \pi t)^{d/2}}\:t^n\ =\ 2^{-2n}\,\pi^{-d/2}\,\left[(x_A-x_B)^2\right]^{n-d/2}\,\Gamma\left(\frac{d}{2}-n\right)\,,
\end{equation}
where $n \in \mathbb{Z}$, and then employ the results from Appendix~\ref{app:DRxspace} to carry out the integration directly in the coordinate space in $d=4-2\epsilon$ dimensions in the DR scheme. Excluding $\Delta\Gamma^{(1)}$, \eqref{eq:G1CovScalt2} now reads
\begin{equation}
\label{eq:G1CovScalt3}
\begin{aligned}
     \Gamma^{(1)}_{\text{UV}} =& -\frac{i}{16\pi^2}\int_{x_A,x_B}\:
\left[(x_A-x_B)^2\right]^{-1+\epsilon}\,\left(R_{MN}\:\partial_{\nu}\phi^M\:\partial^{\nu}\phi^N+U^{A}_{\:\:A}\right)\,\delta(x_B-x_A)\hspace{3cm}\\
&+\frac{i}{64\pi^2\epsilon}\int_{x_A,x_B}\left[(x_A-x_B)^2\right]^{\epsilon}\,\Big\{
   U^{A}_{\:\:M}U^{M}_{\:\ A}+
R^{A}_{\:\:MPN}\:\partial_{\mu}\phi^M\:\partial^{\mu}\phi^N\:
R^{P}_{\:\:SAT}\:\partial_{\nu}\phi^S\:
\partial^{\nu}\phi^T\\
&\hspace{2cm}+R^{A}_{\:\:MPN}\:\partial_{\nu}\phi^M\:\partial^{\nu}\phi^N\:U^{P}_{\:\:A} + U^{A}_{\ P}R^{P}_{\ MAN}\:
\partial_{\nu}\phi^M\:\partial^{\nu}\phi^N\:\Big\}\,\delta(x_B-x_A)\:.
\end{aligned}
\end{equation}
As explained in detail in Appendix~\ref{app:DRxspace}, because of the scaling factor $\left[(x_A-x_B)^2\right]^{-1+\epsilon}$ in~\eqref{eq:G1CovScalt3}, the term proportional to the field-space Ricci tensor $R_{MN}$ and the double derivative of the potential $U^A_{\ A}$ vanishes. The second integral on the RHS of~\eqref{eq:G1CovScalt3} does not involve any differential operators and possesses the correct contributing factor $\left[(x_A-x_B)^2\right]^{\epsilon}$. Thus, the only non-trivial UV structure of the covariant effective action is given by
\begin{equation}
\begin{aligned}
  \label{eq:G1UVCovScalar}
 \Gamma^{(1)}_{\text{UV}}\ &=\: -\frac{1}{64\pi^2 \epsilon}\int_{x_A}\Big(\,U^{A}_{\:\:M}U^{M}_{\ A}+
 R^{A}_{\ MBN}\:\partial_{\mu}\phi^M\: \partial^{\mu}\phi^N\:U^{B}_{\ A}+U^{A}_{\ B}
 R^{B}_{\ MAN}\:\partial_{\mu}\phi^M\:\partial^{\mu}\phi^N\\ 
&\hspace{3cm}+R^{A}_{\ MBN}
\:\partial_{\mu}\phi^M\:\partial^{\mu}\phi^N\:
R^{B}_{\ SAT}\:\partial_{\nu}\phi^S\:\partial^{\nu}\phi^T\Big)\;.
\end{aligned}
\end{equation}
The result for $\Gamma^{(1)}_{\text{UV}}$ in~\eqref{eq:G1UVCovScalar} has already been presented in the literature~\cite{Alonso:2016oah,Alonso:2015fsp,Helset:2022pde} and more recently in \cite{Jenkins:2023rtg, Jenkins:2023bls}, through more computationally extensive methods. Thus, this exercise reaffirms not only the validity of our approach, but its efficiency as well. The authors of \cite{Jenkins:2023rtg, Jenkins:2023bls} also presented a general expression for the divergent one-loop contribution in $d=4-2\epsilon$ for quadratic actions involving scalar and gauge fields. This result was first derived by 't Hooft in \cite{tHooft:1973bhk} and was extended by~\cite{Jenkins:2023rtg, Jenkins:2023bls} to ensure field-space covariance. 

We end this section by offering a few comments on the efficiency of our heat-kernel method. This may be exemplified by noticing that by formally setting ${V \to  R(\partial\phi )^2 + U}$ in the heat kernel  given in~\eqref{eq:HKtxy}, we~reproduce  the full covariant structure of the scalar effective action {\em with} curvature in~\eqref{eq:G1ScalarCurv}. Interestingly enough, we find that the one-loop SG-QEA for the model under consideration has no further UV poles from higher-order EFT operators beyond those listed in~\eqref{eq:G1UVCovScalar}. This means that the UV infinities in this non-renormalisable theory have been organised in a finite number of EFT operators. This should be contrasted with ordinary formulations of QFT models with non-renormalisable derivative interactions, for which one gets infinite number of UV-divergent EFT operators already at the one-loop level. We may even conjecture that this re-organisation principle of the {\em finite number} of UV infinities will persist beyond the one-loop order, even though new UV-divergent EFT operators will emerge at each higher loop order. We believe that this is a unique feature of such a geometric approach to~QFT.


\section{Covariant One-Loop Fermionic Effective Action}\label{sec:CovFermionAction}

In this section we consider the minimal SG-QFT model introduced 
in~\cite{Gattus:2023gep,Gattus:2024spj} that features a non-trivial fermionic curvature both in two and four spacetime dimensions. After we have briefly described this minimal model, we will compute its one-loop  
SG-QEAs by generalising the concept of the Clifford algebra to field space. The so-generalised field-space algebra will help us to bosonise the fermionic part of the effective action, similar to what one usually does for ordinary fermionic theories without curvature (cf.~Subsection~\ref{subsec:Yukawa}).

\subsection{Minimal Fermionic Model with Field-Space Curvature}\label{subsec:MinimalFM}

Let us consider a minimal SG-QFT model with non-zero fermionic curvature, which includes one scalar field~$\phi$ and one Dirac fermion, represented by a column vector $\psi$, in $d$ spacetime dimensions. The Lagrangian of this minimal model is given by \cite{Gattus:2023gep,Gattus:2024spj}
\begin{equation}
   \label{eq:ModelI}
\begin{aligned}
\mathcal{L}\ =\ \frac{1}{2} k(\phi) (\partial_{\mu}\phi)\, (\partial^{\mu} \phi)\: +\: \frac{i}{2} \left(g_0 +g_1\overline{\psi}\psi \right)\left[\overline{\psi} \gamma^{\mu} (\partial_{\mu} \psi)\,-\, (\partial_{\mu} \overline{\psi}) \gamma^{\mu} \psi\right]\: -\: U\;,
\end{aligned}
\end{equation}
where $g_{0,1}$ are constants and $\gamma^\mu$ are the Dirac matrices, satisfying the Clifford algebra: 
\begin{equation}
  \label{eq:Clifford}
\{ \gamma^\mu\,, \gamma^\nu\}\: =\: 2\, \eta^{\mu\nu}\, {\bf 1}_d\,,
\end{equation}  
with $\eta_{\mu\nu} = \text{diag}(1\,,-{\bf 1}_{d-1})$.
In addition, $U = U({\bf \Phi})$ is the zero-grading scalar potential that includes the Yukawa and possibly interactions of higher powers in the chart components of ${\bf \Phi} = (\phi\,, \psi^{\tran}\!, \overline{\psi}{})^\tran$. Note that the model function $\zeta^\mu_a$ derived from~\eqref{eq:ModelI} takes on the factorisable form: ${\zeta^\mu_{a} = \zeta_{b}\,^b\Gamma^\mu_{a}}$, where
\begin{equation}
   \label{eq:zetaMI}
\zeta_a\ =\ \Big\{0\,, \big(g_0 + g_1\overline{\psi}\psi\big)\,\overline{\psi}\,, \big(g_0 + g_1\overline{\psi}\psi\big)\,\psi^{\tran}\Big\}
\end{equation}
and $\Gamma^\mu = \{ ^b\Gamma^\mu_{a} \}$ is defined after \eqref{eq:Sigmamu}. Using the method of~\cite{Finn:2020nvn} briefly outlined in Section~\ref{sec:SGQFT}, we may derive the field-space super\-metric~$\boldsymbol{G} = \{ {}_a G_b\}$ in the superspace of $\boldsymbol{\Phi}$,
\begin{equation}
   \label{eq:G_MFM}
    \boldsymbol{G}\, =\,\left(\begin{array}{crl}
k & 0 &\ 0\\
0 & 0 &\ f^\tran\\
0 & -f &\ 0
\end{array}\!\right),
\end{equation}
with $f\, =\, \left(g_0+ g_1\overline{\psi}\psi\right) {\bf 1}_d\, +\, g_1\psi\overline{\psi}$. From~\eqref{eq:G_MFM}, it is obvious that the scalar and fermionic fields form two irreducible supermanifold subspaces in this minimal SG-QFT model of~\eqref{eq:ModelI}. In this section, we will only focus on the fermionic part of the one-loop effective action, since its scalar part has already been analysed in previous sections. 

With this latter simplification, we may compute a key quantity by virtue of~\eqref{eq:lambdamu}, that is the mixed Lorentz and field-space tensor ${}_a\lambda^{\mu}_b$, 
\begin{equation}
\begin{aligned}
{}_a\lambda^{\mu}_b\: =&\: g_0 \left(\!
\begin{array}{cc}
   0  & \gamma^{\mu}_{\:ba}\\
    \gamma^{\mu}_{\:ab} &  0
\end{array}\!\right) \\
&-\frac{g_1}{2}
\left(\!
\begin{array}{cc}
\bar{\psi}_{a} (\bar{\psi}\gamma^{\mu})_{b}
+\bar{\psi}_{b} (\bar{\psi}\gamma^{\mu})_{a}  &  \bar{\psi}_{a} (\gamma^{\mu}\psi)_{b} + (\bar{\psi}\gamma^{\mu})_{a}\psi_{b} -2 (\bar{\psi}\psi) \gamma^{\mu}_{\:ba}\\
\bar{\psi}_{b} (\gamma^{\mu}\psi)_{a} + (\bar{\psi}\gamma^{\mu})_{b}\psi_{a} -2 (\bar{\psi}\psi) \gamma^{\mu}_{\:ab} & 
-\,\bar{\psi}_{a} (\bar{\psi}\gamma^{\mu})_{b}
-\bar{\psi}_{b} (\bar{\psi}\gamma^{\mu})_{a}
\end{array}\!\right). 
\end{aligned}
\end{equation}
As we will see in the next subsection, the mixed tensor ${}_a\lambda^{\mu}_b$ will be an essential building block, in order to construct a field-space generalised Clifford algebra.

\subsection{Field-Space Generalised Clifford Algebra and Bosonisation}\label{subsec:Clifford}

In order to calculate the fermionic part of the one-loop SG-QEA given in~\eqref{eq:Gamma1loop}, we ignore in the covariant inverse propagator $_aS_b$ in~\eqref{eq:aSb} all terms that depend on the model function $_mk_n$, i.e.~$k$ for the minimal fermionic model of interest to us. Thus, by means of the supermetric~\eqref{eq:G_MFM} appropriately reduced to the fermionic subspace, or otherwise by field-space covariance, it is not difficult to write down the mixed-rank inverse propagator \cite{Gattus:2023gep,Gattus:2024spj}, 
\begin{equation}
  \label{eq:aSb_MFM}
^{a}S_{b}\ =\  \:^a\lambda^{\mu}_{m}(iD_\mu)^m_{\:\:b}\:+\:(-1)^{bm}\:^a\lambda^{\mu}_{m;b}\:i\partial_\mu\Phi^m\: - \:^{a}U_{b}\,,
\end{equation}
which enters the one-loop effective action. 

By simple comparison of~\eqref{eq:aSb_MFM} with the inverse of the flat-space fermion propagator in~\eqref{eq:Fpropx}, it is easy to see that
${}^a\lambda^{\mu}_b$ will be the analogue of a $\gamma^{\mu}$ matrix. However, in order to bosonise the action as done in~\eqref{eq:DiracSq} but for the minimal model at hand, we introduce another set of such tensors, denoted below as~$^a\bar{\lambda}^{\mu}_{b}$, which are chosen so as to obey the Clifford algebra:
\begin{equation}
  \label{eq:Cliffordlambda}
\:^a\lambda^{\mu}_{m}\:^m\bar{\lambda}^{\nu}_{b}\: +\: {}^a\lambda^{\nu}_{m}\:^m\bar{\lambda}^{\mu}_{b}\ =\ 2\eta^{\mu\nu}\:^a\delta_b\,.
\end{equation}
If we use the matrix representation in the fermionic field space: $\lambda^\mu \equiv \{ ^a\lambda^\mu_b \}$ , then $\bar{\lambda}^\mu \equiv \{ ^a\bar{\lambda}^\mu_b \}$ turn out to be their adjugate or inverse matrices, i.e.
\begin{equation}
\bar{\lambda}^{\mu}\ =\ \big\{(\lambda^{0})^{-1}\,,-(\lambda^{i})^{-1}\big\}\; ,
\end{equation}
where the index $i$ labels the $d-1$ spatial dimensions of the theory. Hence, the so-constructed field-dependent quantities, $\lambda^\mu$ and $\bar\lambda^\mu$, are analogous to the usual four-vector matrix expressions, $\sigma^\mu = ({\bf 1}_2\,,\boldsymbol{\sigma})$ and $\bar{\sigma}^\mu = ({\bf 1}_2\,,-\boldsymbol{\sigma})$, which are defined through the Pauli matrices and obey the same Clifford algebra as in~\eqref{eq:Cliffordlambda} in four dimensions. 

We may now utilise the Clifford algebra to compute the one-loop effective action through the process of {\em bosonisation}. To this end, the following important property was found to hold for the minimal SG-QFT model under consideration:
\begin{equation} 
  \label{eq:sdet-lambda}
\operatorname{sdet}\big({}^a\lambda^{\mu}_{m}\,(iD_\mu)^m_{\:\:b}\big)\ =\ \operatorname{sdet}\big({}^a\bar{\lambda}^{\nu}_{m}\,(iD_\nu)^m_{\:\:b}\big)\ =\ (p^2)^d\: +\: f(\psi,\!\bar{\psi})\,,
\end{equation}
where $f(\psi,\!\bar{\psi})$ is some field-space and Lorentz invariant function involving fermion fields and their derivatives, so that $f(0,\!0) = 0$. Moreover, the 
differential operator $^a(D_\mu)_b$ [cf.~\eqref{eq:aDmub}]
satisfies the property:
\begin{equation}
    ^a(D_\mu)_{m}\:^m(D_\nu)_{b}\ =\ (-1)^{b(a+1)}\:_b(D_\nu)^{m}\:_m(D_\mu)^{a}\, .
\end{equation} 
To ease our computation, we introduce the shorthands,
\begin{equation}\label{eq:fermion_V}
     \lambda^{\mu}D_\mu\ \equiv\ ^a\lambda^{\mu}_{m}(D_\mu)^m_{\:\:b}\,, \qquad
V\ \equiv\ \:^{a}V_{b}\ \equiv\ (-1)^{bm}\:^a\lambda^{\mu}_{m;b}\:i\partial_\mu\Phi^m- \:^{a}U_{b}\,.
\end{equation}
With the above abbreviations, the one-loop SG-QEA~$\Gamma^{(1)}$ can be successively written as follows:
\begin{equation}
   \label{eq:G1cov_MFM}
\begin{aligned}
    \Gamma^{(1)}\: =&\ \frac{i}{2} 
    \ln \operatorname{sdet}\big({}^{a}S_{b}\big)\: =\: \frac{i}{2}\, \ln\operatorname{sdet}\Big\{\big(i\lambda^\mu D_\mu\big)\, \big[{\bf 1} +\:\big(i\lambda^\mu D_\mu\big)^{-1}\,V\big]\Big\}\\
    =&\ \frac{i}{2}\, \ln\operatorname{sdet}\big(i\lambda^\mu D_\mu\big)\: +\: 
    \frac{i}{2}\, \ln\operatorname{sdet}
    \big[{\bf 1} +\:\big(i\lambda^\mu D_\mu\big)^{-1}\,V\big]\;. 
\end{aligned}    
\end{equation}
The last equation on the RHS of~\eqref{eq:G1cov_MFM}
will be used to compute the covariant one-loop effective action~~$\Gamma^{(1)}$. 

Like in the minimal Yukawa model in Subsection~\ref{subsec:Yukawa}, we employ~\eqref{eq:sdet-lambda} to {\em bosonise} the fermionic kernel in~\eqref{eq:G1cov_MFM} as,
\begin{equation}
  \label{eq:sdetlambdaD}
    \ln \operatorname{sdet}\left(i\lambda^{\mu}D_\mu\right)\ =\ \frac{1}{2}\ln\operatorname{sdet}\left(-\lambda^{\mu}D_\mu\bar{\lambda}^{\nu}D_\nu\right)\, .
\end{equation}
By virtue of the Clifford algebra~\eqref{eq:Cliffordlambda}, the differential operator $\lambda^{\mu}D_\mu\bar{\lambda}^{\nu}D_\nu$ can then be written  as
\begin{equation}
   \label{eq:lDlbarD}
    \lambda^{\mu}D_\mu\bar{\lambda}^{\nu}D_\nu\ =\  D^2\: +\:  \lambda^{\mu} [D_\mu\,, \bar{\lambda}^{\nu}]\,D_\nu\: +\: 
    \Sigma^{\mu\nu}\, [D_\mu\,, D_\nu]\; ,
\end{equation}
with $\Sigma^{\mu\nu} \equiv \frac{1}{4}\, ( \lambda^{\mu}\bar{\lambda}^{\nu} - \lambda^{\nu}\bar{\lambda}^{\mu} )$.
As a consequence, $\lambda^{\mu}D_\mu\bar{\lambda}^{\nu}D_\nu$ contains the covariant field-space Laplacian $D^2$ defined in~\eqref{eq:HKlead} and subleading linear differential operators proportional to  $\partial_\mu$ or~$\partial_\nu$. As before, we 
expand the covariant expression in~\eqref{eq:sdetlambdaD} about the leading heat kernel $D^2$ which will enable us to determine the UV structure of the theory. With this aim in mind,  we may rewrite~\eqref{eq:sdetlambdaD} as 
\begin{equation}
   \frac{1}{2}\ln\operatorname{sdet}\left(-\lambda^{\mu}D_\mu\bar{\lambda}^{\nu}D_\nu\right)\ =\ \frac{1}{2}\ln\operatorname{sdet}\left(-D^2 - A \right)
\end{equation}
where we defined the subleading differential operator in~\eqref{eq:lDlbarD} as
\begin{equation}
A \ \equiv\ \lambda^{\mu} [D_\mu\,, \bar{\lambda}^{\nu}]\,D_\nu\: +\: M\; .
\end{equation}
Note that the above expression contains a \textit{magnetic-moment}-type operator~$M$ which is generated by a non-zero field-space curvature. More explicitly, in the field-space, the operator~$M$ is defined as
\begin{equation}
   \label{eq:Moperator}
    {}^AM_B\ \equiv\  {}^A(\Sigma^{\mu\nu}\, {\cal R}_{\mu\nu})_B\;,
\end{equation}
where 
\begin{equation}
   \label{eq:Rmunu}
     \, {}^A({\cal R}_{\mu\nu})_B\ \equiv\ {}^{A}[D_\mu\,, D_\nu]_B\ =\ (-1)^{MN}\, R^{A}_{\:\:BMN}\:\partial_\mu \Phi^M\:\partial_\nu \Phi^N\;
\end{equation}
is a {\em differential two-form rank-2 field-space tensor}. In spite of its similarity in notation, ${\cal R}_{\mu\nu}$, which is sourced from a non-zero curvature on the field-space, differs crucially from the mixed rank-2 space-time Riemann tensor reported in \cite{Barvinsky:1985an}, as the latter can only arise if spacetime possesses a non-zero Riemannian curvature. 

We should stress here that the generation of the magnetic-moment operator~$M$ is a novel and distinct feature of SG-QFTs with non-zero fermionic curvature. We observe that the operator $M$ describes magnetic-moment transitions in a fermionic system, potentially leading to interesting phenomenological implications which can be analysed elsewhere.

We may now put everything together to obtain a computationally amenable form for the SG-QEA in~\eqref{eq:G1cov_MFM}, i.e.
\begin{equation}
    \label{eq:G1covMFM2}
  \Gamma^{(1)}\: =\:  
    -\frac{i}{4}\, \operatorname{tr}\ln
    \left(-D^2 - A \right)\ -\
    \frac{i}{2}\, \operatorname{tr}\ln
    \big[{\bf 1} +\big(\!\!-\!i\lambda^\mu D_\mu\big)^{-1}\,
    V\big]\,.    
\end{equation}
Since the supertrace acts on a fermionic subspace only, it can be converted to a conventional trace by multiplying it with a minus sign. Finally, the last term in~\eqref{eq:G1covMFM2} can be expanded in powers of the operator~$(i\lambda^\mu D_\mu)^{-1}$. 
In~particular, with the aid of the Clifford algebra~\eqref{eq:Cliffordlambda}, we may express  
$(i\lambda^\mu D_\mu)^{-1}$ in terms of~$D^2$ as 
\begin{equation}
  (i\lambda^\mu D_\mu)^{-1} =\  i\bar{\lambda}^{\mu}D_\mu\, \big(\!-\!D^2-A\big)^{-1}\,,  
\end{equation}
by which the second operator $(-D^2-A)^{-1}$ can be more easily represented by a Schwinger proper-time integral. 

We are now in a position to explicitly compute the UV structure of the SG-QEA for this theory up to fourth order in spacetime derivatives. To do so, we expand the RHS of~\eqref{eq:G1covMFM2} to second order in the Schwinger's parameters $t$ and $s$, 
\begin{equation}
  \label{eq:G1_I123}
\begin{aligned}
\Gamma^{(1)}\: =\: \frac{i}{4}\operatorname{tr}(I_1-2I_2-I_3+\ldots)\,,
\end{aligned}
\end{equation}
where
\begin{eqnarray}
   \label{eq:I1}
I_1 \!&=&\! \smashoperator[r]{\int\limits_{\substack{%
    x_A,x_B \hfill}}}\ 
\int_{0}^{\infty}\frac{dt}{t} \langle x_A|  e^{tD^2}|x_B\rangle\bra{x_B} \Big({\bf 1} +tA+\frac{t^2}{2}B\Big)\ket{x_A}\ \equiv\ I_1^0\:+\:I_1^A\:+\:I_1^B\,,\\
   \label{eq:I2}
I_2 \!&=&\! 
\hspace{-3mm}\smashoperator[r]{\int\limits_{\substack{%
    x_A,x_B,x_C \hfill}}}\quad\int_{0}^{\infty}dt \bra{x_A}i \bar{\lambda}^\mu D_\mu\ket{x_B} \langle x_B|  e^{tD^2}|x_C\rangle\bra{x_C}\Big({\bf 1}+tA+\frac{t^2}{2}B\Big) V\ket{x_A}\,,\nonumber\\
    \!&\equiv&\! I_2^0\:+\:I_2^A\:+\:I_2^B \\
   \label{eq:I3}
I_3\!&=&\!
 \hspace{-3mm}\smashoperator[r]{\int\limits_{\substack{%
    x_A,x_B,x_C\\x_D,x_E,x_F\hfill}}}\quad    \int_{0}^{\infty}dt\bra{x_A}i \bar{\lambda}^\mu D_\mu\ket{x_B}\langle x_B|  e^{tD^2}|x_C\rangle\bra{x_C}\Big({\bf 1} +tA+\frac{t^2}{2}B\Big) V\ket{x_D}\nonumber\\
    &&\times\,\int_{0}^{\infty}ds\bra{x_D}i \bar{\lambda}^\nu D_\nu\ket{x_E}\langle x_E|  e^{tD^2}|x_F\rangle\bra{x_F}\Big({\bf 1} + sA+\frac{s^2}{2}B\Big) V\ket{x_A}\,,
\end{eqnarray}
and $B \equiv A^2 - [D^2, A]$.  The integral $I_1$ comes from the first trace term in~\eqref{eq:G1covMFM2}, while $I_2$ and $I_3$ arise from the second trace term in~\eqref{eq:G1covMFM2} 
after we perform a logarithmic expansion and keep only powers of~$(-i\lambda^\mu D_\mu)^{-1}V$ up to first and second order, respectively. We ignore further contributions to $\Gamma^{(1)}$ that arise from terms beyond the second order in $t$ and $s$. However, as we will outline below, only a subset of these enter the UV structure of the theory. To better organise the exposition of our results, we have decomposed the sum of integrals within~$I_{1,2}$, in groups that have no dependence on the operators $A$
and $B$ and those that they do, i.e.~$I_{1,2}\equiv I_{1,2}^0\:+\:I_{1,2}^A\:+\:I_{1,2}^B$.
Finally, we note that the field-space matrices $\lambda^\mu$ and $\bar{\lambda}^\mu$ (as defined after~\eqref{eq:Cliffordlambda}) obey similar trace identities like the $\gamma^\mu$ matrices, e.g.~the trace of odd number of $\lambda^\mu$- or $\bar{\lambda}^\mu$-matrices vanishes.

To gain valuable insight into our analytical computation, we will first consider the leading term in the covariant heat kernel expansion \eqref{eq:HKD2}. After performing the integration over the Schwinger's proper time~$t$ using the  integral identity \eqref{eq:Int_tn}, the integrals $I_{1,2}$ given in~\eqref{eq:I1}--\eqref{eq:I2} become
\begin{eqnarray}
  \label{eq:I1x}
I_1 \!&=&\! 
\hspace{-3mm}\smashoperator[r]{\int\limits_{\substack{%
    x_A,x_B \hfill}}} 
\bra{x_B}\Big(\frac{1}{\pi^2}|x_A-x_B|^{-4+2\epsilon}
+\frac{1}{4\pi^2}|x_A-x_B|^{-2+2\epsilon}A\nonumber\\
&&-\,\frac{1}{2}\frac{1}{16\pi^2\epsilon}|x_A-x_B|^{2\epsilon}B\Big)\ket{x_A}\, ,
\qquad\\[3mm]
   \label{eq:I2x}
I_2 \!&=&\! 
\hspace{-4mm}\smashoperator[r]{\int\limits_{\substack{%
    x_A,x_B,x_C \hfill}}} \bra{x_A}i \bar{\lambda}^\mu D_\mu\ket{x_B}\bra{x_C}\Big(\frac{1}{4\pi^2}|x_B-x_C|^{-2+2\epsilon} -\,\frac{1}{16\pi^2\epsilon}|x_B-x_C|^{2\epsilon}A\nonumber\\
&&+\,\frac{1}{2}\frac{1}{64\pi^2\epsilon}|x_B-x_C|^{2+2\epsilon}B\Big)\,V\ket{x_A}\,,
\end{eqnarray}
where $|x|=[(x)^2]^{1/2}$.

From our analysis in Appendix~\ref{app:DRxspace}, we know that the only terms contributing to the UV poles of $I_1$ and $I_2$, which were obtained by a single integration over the Schwinger proper time~$t$, are those scaling as $|x_A-x_B|^{2\epsilon}$. These terms can also appear, for example, when a higher order term, proportional to~$|x_A-x_B|^{2+2\epsilon}$, happens to be differentiated twice. In view of this observation, it is easy to see that in the integral $I_1$ in~\eqref{eq:I1x}, the only potentially non-vanishing term is the one with the $B$ operator, i.e. $I_1^B$, provided no derivative acts on~$|x_A-x_M|^{2\epsilon}$. Instead, we have $(I^{0,A}_1)_{\text{UV}} = 0$. With this information in mind, we once again turn our attention to the expression \eqref{eq:I1} with the full covariant heat kernel and write explicitly the functional form for the $B$ operator as
\begin{equation}
\begin{aligned}
      I_1^B \!&=\! \smashoperator[r]{\int\limits_{\substack{%
    x_A,x_B \hfill}}}\ 
\int_{0}^{\infty}\frac{dt}{t}\frac{t^2}{2} \langle x_A|  e^{tD^2}|x_B\rangle\bra{x_B} \frac{1}{2}\left[(D^\mu N^\beta) + 2 \Sigma^{\mu \nu}(D_\nu N^\beta)\right]\big(\!-\mathcal{R}_{\mu \beta}+\{D_\mu , D_\beta\} \big)\\
&\quad+\frac{1}{2}\lambda^\mu D_\mu(\bar{\lambda}^\nu N^\beta)\big(\!-\mathcal{R}_{\nu \beta}+\{D_\nu , D_\beta\} \big) + (D^2)^2 +2MD^2 +\lambda^\mu D_\mu\left[(\bar{\lambda}^\nu (D_\nu M)\right]|x_A\rangle
\end{aligned}
\end{equation}
where $N^\mu \equiv \lambda^\alpha (D_\alpha \bar{\lambda}^\mu)$ and ${\cal R}_{\mu \nu}$ is defined in~\eqref{eq:Rmunu}. We make use of the covariant heat diffusion equation \eqref{eq:heat} in order to derive the following set of integral identities
\begin{equation}
\begin{aligned}
    \int_{0}^{\infty}\frac{dt}{t}\,\frac{t^2}{2}\: \langle x_A|\, D^2\: e^{tD^2}\,|x_B\rangle\: \delta(x_A-x_B)\ &\sim\ \frac{1}{\varepsilon_{\rm IR}}\ ,\\
    \int_{0}^{\infty}\frac{dt}{t}\,\frac{t^2}{2}\: \langle x_A|\, (D^2)^2\: e^{tD^2}\,|x_B\rangle\: \delta(x_A-x_B)\ &=\ 0\, ,
\end{aligned}
\end{equation}
where $\varepsilon_{\mathrm{IR}}$ denotes the infrared pole in the DR scheme. The covariant identities listed above are consistent with our consideration on the scaling factor $|x_A-x_B|^{2\epsilon}$ and its derivatives, as explicitly discussed in~Appendix~\ref{app:DRxspace}. 
Hence, upon integration over the remaining space-time variable $x_B$ by means of the $\delta$-function, in the DR scheme the UV-divergent part of the integral $I_1$ is  
\begin{equation}
\begin{aligned}\label{eq:I1UV}
   (I_1^B)_{\text{UV}}\ =\ \frac{1}{64\pi^2 \epsilon}\int_{x_A} 
   &{}^A\big\{\left[D^\mu \left(\lambda^\alpha (D_\alpha \bar{\lambda}^\beta)\right)+2 \Sigma^{\mu \nu}D_\nu\left( \lambda^\alpha (D_\alpha \bar{\lambda}^\beta)\right)\right]\mathcal{R}_{\mu \beta}-2\lambda^\mu D_\mu\left(\bar{\lambda}^\nu (D_\nu M)\right)\\
   &\quad\! +\lambda^\mu D_\mu \left(\bar{\lambda}^\nu \lambda^\alpha (D_\alpha \bar{\lambda}^\beta)\right)\mathcal{R}_{\nu \beta}\big\}_A\;,
\end{aligned}
\end{equation}
whilst it is $(I^{0,A}_1)_{\text{UV}} = 0$ as argued above.
Observe that there are new fermionic EFT operators emerging as a consequence of a non-zero field-space curvature. In fact, all terms in~\eqref{eq:I1UV} vanish in the flat field-space limit where $\lambda^\mu \rightarrow \gamma^\mu$ and any fermionic field-dependence disappears.

The integral $I_2$ in~\eqref{eq:I2x} differs crucially from the integral~$I_1$. It has an extra differential operator $D_\mu$ and the potential-like term $V$ which contains the usual potential $U$ and a term that depends on a covariant field-space derivative acting on the $\lambda^\mu$ matrix.  Since the operators $A$ and $B$ both have an even number of $\lambda^\mu$ matrices, the term involving $U$ vanishes since it involves the trace of an odd number of $\lambda^\mu$ matrices. The term depending on the operator~$A$, denoted as~$I^A_2$, is accompanied with the proper scaling factor, $|x_B-x_C|^{2\epsilon}$, leading to the following covariant UV-divergent contribution:
\begin{equation}
\begin{aligned}
    (I_2^A)_{\text{UV}} = \frac{i}{32 \pi^2 \epsilon}\int_{x_A}{}^{A}\left[(D^\mu V)\bar{\lambda}^\nu \mathcal{R}_{\mu \nu}+\lambda^\mu D_\mu (\bar{\lambda}^\nu V)\bar{\lambda}^\alpha \mathcal{R}_{\nu \alpha}+2\Sigma^{\mu\nu}(D_\nu V)\bar{\lambda}^\alpha \mathcal{R}_{\mu \alpha}\right]_A
\end{aligned}
\end{equation} 
The term with the $B$ operator gives a UV-divergent contribution when acting on the covariant derivative of  $\lambda^\mu$:
\begin{equation}
\begin{aligned}
    (I_2^B)_{\text{UV}} &= \frac{i}{64 \pi^2 \epsilon} \int_{x_A} {}^{A}\Big\{\lambda^\mu D_\mu\left[\bar{\lambda}^\nu \lambda^\alpha D_\alpha (\bar{\lambda}^\beta D_\beta V)\right]\bar{\lambda}_\nu +\lambda^\mu D_\mu \left[\bar{\lambda}^\nu D_\nu \left(\lambda^\alpha D_\alpha(\bar{\lambda}^\beta V)\right)\right]\bar{\lambda}_\beta \\
     &\hspace{2.3cm} + D^\mu \left[\lambda^\nu D_\nu(\bar{\lambda}^\alpha D_\alpha V)\right]\bar{\lambda}_\mu+2 \Sigma^{\mu \nu}D_\nu \left[\lambda^\alpha D_\alpha (\bar{\lambda}^\beta D_\beta V)\right]\bar{\lambda}_\mu \\
    &\hspace{2.3cm}   +  \lambda^\mu D_\mu \left[\bar{\lambda}^\nu D_\nu\left(\lambda^\alpha \bar{\lambda}^\beta D_\beta V\right)\right]\bar{\lambda}_\alpha \Big\}_A
\end{aligned}
\end{equation}
As in \eqref{eq:I1UV}, we obtain new EFT operators that vanish in the flat field-space limit.

Let us finally turn our attention to the integral~$I_3$ in~\eqref{eq:I3x}. After carrying out two integrations over the Schwinger's proper times $t$ and $s$, ~$I_3$ can be written as 
\begin{eqnarray}
  \label{eq:I3x}
 I_3 \!&=&\!
 \hspace{-5mm}
 \smashoperator[r]{\int\limits_{\substack{%
    x_A,x_B,x_C\\x_D,x_E,x_F\hfill}}}  \bra{x_A}i \bar{\lambda}^\mu D_\mu\ket{x_B}  \bra{x_C}\Big(\frac{1}{4\pi^2}|x_B-x_C|^{-2+2\epsilon}-\frac{1}{16\pi^2\epsilon}|x_B-x_C|^{2\epsilon}A\nonumber\\
    &&+\, \frac{1}{2}\frac{1}{64\pi^2\epsilon}|x_B-x_C|^{2+2\epsilon}B\Big) V\ket{x_D}\bra{x_D}i \bar{\lambda}^\nu D_\nu\ket{x_E}\,\bra{x_F}\Big(\frac{1}{4\pi^2}|x_E-x_F|^{-2+2\epsilon}\nonumber\\[2mm]
    &&-\,\frac{1}{16\pi^2\epsilon}|x_E-x_F|^{2\epsilon}A\, +\,\frac{1}{2}\frac{1}{64\pi^2\epsilon}|x_E-x_F|^{2+2\epsilon}B\Big)V\ket{x_A}\;,
\end{eqnarray}
As~a~consequence of the double insertion of the heat kernel, we get different forms of spacetime integrands that do not involve a residual $\delta$-function, like $\delta (x_A - x_B)$, in the last step of integration over $x_B$. The complete set of all terms that occur in $I_3$ is quite lengthy to list them all here. But, in terms of the spacetime integrations involved and the number of $D_\mu$-derivatives, they can be organised as follows: (i)~one integral with at least two derivatives, (ii)~two integrals with four and eight derivatives, (iii)~three integrals with six derivatives, and (iv)~one integral with at least ten derivatives. The terms that feature at least two derivatives give rise to a UV-divergent contribution to $I_3$, which was computed to be
\begin{equation} 
   \label{eq:I3UV}
    \left(I_3\right)_{\text{UV}}\: =\: -\frac{1}{32 \pi^2 \epsilon}\int_{x_A} {}^A\left[\bar{\lambda}^\mu V \bar{\lambda}_\mu (D^2 V)\, +\, (D^2 \bar{\lambda}^\mu)V \bar{\lambda}_\mu V\, +\, 2(D_\nu \bar{\lambda}^\mu) V \bar{\lambda}_\mu (D^\nu V)\right]_A\,.
\end{equation}
Notice that only the first term in the integrand survives in the flat field-space limit which stems from the wavefunction renormalisation of the scalar fields in the theory. Instead, the other two terms in the integrand of~\eqref{eq:I3UV}
represent new EFT operators that can only emerge from a curved fermionic field space.

In addition to the above EFT interactions, other UV-divergent EFT operators with at least six spacetime covariant derivatives can also be generated. These sixth-order EFT operators can be worked out in a methodical and systematic way within our approach. But listing all these EFT operators is an exercise that goes beyond the scope of the present work. Nonetheless, it is important to stress here that the one-loop SG-QEA for the model under consideration has no further UV poles beyond the sixth order of spacetime derivatives. 
Hence, exactly as was the case for the scalar QFT model discussed in Section~\ref{sec:CovScalarCurv}, we observe that 
the UV infinities in this non-renormalisable fermionic theory have re-organised themselves in a finite number of EFT operators, as opposed to ordinary formulations of QFT models with non-renormalisable derivative interactions, for which  infinite number of UV-divergent EFT operators get generated at one-loop level. Consequently, this unique feature seems to persist, at least within the framework of minimal~SG-QFTs with non-zero fermionic curvature.

\subsection{Beyond the Minimal Supergeometric Framework}

In this and previous sections, we have been analysing minimal SG-QFTs that feature either a scalar or a fermionic curvature, without including gauge interactions. Going beyond this minimal framework is not a trivial task, as the complexity of the problem increases drastically. For this reason, we will only briefly outline potential approaches that could be followed to compute SG-QEAs for specific non-minimal scenarios. 

If a SG-QFT contains both scalar and fermionic fields but {\em without} scalar-fermion mixing in the kinetic terms of the covariant inverse propagator $^aS_b$ given in~\eqref{eq:Sab}, 
one may then still be able to use the techniques that we have developed in this paper. In this case, the grading-one (fermionic) part of~$^aS_b$ can be bosonised independently from the grading-zero (scalar) part. Specifically, the Laplacian describing the scalar kinetic term may be written in the local frame~as 
\begin{equation}
   \label{eq:DkD}
(D_\mu)^a_{\:\:m}\:^m k_n\:(D^\mu)^n_{\:\:b}\ =\ ^a e_{\hat{a}}\: ^{\hat{a}}(\widehat{D}_\mu)_{\hat{m}}\,
^{\hat{m}}k_{\hat{n}}\, {}^{\hat{n}}(\widehat{D}^\mu)_{\hat{b}}
\: ^{\hat{b}}e^{\st}_b\,,  
\end{equation}
where $^{\hat{m}}k_{\hat{n}}$ is a diagonal matrix in its scalar block entry, that is~$\{^{\widehat{M}}k_{\widehat{N}}\} = \text{diag}\,({\bf 1}_N\,, {\bf 0}_{2dM})$. The~latter is a consequence of the approach introduced in~\cite{Finn:2019aip} to determine the field-space vielbeins $_A e^{\widehat{B}}$ and the local supermetric $_{\widehat{A}}H_{\widehat{B}}$ (see also our discussion in Section~\ref{sec:SGQFT}). As discussed in Section~\ref{sec:CovScalarCurv} (with more details given
in~Appendix~\ref{app:spin_connection}), this would lead to an
obvious modification of the leading term in the heat kernel in~\eqref{eq:HKD2}, where the pre-factor $K^A_{\ B}$ in~\eqref{eq:KAB} will now read: 
\begin{equation}
  \label{eq:KsfAB}
^A{\cal K}_B \ =\ ^A e_{\widehat{A}} (x_A)\: ^{\widehat{A}}k_{\widehat{B}}\: {}^{\widehat{A}}e^{\st}_B (x_B)\, .
\end{equation}
Observe that in the coincident limit $x_A \to x_B$, we have: $^A{\cal K}_B \to {}^A k_B$, which is in general field-dependent and it does no longer equate to ${}^A \delta_B$.  

If the fermionic kinetic term happens to have a non-factorisable form for $\zeta^\mu_a$, i.e.~${\zeta^\mu_{a} \neq \zeta_{b}\,^b\Gamma^\mu_{a}}$, we can always find the {\em closest} minimal SG-QFT model, for which   ${\zeta^\mu_{a} = \zeta_{b}\,^b\Gamma^\mu_{a}}$, so that its kinetic term can be bosonised using the Clifford algebra of the $\lambda^\mu$'s in~\eqref{eq:Cliffordlambda}. Then, the SG-QEA for such a non-minimal model will consist in computing deviations from the minimal one, through the linear decomposition: 
\begin{equation}
   \label{eq:L}
 ^a\tilde{\lambda}^\mu_b\ =\ 
 {}^a\lambda^\mu_b\: +\: {}^a n^\mu_b\, .
\end{equation}
Evidently, this approach inevitably leads to the introduction of an extra covariant mixed tensor ${}^a n^\mu_b$ which 
acts as a remainder of~$^a\tilde{\lambda}^\mu_b$. Unlike ${}^a\lambda^\mu_b$, the field-dependent tensors, ${}^a\lambda^\mu_b$ and ${}^a n^\mu_b$, do not satisfy in general the Clifford algebra. It should be clarified here that the role of the remainder tensor ${}^a n^\mu_b$ will be to induce new covariant EFT operators beyond those obtained in the minimal SG-QFT model discussed in Subsection~\ref{subsec:Clifford}.  

We note that SG-QFTs may also include gauge fields as extra bosonic chart variables following the standard VDW formalism~\cite{Vilkoviskii:1984un,DeWitt:1985sg}. In a gauge-invariant extension of an SG-QFT, the new aspect is that 
the generators of the gauge group become the Killing vectors of the field-space manifold (e.g.~see~\cite{Helset:2022tlf} and references therein). To what extent this property can carry over to super-manifolds is an open question that would require further investigation in another dedicated study.  
In this respect, we find encouraging the earlier works which considered the application of the VDW formalism to Yang--Mills theories~\cite{Vilkoviskii:1984un,DeWitt:1985sg,REBHAN1987832} and the SM~\cite{Bounakis:2017fkv}, but on flat field-space manifold. They have shown that one can define a projected field-space metric, such that the resulting VDW effective action is independent of the gauge-fixing condition through two loop order. We~therefore envisage to be able to employ the techniques developed in these previous works, in order to construct the projected field-space metric in a unique manner.

We should reiterate here that for the non-renormalisable extensions of SG-QFTs discussed above and in the previous section, the metric of the field space can unambiguously be determined from the classical action $S$ as outlined in Section~\ref{sec:SGQFT}. This can be done in a systematic and self-consistent manner by keeping track of the powers of $\hbar^n$, e.g.~as $\hbar^n\,\Gamma^{(n)}[\boldsymbol{\Phi}]$, where $n = 0,1,2, \dots$ indicates the loop order of the 1PI SG-QEA, according to our discussion in Subsection~\ref{subsec:Master}. For~${n=0}$, we have $S[\boldsymbol{\Phi}] = \Gamma^{(0)}[\boldsymbol{\Phi}]$, and so taking the limit $\hbar \to 0$ for determining the field-space metric should be considered to be smooth in the 1PI formulation of the quantum effective action.
Moreover, any CTs of renormalisation needed to get rid of the UV infinities in the one-loop SG-QEA, which we have computed in this study, should be added to  $\Gamma^{(0)}[\boldsymbol{\Phi}]$, but with an extra factor $\hbar$, in line with the earlier work in~\cite{Ellicott:1987ir}. One may even require that the classical form of the field-space metric be preserved at some reference point $\mu$ of the RG scale. This is exactly the renormalisation scheme that we have tacitly adopted in all our computations.

We conclude this section by commenting on the issue of {\em multiplicative anomalies} which usually haunts algebraic manipulations with matrices of infinite dimensionality and infinite trace. We have encountered such manipulations in the process of {\em bosonisation} of the effective action, before we applied the Schwinger--DeWitt heat-kernel technique. Specifically, for two infinite dimensional matrices or operators, e.g.~$A$ and $B$, the following expression: \begin{equation}
  \label{eq:MultiAnomaly}
    a(A,B)\: =\: \ln \operatorname{det} AB\, -\, \ln\operatorname{det}A\, -\, \ln \operatorname{det}B
\end{equation}
does not need to vanish in general. Only for {\em trace class} operators (which possess a finite trace), one can show that their trace is independent of the chosen orthonormal basis, obeying the desirable algebraic constraint: $a(A,B) = 0$~\cite{Conway:1999}. However, as demonstrated in~\cite{Evans:1998pd} within simple QFT examples, multi\-plicative anomalies do not encode any novel physics but they merely illustrate potential ambiguities in the Schwinger proper time regularisation scheme, especially for mass-dependent regularisation methods, such as the $\zeta$-function regularisation scheme. For this reason, all our computations were done in the massless DR scheme. Moreover, when an IR mass regulator was introduced, we have always verified that all our results agree with well-known results in the flat field-space limit which were derived independently through diagrammatic methods. In this way,  potential ambiguities that might originate from multiplicative anomalies were kept under good control.

\section{Conclusions}\label{sec:Concl}

Supergeometry in Quantum Field Theory offers a new concept in going beyond the standard\- super\-symmetric framework. Unlike conventional supersymmetric theories, SG-QFTs treat scalars and fermions on equal footing, as in\-dependent field variables, and as such, they permit for general scalar-fermion field trans\-formations on the configuration space of a supermanifold. In particular, SG-QFTs do not require equality of the degrees of freedom between bosons and fermions, at least in their minimal setting. Clearly, an on-shell S-matrix amplitude is usually field-reparametrisation independent. Nonetheless, the diffeomorphic structure of the Super-Geometric Quantum Effective Action reduces the appearance of independent UV divergences that occur in the emergent covariant EFT operators in the configuration space. 
Consequently, the computation of the SG-QEA, along with its covariant UV structure at the one-loop order, was one of the central objectives of the present work.

Before embarking with the calculation of the SG-QEA,  we briefly reviewed the old covariant field-space formalism~\cite{Honerkamp:1971sh} for bosonic theories, including other more recent developments within the context of the geometric SMEFT~\cite{Jenkins:2023bls, Alminawi:2023qtf, Craig:2023hhp, Assi:2023zid, Alonso:2016oah, Nagai:2019tgi, Helset:2020yio, Cohen:2021ucp, Talbert:2022unj, Helset:2022tlf, Fumagalli:2020ody, Cohen:2023ekv, Loisa:2024xuk,Buchmuller:1985jz, Grzadkowski:2010es, Dedes:2021abc, Alonso:2015fsp, Alonso:2023upf}. To go beyond the tree level, we devised a new approach to the Schwinger--DeWitt heat-kernel technique based on the Zassenhaus formula where the emergence of all possible covariant EFT operators can be systematically tracked. For a given SG-QFT, 
our intuitive approach provides a comprehensive account of all possible covariant EFT operators that can occur in the configuration space, unlike other more laborious methods that utilise 't~Hooft's matching procedure. Hence, our {\em covariantly}-improved heat-kernel method enables one to identify all covariant EFT operators in a consistent and systematic manner. 

The aforementioned covariant heat-kernel approach was further 
extended to include SG-QFTs that feature non-zero fermionic curvature in two and four spacetime dimensions. In~order to compute the resulting SG-QEA by the process of bosonisation, we had to introduce a novel Clifford algebra which is obeyed by a set of field-dependent $\gamma$-matrices, denoted as $\lambda^\mu$, and their adjugate counterparts, $\bar{\lambda}^\mu$. In this respect, we have attempted to minimise the impact of potential ambiguities due to the so-called multiplicative anomalies by matching the UV divergences of the one-loop SG-QEA to those obtained in the flat field-space limit of the theory. In this way, we find that the EFT interactions resulting from the one-loop SG-QEA are manifestly diffeomorphically invariant in configuration and field space alike. 
Although our explicit computations have been restricted to one-loop level, it is important to note that the extension of our approach to evaluating higher-loops can in principle proceed recursively through the master equation derived in~\eqref{eq:masterSGQEA}. 

The present study opens up new horizons for further theoretical and phenomenological exploration within the framework of SG-QFTs. One obvious direction will be to include gauge and gravitational interactions to chiral fermions which could lead to a complete geometrisation of realistic theories of micro-cosmos, such as the SM and its gravitational sector. Furthermore, one might consider to include additional global or local symmetries in SG-QFTs which will act as isometries~\cite{Vilkovisky:1984st,DeWitt:1985sg} on the supermanifold~\cite{DeWitt:2012mdz}. In this way, 
one may be tempted to investigate whether the famous S-matrix theorem by Haag, Lopusza\'nski and Sohnius~\cite{Haag:1974qh} could be evaded by defining off-shell
 less restrictive forms of scalar-fermion symmetries in a SG-QFT which does not rely on ordinary supersymmetry.

An important advantage of a Riemannian geometry that we have adopted in this work for the field space over other Finslerian-type geometries emanates from their powerful embedding theorems. These embedding theorems due to Nash and Kuiper (NK)~\cite{Nash:1954,Kuiper:1955a,Kuiper:1955b,Nash:1956} ensure that a curved manifold can always be isometrically embedded into an Euclidean (flat) field-space of sufficiently higher dimensionality. The NK theorems hold not only for compact spaces, but also for non-compact hyperbolic spaces, like the field-space of the minimal Hilbert--Einstein action~\cite{Finn:2019aip}, in which the field-space Ricci scalar turns out to be negative. It is crucial to emphasise here that such embeddings lead to renormalisable QFTs, at least as far as their kinetic sector is concerned. In fact, the famous $O(N)$ non-linear chiral model~\cite{Weinberg:1968de,Honerkamp:1971sh} and the embedding 
of its compact curved field-space manifold into an Euclidean field-space by including the $\sigma$, or the Higgs field, as an extra dimension along the radial direction, constitutes one of the classical examples of UV completion (e.g.~see discussion of Section V.C of~\cite{Cornwall:1974km}). On the other hand, the NK theorems do not provide a precise recipe on how to explicitly construct the mapping function for such topological embeddings, even though their existence is guaranteed. Nonetheless, it is interesting to note that although in their infancy, studies of such embeddings of supermanifolds into higher-dimensional Euclidean-type, orthosymplectic geometries~\cite{DeWitt:2012mdz} do exist and this topic remains an active field of research by the mathematics community~\cite{Witten:2012bg,Noja:2018edj,Bettadapura:2018fyz}. 

From a more phenomenological perspective, one may envisage that SG-QFTs offer a new portal to dark-sector dynamics. For instance, dark-sector fermions through their non-linear dynamics can change the dispersion properties of weakly interacting particles, like SM neutrinos and axions. In particular, the emergence of the new field-space-induced magnetic-moment operator $M$, as defined in~\eqref{eq:Moperator}, can induce transitions between fermions with interesting phenomenological consequences. 
Likewise, it would be interesting to  go beyond the minimal fermionic setting that we have studied here and investigate the dynamics of non-minimal SG-QFTs as well. We plan to explore some of the above ideas in future works.

\subsection*{Acknowledgements} 

We thank Fedor Bezrukov for enlightening discussions.  
The work of AP is supported in part by the STFC Research Grant ST/X00077X/1. VG~acknow\-ledges support by the University of Manchester through the President's Doctoral Scholar Award. 

\newpage

\appendix
\renewcommand{\theequation}{\Alph{section}.\arabic{equation}}

\setcounter{equation}{0}
\section{Higher-Point Covariant Interactions}\label{app:HighPointCov}

In deriving higher-point covariant interactions, like $S_{abc\dots}$, the identity given in~\eqref{eq:DmuRabc} by Ecker and Honerkamp~(EH)~\cite{Ecker:1972tii} plays an instrumental role. Since we have not seen, to the best of our knowledge, a sufficiently detailed proof of the EH~identity in the literature, we give in Subsection~\ref{subsec:HEid} of this appendix all intermediate steps. In particular, we show how the EH identity can be generalised to a supermanifold that includes both bosonic and fermionic field coordinates. Then, in Subsection~\ref{subsec:Sabcd} we apply this identity to obtain the four-point correlation function $S_{abcd}$.

\subsection{Proof of the Ecker--Honerkamp Identity}\label{subsec:HEid}

Here we will provide the intermediate steps needed to prove the important identity due to Ecker and Honerkamp~\cite{Ecker:1972tii}:
\begin{equation}
  \tag{\ref{eq:DmuRabc}}
(D_\mu)_{ab;c} \:=\:R_{abcm}\:\partial_\mu\Phi^m\,.
\end{equation}
The EH identity has proven to be fundamental in writing down compact expressions for the supervertices, as discussed in Section \ref{sec:CovInts} and will be further applied for the covariant scalar-fermion four-vertex in the next subsection. 

To start with, we treat the covariant derivative, 
\begin{equation}
(D_\mu)_{ab}\ =\ \left[G_{AB}(x_A) \:\partial_{\mu}^{(A)} +\Gamma_{ABM}(x_A)\:\partial_\mu\Phi^M(x_A)\right]\delta(x_A-x_B)\,,\nonumber
\end{equation}
which was defined by~\eqref{eq:Dmuab} in Section~\ref{sec:CovInts},
as a rank-2 tensor in the configuration space. As such, we use 
the standard rules of covariant derivatives for rank-2 tensors with lower indices to obtain
\begin{equation}
   \label{a15}
    (D_\mu)_{ab;c} \:=\:  (D_\mu)_{ab,c} - (D_\mu)_{am} \Gamma^{m}_{\:\:bc}-(-1)^{b(m+a)}(D_\mu)_{mb} \Gamma^{m}_{\:\:ac}\,.
\end{equation}
Here we reiterate that repeated configuration-space indices imply two things: (i)~summation over their corresponding field-space indices, and (ii)~integration over their respective spacetime coordinates. Our goal is to prove the equivalence between~\eqref{a15} and~\eqref{eq:DmuRabc}. 

To this end, we first write $(D_\mu)_{ab,c}$ explicitly as
\begin{eqnarray}
   \label{a5}
   (D_\mu)_{ab,c}\!&=&\!   G_{AB,C}(x_A)\, \delta(x_A-x_C)\:\partial_{\mu}^{(A)}\delta(x_A-x_B)+\Gamma_{ABM}(x_A)\:\delta(x_A-x_B)\:\delta^{M}_{\:\:C}\:\partial_\mu^{(A)}\delta(x_A-x_C)\nonumber\\
    &&\! +(-1)^{CM}\Gamma_{ABM,C}(x_A)\:\partial_\mu\Phi^M(x_A)\: \delta(x_A-x_B)\delta(x_A-x_C)\,, 
\end{eqnarray}
and then the two metric-connection dependent terms that occur on the RHS of~\eqref{a15}, 
\begin{eqnarray}
  \label{a16}
(D_\mu)_{am} \Gamma^{m}_{\ bc} \!&=&\! 
\int_{x_M}\left[G_{AM}(x_A) \:\partial_{\mu}^{(A)}\delta(x_A-x_M) +\Gamma_{AMN}(x_A)\:\partial_\mu\Phi^N(x_A)\:\delta(x_A-x_M) \right]\nonumber\\
&&\:\times\,\Gamma^{M}_{\:\:BC}(x_M)\:\delta(x_M-x_B)\:\delta(x_M-x_C)\,,\\[3mm]
  \label{a16pt2}
(D_\mu)_{mb} \Gamma^{m}_{\ ac} \!&=&\! 
\int_{x_M}\left[G_{MB}(x_M) \:\partial_{\mu}^{(M)}\delta(x_M-x_B) +\Gamma_{MBN}(x_M)\:\partial_\mu\Phi^N(x_M)\:\delta(x_M-x_B) 
\right]\nonumber\\
&&\:\times\, \Gamma^{M}_{\:\:AC}(x_M)\:\delta(x_M-x_A)\:\delta(x_M-x_C)\;.
\end{eqnarray}

Our next step is to change the spacetime dependence of the metric $G_{AM}$ in~\eqref{a16} from $x_A \to x_M$, which will enable us to contract $G_{AM}$ with the metric connection~$\Gamma^{M}_{\ BC}(x_M)$. This step can be accomplished by making a Taylor series expansion about $x_M$,
\begin{equation}
  \label{a4}
G_{AM}(x_A)\: =\: G_{AM}(x_M)\,+\,(x_A-x_M)_\mu \partial^\mu G_{AM}(x_M)\,+\,\frac{1}{2}(x_A-x_M)^2\:\partial^2 G_{AM}(x_M)\,+\ldots\,, 
\end{equation}
together with the metric identity, 
\begin{equation}
  \label{eq:GABC}
    G_{AB,C}\ =\ \Gamma_{ABC}+(-1)^{AB}\Gamma_{BAC}\,,
\end{equation}
and the following $\delta$-function identities:
\begin{equation}
  \label{eq:delta}
    \begin{aligned}
      z\,\delta(z)&=0\,,  \qquad \qquad \quad \ \:  z\,\delta^\prime(z)=- \delta(z)\,,\\
      z\,\delta^{\prime\prime}(z)&=- 2\,\delta^\prime(z)\,, \qquad z^2\,\delta^{\prime\prime}(z)= 2\,\delta(z)\;,
    \end{aligned}
\end{equation}
where $z= x_A-x_M$ and the symbol prime $(\,{}^\prime\, )$ stands for differentiation with respect to the first argument $x_A$ of the $\delta$-function.

With the help of~\eqref{a4}, \eqref{eq:GABC} and \eqref{eq:delta},
we may then rewrite \eqref{a16} as
\begin{equation}
  \label{a10}
\begin{aligned}
&\int_{x_M}\left[G_{AM}(x_A) \:\partial_{\mu}^{(A)}\delta(x_A-x_M)\, +\,\Gamma_{AMN}(x_A)\:\partial_\mu\Phi^N(x_A)\:
\delta(x_A-x_M) \right]\\
&\qquad\times\,\Gamma^{M}_{\:\:BC}(x_M)\:\delta(x_M-x_B)\:\delta(x_M-x_C)\\
&=-\int_{x_M} \partial_\mu^{(M)}\big[\Gamma_{ABC}(x_M)\:\delta(x_M-x_B)\:\delta(x_M-x_C)\big]\:\delta(x_M-x_A)\\
&\qquad+\left(2\:\Gamma_{AMN}(x_A)\, +\, (-1)^{AM}\Gamma_{MAN}(x_A)\right)\partial_\mu\Phi^N(x_A)\:
\Gamma^{M}_{\:\:BC}(x_A)\:\delta(x_A-x_B)\:\delta(x_A-x_C)\;.
\end{aligned}
\end{equation}
By combining \eqref{a5}, \eqref{a16pt2} and \eqref{a10}, Equation~\eqref{a16} can be brought into the form:
\begin{equation}\label{a19}
\begin{aligned}
  (D_\mu)_{ab;c}\: =&\ \Big( (-1)^{MC}\,\Gamma_{ABM,C}(x_A)\, -\,\Gamma_{ABC,M}(x_A)\Big)\:\partial_\mu\Phi^M(x_A)\:\delta(x_A-x_B)\:\delta(x_A-x_C)  \\ 
  &\ +(-1)^{AM}\Big(\Gamma_{MAN}(x_A)\:\partial_\mu\Phi^N(x_A)\:
  \Gamma^{M}_{\:\:BC}(x_A)\\
  &\ -\,(-1)^{C(M+B)}\,\Gamma_{MAC}(x_A)\:\Gamma^{M}_{\:\:BN}(x_A)\:
  \partial_\mu\Phi^N(x_A)\Big)\, \delta(x_A-x_B)\:\delta(x_A-x_C)\,.
\end{aligned}
\end{equation}
This last expression can in turn be written exactly in the form stated in~\eqref{eq:DmuRabc} by swapping the dummy indices $M$ and $N$ for the last two terms in~\eqref{a19} and also noting that the super-Riemann tensor with lower indices in the configuration space is given by
\begin{equation}
    R_{abcd}\ =\ R_{ABCD}(x_A)\:\delta(x_A-x_B)\, \delta(x_A-x_C)\,\delta(x_A-x_D)\,, 
\end{equation}
where 
\begin{equation}
\begin{aligned}
    R_{ABCD}\ =&\ \: G_{AM}\,R^{M}_{\:\:BCD}\\
    =&\ \: (-1)^{CD}\Gamma_{ABD,C}\: -\: \Gamma_{ABC,D}\: +\:(-1)^{M(A+D)+D(B+C)}\Gamma_{MAD}\,\Gamma^{M}_{\:\:BC}\\
    &\ \: -(-1)^{(M+A)(B+D)+CD}\,\Gamma_{MBD}\,\Gamma^{M}_{\:\:AC}\;.
\end{aligned}
\end{equation}
This completes our proof of the EH identity for supermanifolds.

\subsection{Covariant Scalar-Fermion Four-Vertex}\label{subsec:Sabcd}

We may now use successively~\eqref{eq:Dmuab}, along with the EH identity~\eqref{eq:DmuRabc}, to carry out covariant functional differentiations on the action $S \equiv \int d^dx\, {\cal L}$ in the configuration space. This is the route we followed to obtain from $S_a$ in~\eqref{eq:Sa} the covariant inverse propagator $S_{ab}$ in~\eqref{eq:Sab} and the three-point scalar-fermion interaction $S_{abc}$ in~\eqref{eq:Sabc}. By taking one more 
field derivative and after some tedious super-algebra, we arrive at the scalar-fermion four-point interaction,
\begin{equation}
  \label{eq:Sabcd}
\begin{aligned}
    &S_{abcd}\:=\: \partial_\mu\Phi^m\left((-1)^{p(c+d)}\:_m k_n\:R^{n}_{\:\:abp;cd}+\:_m k_n\:R^{n}_{\:\:abq}R^{q}_{\:\:cdp}+(-1)^{an+dp}\:_m k_{n;[a}\:R^{n}_{\:\:bcp;d]}\right.\\
    &\quad\left.+(-1)^{n(a+d)+d(b+c)}\:_m k_{n;[a\underline{d}}R^{n}_{\:\:bc]p}+(-1)^{p(a+b+c+d)}\:_m k_{p;abcd}\right)\partial^\mu\Phi^p\\
    &\quad+(-1)^{am}(D_\mu)^m_{\:\:[a}\left((-1)^{dp}\:_m k_n\:R^{n}_{\:\:bcp;d]}+(-1)^{bn}\:_m k_{n;b}\:R^{n}_{\:\:cd]p}+(-1)^{p(b+c+d)}\:_m k_{p;bcd]}\right)\partial^\mu\Phi^p\\
    &\quad+\partial_\mu\Phi^m\left((-1)^{cp}\:_m k_n\:R^{n}_{\:\:abp;[c}+(-1)^{an}\:_m k_{n;[a}\:R^{n}_{\:\:bcp}\right)(D^\mu)^p_{\:\:d]}+(-1)^{am+p(b+c)}(D_\mu)^m_{\:\:[a}\:_m k_{p;bc]}(D^\mu)^p_{\:\:d}\\
    &\quad+(-1)^{b(n+a)+d(m+a+b+c)}(D_\mu)^m_{\:\:[d}\:_m k_{n;b}\:R^{n}_{\:\:ac]p}\partial^\mu\Phi^p+(-1)^{b(n+a)+c(m+a+b)}(D_\mu)^m_{\:\:[c}\:_m k_{n;b}\:R^{n}_{\:\:ad]p}\partial^\mu\Phi^p\\
    &\quad+(-1)^{p(b+c)+am+d(m+b+c)}R^{m}_{\:\:[a\underline{d}q}\partial_\mu\Phi^q\:_m k_{p;bc]}\:\partial^\mu\Phi^p+(-1)^{am+n(b+d)+cd}(D_\mu)^m_{\:\:(a}\:_m k_{n;b\underline{d}}(D^\mu)^n_{\:\:c)}\\
    &\quad+(-1)^{am}(D_\mu)^m_{\:\:[a}\:_m k_{n}\:R^{n}_{\:\:bcp}(D^\mu)^p_{\:\:d]}+(-1)^{m(a+d)+d(b+c)}R^{m}_{\:\:[a\underline{d}q}\partial_\mu\Phi^q\:_m k_{n}R^{n}_{\:\:bc]p}\:\partial^\mu\Phi^p\\
    &\quad+i(-1)^{a}\left(\:_a\lambda^{\mu}_{m}\:R^{m}_{\:\:bcp}(D_\mu)^p_{\:\:d}+(-1)^{bm+d(m+c)}\:_a\lambda^{\mu}_{m;[b\underline{d}}(D_\mu)^m_{\:\:c]}+(-1)^{m(b+c)}\:_a\lambda^{\mu}_{m;bc}(D_\mu)^m_{\:\:d}\right)\\
&\quad+i(-1)^{a}\left((-1)^{dp}\:_a\lambda^{\mu}_{m}\:R^{m}_{\:\:bcp;d}+(-1)^{bm}\:_a\lambda^{\mu}_{m;[b}\:R^{m}_{\:\:cd]p}+(-1)^{m(b+c+d)}\:_a\lambda^{\mu}_{p;bcd}\right)\partial_\mu\Phi^p\: -\: U_{abcd}\:.
\end{aligned}
\end{equation}
We should note that in~\eqref{eq:Sabcd} the underlined index $d$, i.e.~$\underline{d}$, albeit being free, it still does not enter the symmetrisation process. It only keeps its placing with respect to the other free indices, $a$, $b$, and $c$, which get symmetrised or cyclically permuted according to~\eqref{eq:operations}. As~we discuss in Section~\ref{sec:CovSFaction}, it is the supersymmetrised part of $S_{abcd}$, i.e.~$S_{\{abcd\}}$, that enters the SG-QEA beyond the one-loop level. As we did for $S_{abc}$ in~\eqref{eq:Sabc}, we have checked that in the homogeneous limit $\partial_\mu {\bf \Phi} \to 0$, the four-supervertex $S_{abcd}$ in~\eqref{eq:Sabcd} reproduces the analytic result presented in~\cite{Gattus:2023gep,Gattus:2024spj}. Clearly, the complexity of the expressions increases drastically, as we are calculating more and more derivatives in the configuration space. 

\setcounter{equation}{0}
\section{Spin Connection in SG-QFTs}\label{app:spin_connection}

In Section \ref{sec:SGQFT}, we have introduced the field-space vielbeins $\:_A e^{\widehat{B}}$, where hatted ($\:\widehat{}\:$) indices refer to local-frame quantities. In what follows, we assume that unhatted indices are raised and lowered by the covariant supermanifold metric $_A G_B$ and its inverse, while hatted indices are raised and lowered by the local field-space metric $_{\widehat{A}} H_{\widehat{B}}$. 
Given that vielbeins have a mixed coordinate and non-coordinate nature, it is possible to define two types of metric connections: (i)~$\Gamma^{A}_{\:\:B C}$ for the usual global tensors and (ii)~$\omega_{B\ \widehat{C}}^{\:\widehat{A}}$ when applied to tensors that carry local-frame indices. The latter connection is termed the \textit{spin connection}. The vielbeins in question satisfy the so called tetrad postulate which is just the requirement that the total covariant derivative of the vielbeins should vanish, i.e.
\begin{equation}\label{eq:tertad_post}
    \:^{\widehat{A}} e^{\st}_{\:\:A\:;B}\: =\:  \:^{\widehat{A}} e^{\st}_{\:\:A\:,B}-\:^{\widehat{A}} e^{\st}_{\:\:M}\:\Gamma^{M}_{\:\:AB}+(-1)^{B(\widehat{A}+A)}\:\omega_{B\:\:\:\:\widehat{B}}^{\:\:\widehat{A}}\:\:^{\widehat{B}} e^{\st}_{\:\:A}\: =\: 0\,.
\end{equation}
One can then solve \eqref{eq:tertad_post} to obtain the definition of the field-space spin connection reported in~\eqref{eq:spinomega}. Hence, $\omega_{B\:\:\:\:\widehat{C}}^{\:\:\widehat{A}}$ is a vector in the unhatted indices, a non-tensor in the hatted index and is antisymmetric with respect to the local-frame indices, 
i.e.~$\omega_{C}^{\:\:\widehat{A}\widehat{B}}=-(-1)^{\widehat{A}\widehat{B}}\omega_{C}^{\:\:\widehat{B}\widehat{A}}$.

Another equivalent definition of the field-space spin connection can be obtained in terms of the standard\- connection with all local-frame indices
\begin{equation}\label{omega_local}
\omega_{B\:\:\:\:\widehat{B}}^{\:\:\widehat{A}}=(-1)^{B(\widehat{A}+\widehat{B})}\:\Gamma^{\widehat{A}}_{\:\:\widehat{B}\widehat{C}}\:^{\widehat{C}}e^{\st}_{\:\:B}\:.
\end{equation}
The spin connection generally encodes the inertial effects associated to a given frame.
Hence, in inertial frames where the standard connection is set to zero, inertial effects are absent and the spin-connection vanishes. From \eqref{omega_local}, one can also conclude that in the absence of torsion the following identity holds
\begin{equation}
\label{d10}  
\omega_{B\ \widehat{B}}^{\ \widehat{A}}\:^{\widehat{B}}e^{\st}_{\:\:C}\: =\: (-1)^{C \widehat{A}+B(\widehat{A}+C)}\, \omega_{C\: \widehat{B}}^{\:\widehat{A}}\:^{\widehat{B}}e^{\st}_{\:\:B}\:.
\end{equation}
We will use the equality above to obtain an expression for the operator $(D_\mu)^A_{\ B}$ in the local frame. Since $(D_\mu)^A_{\ B}$ is a proper tensor in field-space, its local frame counterpart is obtained by making use of the vielbeins as usual, i.e.
\begin{equation}\label{d11}
    (D_\mu)^A_{\ B} = \:^A e_{\widehat{A}}\:\:(D_\mu)^{\widehat{A}}_{\ \widehat{B}}\:\:^{\widehat{B}}e^{\st}_{\:\:B}\: = \:^A e_{\widehat{A}}\:\:  (\partial_\mu \Phi^{\widehat{A}})_{;\widehat{B}}\:\:^{\widehat{B}}e^{\st}_{\:\:B}\; ,
\end{equation}
where we have used definition \eqref{eq:Dmuab} to write the expression on the RHS. 
Inserting the definition of $(\partial_\mu \Phi^{\widehat{A}})_{;\widehat{B}}$ in the local frame in \eqref{d11} and using~\eqref{d10} and~\eqref{omega_local}, we obtain an expression for $(D_\mu)^A_{\ B}$ in terms of the vielbeins and the spin connection,
\begin{equation}
\begin{aligned}
    (D_\mu)^A_{\ B} \ &=\ \:^A e_{\widehat{A}}\left[{}^{\widehat{A}}\delta_{\widehat{B}}\:\partial_\mu + \omega_{\mu\ \widehat{B}}^{\:\widehat{A}}\right]\:^{\widehat{B}}e^{\st}_{\:\:B}\\
    \ &=\ \:^A e_{\widehat{M}}\:^{\widehat{M}}e^{\st}_{\:\:B}\:\partial_\mu + \:^A e_{\widehat{M}}\left(\:^{\widehat{M}}e^{\st}_{\:\:B,C}\right)\partial_\mu \Phi^{C}+\:^A e_{\widehat{A}}\:\;\omega_{\mu\ \widehat{B}}^{\:\widehat{A}}\:^{\widehat{B}}e^{\st}_{\:\:B}\; ,
\end{aligned}
\end{equation}
where in configuration space all the quantities above depend on the spacetime variable $x_A$. In~addition, we have introduced the short-hand notation,
\begin{equation}
\omega^{\:\:\widehat{A}}_{\mu\:\:\:\widehat{B}}\ \equiv\ (-1)^C\:\partial_\mu \Phi^{C}\:\:\omega_{C\ \widehat{B}}^{\:\widehat{A}}\;.
\end{equation}
Alternatively, one can preserve the different spacetime dependence of the vielbeins in configuration space and write the following more compact expression:
\begin{equation}
    (D_\mu)^a_{\ b} \ =\ \:^A e_{\widehat{A}}(x_A)\left[\,{}^{\widehat{A}}\delta_{\widehat{B}}\:\partial_\mu^{(A)}  \delta (x_A-x_B)+\omega_{\mu\ \widehat{B}}^{\:\widehat{A}}(x_A)\:\delta (x_A-x_B)\,\right]\:^{\widehat{B}}e^{\st}_{\:\:B}(x_B)\, .
\end{equation}
To recover the original definition~\eqref{eq:aDmub}, it is sufficient to expand the rightmost vielbein field as prescribed in \eqref{a4} in order to change the spacetime dependence from $x_B \to x_A$. 

One can proceed in a similar fashion to obtain an expression for $D^2$ in the local field-space frame, i.e.
\begin{equation}\label{d17}
     (D^2)^A_{\:B} \ =\ \:^A e_{\widehat{A}}\left[\,{}^{\widehat{A}}\delta_{\widehat{B}}\:\partial^2 + 2\; \omega_{\mu\ \widehat{B}}^{\:\widehat{A}}\: \partial^\mu +\left(\partial^\mu\omega_{\mu\ \widehat{B}}^{\:\widehat{A}}\right)+\omega_{\mu\ \widehat{M}}^{\:\widehat{A}}\:\:
\omega_{\quad\: \widehat{B}}^{\mu\:\widehat{M}}\,\right]\:^{\widehat{B}}e^{\st}_{\:\:B}\; .
\end{equation}
As before, the corresponding configuration space operator reads
\begin{equation}
     (D^2)^a_{\:b} \ =\  (D^2)^A_{\:\:B}(x_A)\:\delta (x_A-x_B)\,,
\end{equation}
which can be equivalently written as
\begin{equation}\label{d19}
\begin{aligned}
    (D^2)^a_{\:b} \ =\ &\:^A e_{\widehat{A}}(x_A)\left[{}^{\widehat{A}}\delta_{\widehat{B}}\:\partial^2 \delta (x_A-x_B)+ 2\; \omega_{\mu\ \widehat{B}}^{\:\widehat{A}}\: \partial^\mu \delta (x_A-x_B) \right.\\    &\left.\hspace{1.8cm}+\left(\partial^\mu\omega_{\mu\ \widehat{B}}^{\:\widehat{A}}
    + \omega_{\mu\ \widehat{M}}^{\:\widehat{A}}\:\:\omega_{\quad\:\widehat{B}}^{\mu\:\widehat{M}}\right)\delta (x_A-x_B)\right]\:^{\widehat{B}}e^{\st}_{\:\:B}(x_B)\, .
\end{aligned}
\end{equation}
To compute the covariant effective action, we require for the operator $D^2$ to be exponentiated. This involves an extra degree of difficulty since now the splitting of the Laplacian operator and the remaining terms in \eqref{d19} has to be done via the Zassenhaus formula.
Upon performing a series expansion in $t$, one recovers  \eqref{eq:HKD2},
\begin{equation}
     \langle x_A|  e^{tD^2}|x_B\rangle\ =\ K^A_{\ B} \: \frac{e^{-(x_A-x_B)^2/{4t}}}{(4\pi t)^{d/2}}\ +\ W^A_{\ B} \;,\nonumber
\end{equation}
where $K^A_{\ B}$ is defined in \eqref{eq:KAB},
and up to order $t^3$,
\begin{equation}
    W^A_{\ B}\: =\: \int_{x_M}\!\frac{e^{-(x_A-x_M)^2/{4t}}}{(4\pi t)^{d/2}}\  {}^A e_{\widehat{A}}(x_A)\,\big< x_M\big|\,t\: {}^{\widehat{A}}\Omega_{\widehat{B}}+\frac{t^2}{2}\left(\:{}^{\widehat{A}}\Omega_{\widehat{M}}\:{}^{\widehat{M}}\Omega_{\widehat{B}} -\big[\partial^2\,, {}^{\widehat{A}}\Omega_{\widehat{B}}\big]\right) \big|x_B
    \big> \: {}^{\widehat{B}}e^{\st}_B(x_B)\,.
\end{equation}
In the above expression, we have used the shorthand notation, 
\begin{equation}
    \:\Omega\ \equiv\ 2\; \omega_{\mu}\,\partial^\mu\: +\:\left(\partial^\mu\omega_{\mu}\right)\:+\:\omega_{\mu}\,\omega^{\mu}\; ,
\end{equation}
where the field-space indices have been suppressed. Finally, it should be commented that the heat-kernel, given in~\eqref{eq:HKD2} for a curved field-space, differs fundamentally in its analytic structure from the one derived for gravitational theories of curved spacetime  (cf.~\cite{Barvinsky:1985an,Barvinsky:2021ijq}).

\setcounter{equation}{0}
\section{The Double Heat Kernel Method}\label{app:DoubleHKM}

In evaluating the first term of the effective action~$\Gamma^{(1)}$ in~\eqref{eq:G1SFren}, we have employed the Zassenhaus formula in the following way: 
\begin{eqnarray}
  \label{eq:DoubleHKM}
 \Gamma^{(1)}\!\!&\supset&\! \frac{i}{2}\operatorname{tr}\ln\Big(\!\!-\partial^2 -m_b^2-M\Big)\ \supset\ 
 -\frac{i}{2}\int_{x,y}\int_0^\infty\! \frac{dt}{t}\,
 \langle x| e^{t(\partial^2 + m^2_b +M)}|y\rangle\,\langle y| 
 e^{tM}\, e^{-\frac{t^2}{2} [\partial^2,M]} |x\rangle
 \nonumber\\
 \!\!&=&\! -\,\frac{i}{2}\int_{x,y}\int_0^\infty\! \frac{dt}{t}\,
 \langle x| e^{t(\partial^2 + m^2_b +M)}|y\rangle\,
 \langle y|\big[\,{\bf 1}\: +\: tM\: +\: \frac{t^2}{2}\, \big(M^2 - [\partial^2,M]\big)\big]|x\rangle
 \:+\ldots\,,\qquad
\end{eqnarray}
where the field-space operator $M$ is given in~\eqref{eq:Defs_M_N} and the ellipses represent powers of $t^3$ and higher in the Schwinger proper time~$t$. 

In the last equation given in~\eqref{eq:DoubleHKM}, the first term within the expectation value, $\langle y|\cdots |x\rangle$, is the unity operator~${\bf 1}$, which leads to an integral that can be computed using the standard heat kernel method. However, the computation of its second term, $\langle y|M|x\rangle$, becomes more elaborate, as it also contains a fermion-field dependent operator~$O_F(x,y)$ in the configuration space:
\begin{equation}
   \label{eq:InvFoperator}
\langle y|M|x\rangle\ \supset\ h^2\,
\langle y|\bar{\psi}\big(i\slashed{\partial}-h\phi-m_f\big)^{-1}\psi| x\rangle\: \equiv\: h^2\,O_F(x,y)\, .
\end{equation}
In this appendix, we give further technical details related to its computation. As we will see below, such a calculation necessitates the introduction of an additional Schwinger time~$s$ (besides $t$) and as such, it was termed the {\em double heat kernel method}.

Upon inserting an identity of states to isolate the kernel, we may compute the fermionic operator $O_F(x,y)$ in~\eqref{eq:InvFoperator} by making use of the bosonisation property~\eqref{eq:DiracSq}:
\begin{equation}
  \label{eq:InvFprop}
\begin{aligned}
  O_F(x,y)\ =&\ \int_{v,z,w}  \langle y|\bar{\psi}| v\rangle \langle v|\left(i\slashed{\partial}+h\phi+m_f\right)| z\rangle\\
  &\ \times\,\langle z|\left[-\partial^2+ih\slashed{\partial}\phi-(h\phi+m_f)^2\right]^{-1}| w\rangle\langle w|\psi 
     |x\rangle\, .
\end{aligned}
\end{equation}
The expectation value of the new bosonic inverse operator in the integrand of~\eqref{eq:InvFprop} can be represented as an integral over a new Schwinger proper time parameter, which we call $s$, i.e. 
\begin{equation}
    \langle z|\left[-\partial^2+ih\slashed{\partial}\phi-(h\phi+m_f)^2\right]^{-1}| w\rangle\: =\: \int_0^\infty ds \langle z|e^{-s\left[-\partial^2+ih\slashed{\partial}\phi-(h\phi+m_f)^2\right]}| w\rangle\,.
\end{equation}
This in turn can be evaluated by means of the Zassenhaus formula,
\begin{equation}
  \label{eq:InvBosZass}
\begin{aligned}
   \int_0^\infty ds\, \langle z|e^{-s\left[-\partial^2-m^2_f +A\right]}| w\rangle&= \int_0^\infty ds\, \frac{e^{sm^2_f -(z-w)^2/{4s}}}{(4\pi s)^{d/2}}\left[1-sA+\frac{1}{2}s^2(A^2-\partial^2A) -\frac{1}{6}s^3 A^3\right.\\
   &\hspace{-5mm}\left.+s^3\left(\frac{1}{3}(\partial A)^2+\frac{1}{2}A\partial^2A \right)+\frac{1}{2s}(z-w)_{\mu}\left(s^2\partial^{\mu} A-s^3 A \partial^{\mu} A\right)\right]\,,
\end{aligned}
\end{equation}
where $A=ih\slashed{\partial}\phi-h^2\phi^2-2h\phi m_f$. We may substitute \eqref{eq:InvBosZass} in \eqref{eq:InvFprop} and notice that the expression in the square brackets is a function of $w$, as well as that $A$ does not contain any differential operators. Then, upon integration over $v$ and $w$, the fermionic operator $O_F(x,y)$ stated in~\eqref{eq:InvFprop} becomes 
\begin{eqnarray}
  \label{eq:InvFerZass}
      O_F(x,y)\!\!&=&\!\!
     \int_z\int_0^\infty ds\: \bar{\psi}(y)\left[ \left(i\slashed{\partial}_y+h\phi(y)+m_f\right)\delta(y-z)\right]\frac{e^{sm^2_f -(z-x)^2/{4s}}}{(4\pi s)^{d/2}}\nonumber\\
&&\times \left[1-sA+\frac{1}{2}s^2(A^2-\partial^2A) -\frac{1}{6}s^3 A^3+s^3\left(\frac{1}{3}(\partial A)^2+\frac{1}{2}A\partial^2A \right)\right.\nonumber\\
&&\left.+\frac{1}{2s}(z-x)_{\mu}\left(s^2\partial^{\mu} A-s^3 A \partial^{\mu} A\right)\right]\psi(x)\,.
\end{eqnarray}

If we insert~\eqref{eq:InvFerZass} in~\eqref{eq:InvFoperator} and the resulting expression of $\langle y|M|x\rangle$ back in the last equation for the effective action in~\eqref{eq:DoubleHKM}, we arrive at a multi-variable integral which has the form:
\begin{equation}
  \label{eq:Ifermion}
\begin{aligned}
     I\: =\: \int_{x,y}\int_0^\infty \frac{dt}{t}\int_0^\infty ds\:\:&t \frac{e^{-(x-y)^2/{4t}+tm^2_b}}{(4\pi t)^{d/2}} \left[f(y)-\frac{i}{2t}(x-y)_\mu \gamma^\mu l(y)\right] \frac{e^{-(y-x)^2/{4s}+sm^2_f }}{(4\pi s)^{d/2}}\\
     &\times\left[s^n g_n(x)+\frac{1}{2s}(y-x)_{\mu}s^m q_m^\mu(x)\right]\,.
\end{aligned}
\end{equation}
Here we should clarify that 
\begin{equation}
    I\: =\: \,\int_{x,y}\int_0^\infty\! dt\:
 \langle x| e^{t(\partial^2 + m^2_b +M)}|y\rangle\,O_F(x,y)\; ,
\end{equation}
and the four functions: $f(x),\ l(x),\ g_n(x),\ q^\mu_m(x)$
that appear in~\eqref{eq:Ifermion} acquire their spacetime dependence implicitly through the scalar field $\phi(x)$ and the fermion fields, $\psi(x)$ and $\bar{\psi}(x)$. Their explicit analytic form is given by
\begin{eqnarray}
    \label{eq:flgq}
    f(x) \!\!&=&\!\! -i\,\slashed{\partial} \bar{\psi}(x)\: +\: \big(m_f + h\phi(x)\big)\bar{\psi}(x)\,,\qquad\qquad
    l(x) \: =\: \bar{\psi}(x)\,,\nonumber\\
    g_n(x) \!\!& = &\!\!
    \begin{cases}
    1 & \:\:\text{for}\: n=0\,,\nonumber\\
    -A(x)\psi(x) & \:\:\text{for}\: n=1\,,\\
    (A^2(x)-\partial^2A(x)\psi(x) & \:\:\text{for}\: n=2\,,\\
    -\frac{1}{6}A^3(x)\psi(x)+\left(\frac{1}{3}(\partial A)^2+\frac{1}{2}A\partial^2A \right)\psi(x) & \:\:\text{for}\: n=3\,,\\
    \end{cases}\\[-3mm]
    && \\
    q_m^\mu(x) \!\!& =&\!\! 
    \begin{cases}
    0 & \:\:\text{for}\: m=0,1\,,\nonumber\\
    \partial^{\mu} A(x)\:\psi(x) & \:\:\text{for}\: m=2\,,\\
    -A(x) \partial^{\mu} A(x)\:\psi(x) & \:\:\text{for}\: m=3\;,\\
    \end{cases}
\end{eqnarray}
as they can be read off from \eqref{eq:InvFerZass}. 

After changing the integration variables $(x,y)\rightarrow(x-y,y)$ and using the properties of Fourier transforms, \eqref{eq:Ifermion} may be recast into the form:
\begin{eqnarray}
   \label{eq:Itransf}
I\!\!&=&\!\! \int_x\: \int_0^\infty \!dt\int_0^\infty \!ds\: \frac{e^{tm^2_b+sm^2_f }}{(4\pi t)^{d/2}\,(4\pi s)^{d/2}}\Bigg\{f(x)\left(\frac{\pi}{a}\right)^{d/2}e^{\frac{1}{4a}\,\partial^2/\partial x^2}\left[s^n g_n(x)-\frac{a^{-d}}{8s}s^m\partial_\mu q^\mu_m(x)\right]\nonumber\\
&&+\left(-\frac{i}{4t}\right)l(x)e^{\frac{1}{4a}\,\partial^2/\partial x^2}\left[\frac{1}{2}\left(\frac{\pi}{a^3}\right)^{d/2}s^n \slashed{\partial}g_n(x)-\frac{1}{16s}\left(\frac{\pi}{a^5}\right)^{d/2}s^m\left(2a+\partial^2\right) \slashed{q}_m(x)\right]\Bigg\},\qquad\quad
\end{eqnarray}
with $a=\frac{1}{4t}+\frac{1}{4s}$. While this last expression may look complicated in $d=4-2\epsilon$ dimensions, only terms~$\mathcal{O}(s^0)$ dictate the UV behaviour of the theory. All terms with higher powers of $s$ lead to UV-finite contributions. After this observation, the UV-divergent part of \eqref{eq:Itransf} was found to be
\begin{equation}
    I_{\text{UV}}=\frac{1}{16\pi^2\epsilon}\int d^d x\left[-i(\slashed{\partial}\bar{\psi})\psi+(h\phi+m_f)\bar{\psi}\psi\right].
\end{equation}
A similar UV-divergent term is found when computing the trace of the operator: $$\psi^\tran\,\big[ (i\slashed{\partial}+h\phi+m_f)^\tran\big]^{-1}\:
\bar{\psi}^\tran\,,$$
where one has to replace $h$ and $m_f$ with their negatives.
Note that terms containing higher powers of $t$ will be UV finite, with the exception of the $\phi^3$ and $\phi^4$ contributions.

\vfill\eject 

\setcounter{equation}{0}
\section{Dimensional Regularisation in Coordinate Space}\label{app:DRxspace}

In this appendix, we will give useful formulae that permit us to evaluate loop integrals in DR, but in the coordinate or $x$-space. Before doing so, we briefly review how DR is applied to loop integrals in momentum or $k$-space. The translation of loop integrals from $k$- to $x$-space goes through the so-called Gegenbauer method, which we will not present here explicitly. The interested reader may consult Appendix~C of the lecture notes by Pascual and Tarrach in~\cite{Pascual:1984zb}. In the following, we will consider two types of coordinate-space integrals for the heat kernels: (i) the massless heat kernel and (ii) the massive heat kernel. 

\subsection{Massless Heat Kernels}\label{subapp:NoMassHK}

To extract the UV behaviour of a theory when complicated differential operators are involved, such as in~\eqref{eq:DeltaG1Scalar}, it might be useful to evaluate firstly the integral over the Schwinger proper time $t$ and then carry out any remaining integrals over position space variables. For massless heat kernels, we encounter the following type of integrals:  
\begin{equation}\label{c1}
\begin{aligned}
    \int_{0}^{\infty}dt\,\frac{e^{-(x_A-x_B)^2/{4t}}}{(4\pi t)^{d/2}}\,t^n \ =\ 2^{-2+2n}\,\pi^{-d/2}\,\left|x_A-x_B\right|^{-d+2+2n}\:\Gamma\left(-1+\frac{d}{2}-n\right)\,,
\end{aligned}
\end{equation}
with $|x|\equiv [(x)^2]^{1/2}$ and $n \in \mathbb{Z}$.
Then, we must evaluate a coordinate-space integral over vanishing spatial separations, i.e.~when $|x_A-x_B|\to 0$. To accomplish this task, one needs to perform this integration directly to $x$-space in the DR, rather than in the often-considered $k$-space. 

Let us therefore start by reiterating the well-known one-loop result for $d$-dimensional momentum integrals in Minkowski space:
\begin{equation}
\label{eq:k_int}
   \int \frac{d^d k}{(2\pi)^d}\,\frac{1}{(k^2-\Delta)^\alpha}\ =\ \frac{i\,(-1)^\alpha}{(4\pi)^{d/2}}\:\frac{\Gamma(\alpha-d/2)}{\Gamma(\alpha)}\left(\frac{1}{\Delta}\right)^{\alpha-d/2},
\end{equation}
with $\alpha \in \mathbb{R}$. The condition of vanishing tadpoles  in $d=4-2\epsilon$ dimensions is imposed by requiring the momentum integral \eqref{eq:k_int} with $\alpha=1$ to go to zero, i.e.
\begin{equation}
    \int \frac{d^d k}{(2\pi)^d}\,\frac{1}{k^2}\ =\ -\frac{i}{(4\pi)^{d/2}}\:
    \frac{\Gamma(1-d/2)}{\Gamma(1)}\:
    \lim_{\Delta\rightarrow0}
    \left(\frac{1}{\Delta}\right)^{1-d/2} =\ 0\,.
\end{equation}

Our aim is now to obtain an expression equivalent to \eqref{eq:k_int}, but in $x$-space. Using Schwinger parameters,  one may represent the power law of an $x^2$ function as
\begin{equation}
    \frac{1}{[x^2]^{d/2-\alpha}}\ =\ \frac{1}{(d/2-\alpha-1)!}\,\int_0^{\infty}du\: u^{d/2-\alpha-1}e^{-ux^2}\ .
\end{equation}
Taking the Fourier transform of the expression above and integrating over the $d$-dimensional $x$-space yields
\begin{equation}
\label{eq:x_FT}
    \int d^dx\, \frac{e^{ik\cdot x}}{[x^2]^{d/2-\alpha}}\: =\: \frac{1}{\Gamma(d/2-\alpha)} \int_0^{\infty}du\: \pi^{d/2}e^{-k^2/{4u}}u^{-\alpha-1}\,,  
\end{equation}
where we have performed the integration over position space by completing the square. After a suitable change of variables, like $t=k^2/{4u}$, one may write the RHS of~\eqref{eq:x_FT} in terms of the integral representation of the usual $\Gamma(x)$-function as
\begin{equation}
\label{c6}
    \int d^dx\, \frac{e^{ik\cdot x}}{[x^2]^{d/2-\alpha}}\: =\: \frac{\pi^d/2}{\Gamma(d/2-\alpha)}\frac{4^{\alpha}}{[k^2]^{\alpha}}\int_0^{\infty}dt\:e^{-t}t^{\alpha-1}\, .
\end{equation}
Finally, integrating both sides of \eqref{c6} with respect to the $d-$dimensional momentum $k$ and making use of \eqref{eq:k_int}, we obtain
\begin{equation}
\label{eq:pos_space}
    \int d^dx\:\frac{\delta^{(d)}(x)}{[x^2]^{d/2-\alpha}}\: =\ i\, (-1)^\alpha 4^{\alpha -d/2}\:\frac{\Gamma(\alpha-d/2)}{\Gamma(d/2-\alpha)}\:\lim_{\Delta\rightarrow0}\left(\frac{1}{\Delta}\right)^{\alpha-d/2},
\end{equation}
which can be used to investigate the presence of $1/\epsilon$-poles directly from position space integrals. Hence, depending on the coefficient $\alpha$, \eqref{eq:pos_space} can either be zero, finite or exhibit IR divergences when $\lim_{\Delta\rightarrow0} (1/\Delta)$ appears. We observe that the integral \eqref{eq:pos_space} vanishes for $0\leq \alpha < 2$, it~is equal to $-i$ for $\alpha =2$, and becomes IR divergent when $\alpha > 2$.

\subsection{Massive Heat Kernels}\label{subapp:MassiveHK}

Let us first remark that even if a mass term is present in the potential $U$, like~in \eqref{eq:R_and_U} and in \eqref{eq:fermion_V}, the theory is effectively massless when a kernel of the form~\eqref{c1} is used. This in turn implies that in this massless scheme, the only non-zero contributions will arise from the wavefunction renormalisation, the correction to the Yukawa coupling and the scalar quartic interactions, as all other counterterms are mass dependent, cf.~\eqref{eq:CTSFmodel}. However, we may access the UV structure of the latter counterterms if we use a massive heat kernel.

For a massive heat kernel, its $x$-space Schwinger proper-time representation yields a more complicated expression than \eqref{c1}. In this case, the proper-time integral can be written in terms of the modified Bessel function of the second kind $K_\nu$ of order $\nu=(d/2-n)$,
\begin{equation}\label{c8}
    \int_{0}^\infty \frac{dt}{t}\: \frac{e^{-t \mu^2-(x-y)^2/{4t}}}{(4 \pi t)^{d/2}}\:t^n\ =\ (\mu^2)^{\frac{1}{4}(d-2n)}\,2^{1-d/2-n}\,\pi^{-d/2}\,|x-y|^{n-d/2}\,K_\nu\left(\sqrt{\mu^2}|x-y|\right)\,.
\end{equation}
To obtain the analytic result in~\eqref{c8}, one must use the specific integral representation of $K_\nu$~\cite{KBessel},
\begin{equation}
    K_\nu (z)\ =\ \frac{1}{2}\left(\frac{z}{2}\right)^\nu \int_0^\infty \exp\left(-t-\frac{z^2}{4t}\right)\,\frac{dt}{t^{\nu+1}}\ ,
\end{equation}
which also satisfies the standard property: $K_\nu (z)=K_{-\nu} (z)$. In this massive case, the simple power law function, $1/|x|^{d -2\alpha}$ in \eqref{eq:pos_space}, gets replaced with $K_\nu(\sqrt{\mu^2}|x-y|)$. To carry out the integral in $x$-space, one must first expand the Bessel function $K_\nu (z)$ for small arguments $z$~\cite{KBessel},
\begin{equation}
    K_\nu(z) = \frac{1}{2}\left[\Gamma(\nu)\left(\frac{z}{2}\right)^{-\nu}\left(1+\frac{z^2}{4(1-\nu)}+\cdots\right)+\Gamma(-\nu)\left(\frac{z}{2}\right)^{\nu}\left(1+\frac{z^2}{4(1+\nu)}+\cdots\right)\right].
\end{equation}
For convenience, one can then define a reduced Bessel function $\hat{K}_\nu(z)$ having the property, 
\begin{eqnarray}
    z^{-\nu}K_\nu(\mu z)\: =\: \frac{1}{2}\,\mu^\nu 2^{-\nu}\hat{K}_\nu(\mu z)\,,
\end{eqnarray}
so that if $\nu<0$,
\begin{equation}
    \lim_{z\rightarrow0}\hat{K}_\nu(\mu z)\: =\: \Gamma(-\nu)\,.
\end{equation}

For illustration, we will now consider an example where modified Bessel functions appear. Let us compute the following integral over Schwinger proper time:
\begin{equation}\label{c13}
I\ =\ \operatorname{tr}\int_{x,y}\int_0^\infty dt\: (i\slashed{\partial}+m_f)\delta(x-y)\:\frac{e^{-(y-x)^2/{4t}+m_f^2t}}{(4\pi t)^{d/2}}\:Y(x)\,,
\end{equation}
where $Y(x)= h \phi(x)$. Such an expression arises, for example, when computing the effective action for the simple Yukawa model, as given in \ref{subsec:Yukawa}. In $d=4-2\epsilon$ dimensions, the UV-divergent part of the integral~$I$ in~\eqref{c13} may explicitly be computed as 
\begin{equation}
\begin{aligned}
    I_{\text{UV}}\ &=\ -\frac{1}{4\pi^2}\int_{x,y}\delta(x-y)\:
    m_f\,(-m_f^2)^{(1-\epsilon)}\,\hat{K}_{1-\epsilon}\big(\sqrt{-m_f^2}|y-x|\big)\,Y(x)\\
    &=\ -\frac{1}{4\pi^2}\int_{x}m_f (-m_f^2)^{(1-\epsilon)}\,\Gamma(-1+\epsilon)\,Y(x)\\
    &=\ -\frac{1}{4\pi^2\epsilon}\int_{x}m_f^3\,h\phi(x)\;.\\
\end{aligned}
\end{equation}
This result exactly reproduces the counterterm $\delta_c$ in \eqref{eq:CTYuk}.

Let us now elucidate how Bessel functions techniques can be used to evaluate the Schwinger integrals in the double heat kernel method in \eqref{eq:I3x}, e.g.~to recover the flat field-space limit. In this limit, we have
\begin{equation}
    {}^{a}\lambda^\mu_b=\left(\begin{array}{cc}
        -\gamma^\mu & 0 \\
        0 & (\gamma^{\mu})^\tran
    \end{array}\right), \qquad (D_\mu)^a_{\:\:b}=\left(\begin{array}{cc}
        -\partial_\mu & 0 \\
        0 & \partial_{\mu}
    \end{array}\right).
\end{equation}
In a flat field-space, the potential like term $V$ in \eqref{eq:fermion_V} reduces to the scalar potential $U$ and both the $A$ and $B$ operators appearing in the definitions of $I_{1,2,3}$ vanish, since the $\lambda^\mu$-matrices are now field-independent. The only term that survives arises from \eqref{eq:I3}, where we choose now to include the mass term in the kernel so that $U=Y$. By doing so, the integral $I_3$ takes on the form,
\begin{equation}
  \label{eq:Iflat}
\begin{aligned}
    I_3^{\text{flat}}\ &=\ 
 \hspace{-3mm}\smashoperator[r]{\int\limits_{\substack{%
    x_A,x_B,\\x_C,x_D}}}(i \slashed{\partial}+m_f)\delta(x_A-x_B)\:   \int_{0}^{\infty}dt\:\frac{e^{-(x_B-x_C)^2/{4t} + m_f^2t}}{(4\pi t)^{d/2}}Y(x_C)\\[-2mm]
    &\quad \times\,(i \slashed{\partial}+m_f)\delta(x_C-x_D)\int_{0}^{\infty}ds\:\frac{e^{-(x_D-x_A)^2/{4s}+m_f^2s}}{(4\pi s)^{d/2}}Y(x_A)\,.
\end{aligned}
\end{equation}
Notice that each of the Schwinger proper-time integrals in~\eqref{eq:Iflat} is evaluated by means of~\eqref{c8}. In 
$d$ dimensions, the expression resulting from~\eqref{eq:Iflat} becomes then an integral over a product of two Bessel functions $K_\nu(z)$, i.e.
\begin{equation}
\begin{aligned}
I_3^{\text{flat}}\ &=\ \frac{2^d\:d}{(4 \pi)^d}(-m_f^2)^\nu\hspace{-3mm}\smashoperator[r]{\int\limits_{\substack{%
    x_A,x_B,\\x_C,x_D}}}\left[(-\partial^\mu_{(A)}\partial_\mu^{(C)}+m_f^2)\delta(x_A-x_B)\delta(x_C-x_D) \right]|x_B-x_C|^{-\nu}|x_D-x_A|^{-\nu}\\
   &\hspace{3cm}\times K_\nu\left(\sqrt{-m_f^2}|x_B-x_C|\right)Y(x_C)K_\nu\left(\sqrt{-m_f^2}|x_D-x_A|\right)Y(x_A)\,,
\end{aligned}
\end{equation}
with $\nu=d/2-1$. The product of two Bessel functions of the same $\nu$ order, $K_\nu(z)\,K_\nu(\zeta)$, can be written as a $t$-integral over a single $\nu$-order Bessel function, $K_\nu\left(z\zeta/t\right)$,
with an appropriate weighted function.  To be specific, we utilise the integral identity~\cite{KBessel},
\begin{equation}
    K_\nu(z)\, K_\nu(\zeta)=\frac{1}{2}\int_0^\infty\frac{dt}{t}\:\exp\left(-\frac{t^2}{2}-\frac{z^2+\zeta^2}{2t}\right)K_\nu\left(\frac{z\zeta}{t}\right)\,.
\end{equation}
Likewise, by making use of the differential recursive relation,
\begin{equation}
    \frac{\partial K_\nu(z)}{\partial z}\ =\ -K_{\nu-1}(z)-\frac{\nu}{z}K_\nu(z)\,,
\end{equation}
and the recurrence identities for consecutive neighbours
\begin{equation}
    K_\nu(z)\ =\ K_{\nu+2}(z)\: -\: \frac{2(\nu+1)}{z}K_{\nu+1}(z)\ =\ K_{\nu-2}(z)\: +\: \frac{2(\nu-1)}{z}K_{\nu-1}(z)\,,
\end{equation}
one can obtain the wavefunction and mass counterterms listed in \eqref{eq:CTYuk}.

\vfill\eject

\vspace{1cm}

\bibliography{references}

\providecommand{\href}[2]{#2}\begingroup\raggedright\begin{thebibliography}{10}

\bibitem{DeWitt:1967ub}
B.S.~DeWitt, \emph{{Quantum Theory of Gravity. 2. The Manifestly Covariant
  Theory}}, \href{https://doi.org/10.1103/PhysRev.162.1195}{\emph{Phys. Rev.}
  {\bfseries 162} (1967) 1195}.

\bibitem{Gaillard:1985uh}
M.K.~Gaillard, \emph{{The Effective One Loop Lagrangian With Derivative
  Couplings}}, \href{https://doi.org/10.1016/0550-3213(86)90264-6}{\emph{Nucl.
  Phys. B} {\bfseries 268} (1986) 669}.

\bibitem{Pilaftsis:1996fh}
A.~Pilaftsis, \emph{{Generalized Pinch Technique and the Background Field
  Method in General Gauges}},
  \href{https://doi.org/10.1016/S0550-3213(96)00686-4}{\emph{Nucl. Phys. B}
  {\bfseries 487} (1997) 467}
  [\href{https://arxiv.org/abs/hep-ph/9607451}{{\ttfamily hep-ph/9607451}}].

\bibitem{Cornwall:2010upa}
J.M.~Cornwall, J.~Papavassiliou and D.~Binosi, \emph{{The Pinch Technique and
  its Applications to Non-Abelian Gauge Theories}}, Cambridge University Press
  (12, 2010).

\bibitem{Binosi:2009qm}
D.~Binosi and J.~Papavassiliou, \emph{{Pinch Technique: Theory and
  Applications}},
  \href{https://doi.org/10.1016/j.physrep.2009.05.001}{\emph{Phys. Rept.}
  {\bfseries 479} (2009) 1} [\href{https://arxiv.org/abs/0909.2536}{{\ttfamily
  0909.2536}}].

\bibitem{Honerkamp:1971sh}
J.~Honerkamp, \emph{{Chiral multiloops}},
  \href{https://doi.org/10.1016/0550-3213(72)90299-4}{\emph{Nucl. Phys. B}
  {\bfseries 36} (1972) 130}.

\bibitem{Ecker:1972tii}
G.~Ecker and J.~Honerkamp, \emph{{Covariant perturbation theory and chiral
  superpropagators}},
  \href{https://doi.org/10.1016/0370-2693(72)90074-3}{\emph{Phys. Lett. B}
  {\bfseries 42} (1972) 253}.

\bibitem{Alvarez-Gaume:1981exa}
L.~Alvarez-Gaume, D.Z.~Freedman and S.~Mukhi, \emph{{The Background Field
  Method and the Ultraviolet Structure of the Supersymmetric Nonlinear Sigma
  Model}}, \href{https://doi.org/10.1016/0003-4916(81)90006-3}{\emph{Annals
  Phys.} {\bfseries 134} (1981) 85}.

\bibitem{Dixon:1990pc}
L.J.~Dixon, V.~Kaplunovsky and J.~Louis, \emph{{Moduli dependence of string
  loop corrections to gauge coupling constants}},
  \href{https://doi.org/10.1016/0550-3213(91)90490-O}{\emph{Nucl. Phys. B}
  {\bfseries 355} (1991) 649}.

\bibitem{Vilkovisky:1984st}
G.A.~Vilkovisky, \emph{{The Unique Effective Action in Quantum Field Theory}},
  \href{https://doi.org/10.1016/0550-3213(84)90228-1}{\emph{Nucl. Phys. B}
  {\bfseries 234} (1984) 125}.

\bibitem{DeWitt:1985sg}
B.S.~DeWitt, \emph{{The Effective Action}},  in \emph{{Les Houches School of
  Theoretical Physics: Architecture of Fundamental Interactions at Short
  Distances: Proceedings, Les Houches 44th Summer School of Theoretical
  Physics: Les Houches, France, July 1-August 8}}, pp.~1023--1058, 1985.

\bibitem{Barvinsky:1985an}
A.O.~Barvinsky and G.A.~Vilkovisky, \emph{{The Generalized Schwinger-Dewitt
  Technique in Gauge Theories and Quantum Gravity}},
  \href{https://doi.org/10.1016/0370-1573(85)90148-6}{\emph{Phys. Rept.}
  {\bfseries 119} (1985) 1}.

\bibitem{Ellicott:1987ir}
P.~Ellicott and D.J.~Toms, \emph{{On the New Effective Action in Quantum Field
  Theory}}, \href{https://doi.org/10.1016/0550-3213(89)90579-8}{\emph{Nucl.
  Phys. B} {\bfseries 312} (1989) 700}.

\bibitem{Burgess:1987zi}
C.P.~Burgess and G.~Kunstatter, \emph{{On the Physical Interpretation of the
  Vilkovisky-de Witt Effective Action}},
  \href{https://doi.org/10.1142/S0217732387001117}{\emph{Mod. Phys. Lett. A}
  {\bfseries 2} (1987) 875}.

\bibitem{Odintsov:1989gz}
S.D.~Odintsov, \emph{{The Parametrization Invariant and Gauge Invariant
  Effective Actions in Quantum Field Theory}}, {\emph{Fortsch. Phys.}
  {\bfseries 38} (1990) 371}.

\bibitem{Kamenshchik:2014waa}
A.Y.~Kamenshchik and C.F.~Steinwachs, \emph{{Question of quantum equivalence
  between Jordan frame and Einstein frame}},
  \href{https://doi.org/10.1103/PhysRevD.91.084033}{\emph{Phys. Rev. D}
  {\bfseries 91} (2015) 084033}
  [\href{https://arxiv.org/abs/1408.5769}{{\ttfamily 1408.5769}}].

\bibitem{Burns:2016ric}
D.~Burns, S.~Karamitsos and A.~Pilaftsis, \emph{{Frame-Covariant Formulation of
  Inflation in Scalar-Curvature Theories}},
  \href{https://doi.org/10.1016/j.nuclphysb.2016.04.036}{\emph{Nucl. Phys. B}
  {\bfseries 907} (2016) 785}
  [\href{https://arxiv.org/abs/1603.03730}{{\ttfamily 1603.03730}}].

\bibitem{Karamitsos:2017elm}
S.~Karamitsos and A.~Pilaftsis, \emph{{Frame Covariant Nonminimal Multifield
  Inflation}},
  \href{https://doi.org/10.1016/j.nuclphysb.2017.12.015}{\emph{Nucl. Phys. B}
  {\bfseries 927} (2018) 219}
  [\href{https://arxiv.org/abs/1706.07011}{{\ttfamily 1706.07011}}].

\bibitem{Finn:2019aip}
K.~Finn, S.~Karamitsos and A.~Pilaftsis, \emph{{Frame Covariance in Quantum
  Gravity}}, \href{https://doi.org/10.1103/PhysRevD.102.045014}{\emph{Phys.
  Rev. D} {\bfseries 102} (2020) 045014}
  [\href{https://arxiv.org/abs/1910.06661}{{\ttfamily 1910.06661}}].

\bibitem{Falls:2018olk}
K.~Falls and M.~Herrero-Valea, \emph{{Frame (In)equivalence in Quantum Field
  Theory and Cosmology}},
  \href{https://doi.org/10.1140/epjc/s10052-019-7070-3}{\emph{Eur. Phys. J. C}
  {\bfseries 79} (2019) 595}
  [\href{https://arxiv.org/abs/1812.08187}{{\ttfamily 1812.08187}}].

\bibitem{Jenkins:2023bls}
E.E.~Jenkins, A.V.~Manohar, L.~Naterop and J.~Pag\`es, \emph{{Two loop
  renormalization of scalar theories using a geometric approach}},
  \href{https://doi.org/10.1007/JHEP02(2024)131}{\emph{JHEP} {\bfseries 02}
  (2024) 131} [\href{https://arxiv.org/abs/2310.19883}{{\ttfamily
  2310.19883}}].

\bibitem{Alminawi:2023qtf}
M.~Alminawi, I.~Brivio and J.~Davighi, \emph{{Jet Bundle Geometry of Scalar
  Field Theories}},  \href{https://arxiv.org/abs/2308.00017}{{\ttfamily
  2308.00017}}.

\bibitem{Craig:2023hhp}
N.~Craig and Y.-T.~Lee, \emph{{Effective Field Theories on the Jet Bundle}},
  \href{https://doi.org/10.1103/PhysRevLett.132.061602}{\emph{Phys. Rev. Lett.}
  {\bfseries 132} (2024) 061602}
  [\href{https://arxiv.org/abs/2307.15742}{{\ttfamily 2307.15742}}].

\bibitem{Assi:2023zid}
B.~Assi, A.~Helset, A.V.~Manohar, J.~Pag\`es and C.-H.~Shen, \emph{{Fermion
  geometry and the renormalization of the Standard Model Effective Field
  Theory}}, \href{https://doi.org/10.1007/JHEP11(2023)201}{\emph{JHEP}
  {\bfseries 11} (2023) 201}
  [\href{https://arxiv.org/abs/2307.03187}{{\ttfamily 2307.03187}}].

\bibitem{Alonso:2016oah}
R.~Alonso, E.E.~Jenkins and A.V.~Manohar, \emph{{Geometry of the Scalar
  Sector}}, \href{https://doi.org/10.1007/JHEP08(2016)101}{\emph{JHEP}
  {\bfseries 08} (2016) 101}
  [\href{https://arxiv.org/abs/1605.03602}{{\ttfamily 1605.03602}}].

\bibitem{Nagai:2019tgi}
R.~Nagai, M.~Tanabashi, K.~Tsumura and Y.~Uchida, \emph{{Symmetry and geometry
  in a generalized Higgs effective field theory: Finiteness of oblique
  corrections versus perturbative unitarity}},
  \href{https://doi.org/10.1103/PhysRevD.100.075020}{\emph{Phys. Rev. D}
  {\bfseries 100} (2019) 075020}
  [\href{https://arxiv.org/abs/1904.07618}{{\ttfamily 1904.07618}}].

\bibitem{Helset:2020yio}
A.~Helset, A.~Martin and M.~Trott, \emph{{The Geometric Standard Model
  Effective Field Theory}},
  \href{https://doi.org/10.1007/JHEP03(2020)163}{\emph{JHEP} {\bfseries 03}
  (2020) 163} [\href{https://arxiv.org/abs/2001.01453}{{\ttfamily
  2001.01453}}].

\bibitem{Cohen:2021ucp}
T.~Cohen, N.~Craig, X.~Lu and D.~Sutherland, \emph{{Unitarity violation and the
  geometry of Higgs EFTs}},
  \href{https://doi.org/10.1007/JHEP12(2021)003}{\emph{JHEP} {\bfseries 12}
  (2021) 003} [\href{https://arxiv.org/abs/2108.03240}{{\ttfamily
  2108.03240}}].

\bibitem{Talbert:2022unj}
J.~Talbert, \emph{{The geometric \ensuremath{\nu}SMEFT: operators and
  connections}}, \href{https://doi.org/10.1007/JHEP01(2023)069}{\emph{JHEP}
  {\bfseries 01} (2023) 069}
  [\href{https://arxiv.org/abs/2208.11139}{{\ttfamily 2208.11139}}].

\bibitem{Helset:2022tlf}
A.~Helset, E.E.~Jenkins and A.V.~Manohar, \emph{{Geometry in scattering
  amplitudes}}, \href{https://doi.org/10.1103/PhysRevD.106.116018}{\emph{Phys.
  Rev. D} {\bfseries 106} (2022) 116018}
  [\href{https://arxiv.org/abs/2210.08000}{{\ttfamily 2210.08000}}].

\bibitem{Fumagalli:2020ody}
J.~Fumagalli, M.~Postma and M.~Van Den~Bout, \emph{{Matching and running
  sensitivity in non-renormalizable inflationary models}},
  \href{https://doi.org/10.1007/JHEP09(2020)114}{\emph{JHEP} {\bfseries 09}
  (2020) 114} [\href{https://arxiv.org/abs/2005.05905}{{\ttfamily
  2005.05905}}].

\bibitem{Cohen:2023ekv}
T.~Cohen, X.~Lu and D.~Sutherland, \emph{{On Amplitudes and Field
  Redefinitions}},  \href{https://arxiv.org/abs/2312.06748}{{\ttfamily
  2312.06748}}.

\bibitem{Loisa:2024xuk}
E.~Loisa and J.~Talbert, \emph{{Froggatt-Nielsen Meets the SMEFT}},
  \href{https://arxiv.org/abs/2402.16940}{{\ttfamily 2402.16940}}.

\bibitem{Buchmuller:1985jz}
W.~Buchmuller and D.~Wyler, \emph{{Effective Lagrangian Analysis of New
  Interactions and Flavor Conservation}},
  \href{https://doi.org/10.1016/0550-3213(86)90262-2}{\emph{Nucl. Phys. B}
  {\bfseries 268} (1986) 621}.

\bibitem{Grzadkowski:2010es}
B.~Grzadkowski, M.~Iskrzynski, M.~Misiak and J.~Rosiek, \emph{{Dimension-Six
  Terms in the Standard Model Lagrangian}},
  \href{https://doi.org/10.1007/JHEP10(2010)085}{\emph{JHEP} {\bfseries 10}
  (2010) 085} [\href{https://arxiv.org/abs/1008.4884}{{\ttfamily 1008.4884}}].

\bibitem{Dedes:2021abc}
A.~Dedes and K.~Mantzaropoulos, \emph{{Universal scalar leptoquark action for
  matching}}, \href{https://doi.org/10.1007/JHEP11(2021)166}{\emph{JHEP}
  {\bfseries 11} (2021) 166}
  [\href{https://arxiv.org/abs/2108.10055}{{\ttfamily 2108.10055}}].

\bibitem{Alonso:2015fsp}
R.~Alonso, E.E.~Jenkins and A.V.~Manohar, \emph{{A Geometric Formulation of
  Higgs Effective Field Theory: Measuring the Curvature of Scalar Field
  Space}}, \href{https://doi.org/10.1016/j.physletb.2016.01.041}{\emph{Phys.
  Lett. B} {\bfseries 754} (2016) 335}
  [\href{https://arxiv.org/abs/1511.00724}{{\ttfamily 1511.00724}}].

\bibitem{Alonso:2023upf}
R.~Alonso, \emph{{A primer on Higgs Effective Field Theory with Geometry}},
  \href{https://arxiv.org/abs/2307.14301}{{\ttfamily 2307.14301}}.

\bibitem{Isidori:2023pyp}
G.~Isidori, F.~Wilsch and D.~Wyler, \emph{{The Standard Model effective field
  theory at work}},  \href{https://arxiv.org/abs/2303.16922}{{\ttfamily
  2303.16922}}.

\bibitem{Finn:2020nvn}
K.~Finn, S.~Karamitsos and A.~Pilaftsis, \emph{{Frame covariant formalism for
  fermionic theories}},
  \href{https://doi.org/10.1140/epjc/s10052-021-09360-w}{\emph{Eur. Phys. J. C}
  {\bfseries 81} (2021) 572}
  [\href{https://arxiv.org/abs/2006.05831}{{\ttfamily 2006.05831}}].

\bibitem{Gattus:2023gep}
V.~Gattus and A.~Pilaftsis, \emph{{Minimal supergeometric quantum field
  theories}}, \href{https://doi.org/10.1016/j.physletb.2023.138234}{\emph{Phys.
  Lett. B} {\bfseries 846} (2023) 138234}
  [\href{https://arxiv.org/abs/2307.01126}{{\ttfamily 2307.01126}}].

\bibitem{Gattus:2024spj}
V.~Gattus and A.~Pilaftsis, \emph{{Supergeometric Approach to Quantum Field
  Theory}},  \href{https://arxiv.org/abs/2404.13107}{{\ttfamily 2404.13107}}.

\bibitem{DeWitt:2012mdz}
B.S.~DeWitt, \emph{{Supermanifolds}}, Cambridge Monographs on Mathematical
  Physics, Cambridge Univ. Press, Cambridge, UK (5, 2012),
  \href{https://doi.org/10.1017/CBO9780511564000}{10.1017/CBO9780511564000}.

\bibitem{tHooft:1973bhk}
G.~'t~Hooft, \emph{{An algorithm for the poles at dimension four in the
  dimensional regularization procedure}},
  \href{https://doi.org/10.1016/0550-3213(73)90263-0}{\emph{Nucl. Phys. B}
  {\bfseries 62} (1973) 444}.

\bibitem{dewitt1992supermanifolds}
B.~DeWitt, \emph{Supermanifolds}, Cambridge University Press (1992).

\bibitem{Kluth:2023sey}
Y.~Kluth, P.~Millington and P.~Saffin, \emph{{Renormalization group flows from
  the Hessian geometry of quantum effective actions}},
  \href{https://arxiv.org/abs/2311.17199}{{\ttfamily 2311.17199}}.

\bibitem{Schwinger:1963re}
J.S.~Schwinger, \emph{{Quantized gravitational field}},
  \href{https://doi.org/10.1103/PhysRev.130.1253}{\emph{Phys. Rev.} {\bfseries
  130} (1963) 1253}.

\bibitem{Yepez:2011bw}
J.~Yepez, \emph{{Einstein's vierbein field theory of curved space}},
  \href{https://arxiv.org/abs/1106.2037}{{\ttfamily 1106.2037}}.

\bibitem{REBHAN1987832}
A.~Rebhan, \emph{The vilkovisky-dewitt effective action and its application to
  yang-mills theories},
  \href{https://doi.org/https://doi.org/10.1016/0550-3213(87)90241-0}{\emph{Nuclear
  Physics B} {\bfseries 288} (1987) 832}.

\bibitem{Pilaftsis:2022las}
A.~Pilaftsis, K.~Finn, V.~Gattus and S.~Karamitsos, \emph{{Geometrising the
  Micro-Cosmos on a Supermanifold}},
  \href{https://doi.org/10.22323/1.406.0080}{\emph{PoS} {\bfseries CORFU2021}
  (2022) 080} [\href{https://arxiv.org/abs/2204.00123}{{\ttfamily
  2204.00123}}].

\bibitem{Kim:2006th}
C.~Kim, \emph{{Effective action for the scalar field theory with higher
  vertices}}, \href{https://doi.org/10.1103/PhysRevD.74.067702}{\emph{Phys.
  Rev. D} {\bfseries 74} (2006) 067702}
  [\href{https://arxiv.org/abs/hep-th/0609178}{{\ttfamily hep-th/0609178}}].

\bibitem{Cornwall:1974vz}
J.M.~Cornwall, R.~Jackiw and E.~Tomboulis, \emph{{Effective Action for
  Composite Operators}},
  \href{https://doi.org/10.1103/PhysRevD.10.2428}{\emph{Phys. Rev. D}
  {\bfseries 10} (1974) 2428}.

\bibitem{Zinn-Justin:2002ecy}
J.~Zinn-Justin, \emph{{Quantum field theory and critical phenomena}},
  {\emph{Int. Ser. Monogr. Phys.} {\bfseries 113} (2002) 1}.

\bibitem{Barvinsky:2021ijq}
A.O.~Barvinsky and W.~Wachowski, \emph{{Heat kernel expansion for higher order
  minimal and nonminimal operators}},
  \href{https://doi.org/10.1103/PhysRevD.105.065013}{\emph{Phys. Rev. D}
  {\bfseries 105} (2022) 065013}
  [\href{https://arxiv.org/abs/2112.03062}{{\ttfamily 2112.03062}}].

\bibitem{Alexandrov:1996gu}
S.~Alexandrov and D.~Vassilevich, \emph{{Heat kernel for nonminimal operators
  on a Kahler manifold}}, \href{https://doi.org/10.1063/1.531736}{\emph{J.
  Math. Phys.} {\bfseries 37} (1996) 5715}
  [\href{https://arxiv.org/abs/hep-th/9601090}{{\ttfamily hep-th/9601090}}].

\bibitem{Avramidi:2000isc}
I.~Avramidi and T.~Branson, \emph{{Heat kernel asymptotics of operators with
  nonLaplace principal part}},
  \href{https://doi.org/10.1142/S0129055X01000892}{\emph{Rev. Math. Phys.}
  {\bfseries 13} (2001) 847}
  [\href{https://arxiv.org/abs/math-ph/9905001}{{\ttfamily math-ph/9905001}}].

\bibitem{Gusynin}
V.~Gusynin, \emph{Asymptotics of the heat kernel for nonminimal differential
  operators}, {\emph{Ukr Math J} {\bfseries 43} (1991) 1432–1441}.

\bibitem{Moss:2013cba}
I.G.~Moss and D.J.~Toms, \emph{{Invariants of the heat equation for non-minimal
  operators}}, \href{https://doi.org/10.1088/1751-8113/47/21/215401}{\emph{J.
  Phys. A} {\bfseries 47} (2014) 215401}
  [\href{https://arxiv.org/abs/1311.5445}{{\ttfamily 1311.5445}}].

\bibitem{Gusynin:1990ek}
V.P.~Gusynin, E.V.~Gorbar and V.V.~Romankov, \emph{{Heat kernel expansion for
  nonminimal differential operators and manifolds with torsion}},
  \href{https://doi.org/10.1016/0550-3213(91)90568-I}{\emph{Nucl. Phys. B}
  {\bfseries 362} (1991) 449}.

\bibitem{Ahmadiniaz:2022yam}
N.~Ahmadiniaz, J.P.~Edwards, C.~Lopez-Arcos, M.A.~Lopez-Lopez, C.M.~Mata,
  J.~Nicasio et~al., \emph{{Summing Feynman diagrams in the worldline
  formalism}}, \href{https://doi.org/10.22323/1.416.0052}{\emph{PoS} {\bfseries
  LL2022} (2022) 052} [\href{https://arxiv.org/abs/2208.06585}{{\ttfamily
  2208.06585}}].

\bibitem{Edwards:2022dbd}
J.P.~Edwards, C.M.~Mata and C.~Schubert, \emph{{One-loop amplitudes in the
  worldline formalism}},
  \href{https://doi.org/10.1088/1402-4896/ac6a1e}{\emph{Phys. Scripta}
  {\bfseries 97} (2022) 064002}
  [\href{https://arxiv.org/abs/2201.12457}{{\ttfamily 2201.12457}}].

\bibitem{Manzo:2024gto}
L.~Manzo, \emph{{Worldline approach for spinor fields in manifolds with
  boundaries}},  \href{https://arxiv.org/abs/2403.00218}{{\ttfamily
  2403.00218}}.

\bibitem{Abel:2023ieo}
S.~Abel and L.~Heurtier, \emph{{Exact Schwinger Proper Time Renormalisation}},
  \href{https://arxiv.org/abs/2311.12102}{{\ttfamily 2311.12102}}.

\bibitem{Alonso:2022ffe}
R.~Alonso and M.~West, \emph{{On the effective action for scalars in a general
  manifold to any loop order}},
  \href{https://doi.org/10.1016/j.physletb.2023.137937}{\emph{Phys. Lett. B}
  {\bfseries 841} (2023) 137937}
  [\href{https://arxiv.org/abs/2207.02050}{{\ttfamily 2207.02050}}].

\bibitem{Dupays:2021osq}
L.~Dupays and J.-C.~Pain, \emph{{Closed forms of the Zassenhaus formula}},
  \href{https://doi.org/10.1088/1751-8121/acc8a3}{\emph{J. Phys. A} {\bfseries
  56} (2023) 255202} [\href{https://arxiv.org/abs/2107.01204}{{\ttfamily
  2107.01204}}].

\bibitem{Kimura:2017xxz}
T.~Kimura, \emph{{Explicit Description of the Zassenhaus Formula}},
  \href{https://doi.org/10.1093/ptep/ptx044}{\emph{PTEP} {\bfseries 2017}
  (2017) 041A03} [\href{https://arxiv.org/abs/1702.04681}{{\ttfamily
  1702.04681}}].

\bibitem{Helset:2022pde}
A.~Helset, E.E.~Jenkins and A.V.~Manohar, \emph{{Renormalization of the
  Standard Model Effective Field Theory from geometry}},
  \href{https://doi.org/10.1007/JHEP02(2023)063}{\emph{JHEP} {\bfseries 02}
  (2023) 063} [\href{https://arxiv.org/abs/2212.03253}{{\ttfamily
  2212.03253}}].

\bibitem{Jenkins:2023rtg}
E.E.~Jenkins, A.V.~Manohar, L.~Naterop and J.~Pag\`es, \emph{{An algebraic
  formula for two loop renormalization of scalar quantum field theory}},
  \href{https://doi.org/10.1007/JHEP12(2023)165}{\emph{JHEP} {\bfseries 12}
  (2023) 165} [\href{https://arxiv.org/abs/2308.06315}{{\ttfamily
  2308.06315}}].

\bibitem{Vilkoviskii:1984un}
G.A.~Vilkoviskii, \emph{{The gospel according to DeWitt}}, .

\bibitem{Bounakis:2017fkv}
M.~Bounakis and I.G.~Moss, \emph{{Gravitational corrections to Higgs
  potentials}}, \href{https://doi.org/10.1007/JHEP04(2018)071}{\emph{JHEP}
  {\bfseries 04} (2018) 071}
  [\href{https://arxiv.org/abs/1710.02987}{{\ttfamily 1710.02987}}].

\bibitem{Conway:1999}
J.B.~Conway, \emph{{A Course in Operator Theory}}, American Mathematical
  Society (1999).

\bibitem{Evans:1998pd}
T.S.~Evans, \emph{{Regularization schemes and the multiplicative anomaly}},
  \href{https://doi.org/10.1016/S0370-2693(99)00503-1}{\emph{Phys. Lett. B}
  {\bfseries 457} (1999) 127}
  [\href{https://arxiv.org/abs/hep-th/9803184}{{\ttfamily hep-th/9803184}}].

\bibitem{Haag:1974qh}
R.~Haag, J.T.~Lopuszanski and M.~Sohnius, \emph{{All Possible Generators of
  Supersymmetries of the S-Matrix}},
  \href{https://doi.org/10.1016/0550-3213(75)90279-5}{\emph{Nucl. Phys. B}
  {\bfseries 88} (1975) 257}.

\bibitem{Nash:1954}
J.~Nash, \emph{{C1 Isometric Imbeddings}}, {\emph{Annals of Mathematics}
  {\bfseries 60} (1954) 383}.

\bibitem{Kuiper:1955a}
N.H.~Kuiper, \emph{{On C1-Isometric Imbeddings. I}},
  \href{https://doi.org/https://doi.org/10.1016/S1385-7258(55)50075-8}{\emph{Indagationes
  Mathematicae (Proceedings)} {\bfseries 58} (1955) 545}.

\bibitem{Kuiper:1955b}
N.H.~Kuiper, \emph{{On C1-Isometric Imbeddings. II}},
  \href{https://doi.org/https://doi.org/10.1016/S1385-7258(55)50093-X}{\emph{Indagationes
  Mathematicae (Proceedings)} {\bfseries 58} (1955) 683}.

\bibitem{Nash:1956}
J.~Nash, \emph{{The Imbedding Problem for Riemannian Manifolds}}, {\emph{Annals
  of Mathematics} {\bfseries 63} (1956) 20}.

\bibitem{Weinberg:1968de}
S.~Weinberg, \emph{{Nonlinear realizations of chiral symmetry}},
  \href{https://doi.org/10.1103/PhysRev.166.1568}{\emph{Phys. Rev.} {\bfseries
  166} (1968) 1568}.

\bibitem{Cornwall:1974km}
J.M.~Cornwall, D.N.~Levin and G.~Tiktopoulos, \emph{{Derivation of Gauge
  Invariance from High-Energy Unitarity Bounds on the S Matrix}},
  \href{https://doi.org/10.1103/PhysRevD.10.1145}{\emph{Phys. Rev. D}
  {\bfseries 10} (1974) 1145}.

\bibitem{Witten:2012bg}
E.~Witten, \emph{{Notes On Supermanifolds and Integration}},
  \href{https://doi.org/10.4310/PAMQ.2019.v15.n1.a1}{\emph{Pure Appl. Math.
  Quart.} {\bfseries 15} (2019) 3}
  [\href{https://arxiv.org/abs/1209.2199}{{\ttfamily 1209.2199}}].

\bibitem{Noja:2018edj}
S.~Noja, \emph{{Non-Projected Supermanifolds and Embeddings in Super
  Grassmannians}},
  \href{https://doi.org/10.3390/universe4110114}{\emph{Universe} {\bfseries 4}
  (2018) 114} [\href{https://arxiv.org/abs/1808.09817}{{\ttfamily
  1808.09817}}].

\bibitem{Bettadapura:2018fyz}
K.~Bettadapura, \emph{{Embeddings of Complex Supermanifolds}},
  \href{https://doi.org/10.4310/ATMP.2019.v23.n6.a1}{\emph{Adv. Theor. Math.
  Phys.} {\bfseries 23} (2019) 1427}
  [\href{https://arxiv.org/abs/1806.02763}{{\ttfamily 1806.02763}}].

\bibitem{Pascual:1984zb}
P.~Pascual and R.~Tarrach, \emph{{QCD: Renormalization for the Practitioner}},
  vol.~194, Springer-Verlag (1984).

\bibitem{KBessel}
M.~Abramowitz, \emph{Handbook of Mathematical Functions, With Formulas, Graphs,
  and Mathematical Tables,}, Dover Publications, Inc., USA (1974).

\end{thebibliography}\endgroup
\newpage

\end{document}